\documentclass[
twocolumn,
secnumarabic,
nofootinbib,
amsmath,amssymb,
aps,
pra,
longbibliography
]{revtex4-2}

\usepackage{amsmath}
\usepackage{centernot}
\usepackage{amsmath, amssymb}
\usepackage{amsthm}
\usepackage{mathrsfs}
\usepackage{braket,mleftright}
\usepackage{dsfont}
\usepackage{bbm}
\usepackage{float}
\usepackage{comment}
\usepackage{graphicx}
\usepackage{dcolumn}
\usepackage{bm}
\usepackage[colorlinks=true, urlcolor=blue, linkcolor=blue, citecolor=blue]{hyperref}
\usepackage{xcolor}
\usepackage{url}
\usepackage{lipsum}
\usepackage{mathtools}
\usepackage{easybmat}
\usepackage{diagbox}

\begin{document}

\title{Localization and Topological Properties of SU(3) Fermions in \\ non-Abelian Gauge Fields:
Color-Orbit Coupling and Color-Flip Fields}

\author{Bar Alluf}
\affiliation{School of Physics, Georgia Institute of Technology, Atlanta, 30332, USA}

\author{C. A. R. {S\'a} de Melo}
\affiliation{School of Physics, Georgia Institute of Technology, Atlanta, 30332, USA}
\date{\today}

\begin{abstract}
The interplay between disorder, gauge fields, and internal degrees of freedom fundamentally affects localization and topological properties of quantum many-body systems.
Motivated by recent experimental realizations of synthetic non-Abelian gauge fields for SU(3) colored fermions, we investigate their localization and topological properties in one-dimensional bichromatic optical lattices consisting of strong and weak laser beams.
Describing the non-Abelian gauge field in terms of color-orbit coupling and color-flip (Rabi) fields, we obtain a tight-binding description of trapped SU(3) colored fermions corresponding to a generalized three-color Aubry-Andr\'e model.
We show that the presence of color-orbit coupling and color-flip (Rabi) fields explicitly breaks the conventional self-duality of a simple three-color Aubry-Andr\'e system.
This duality breaking generates mobility regions across the energy spectrum, demonstrating that non-Abelian fields 
can either enhance or hinder color localization.
Through exact diagonalization, density of states evaluations, and finite-size scaling of the inverse participation ratio, we obtain phase diagrams identifying regions of extended or localized bulk states. 
Furthermore, due to the presence of color-orbit coupling and color-flip (Rabi) fields, we observe the existence of edge-states which possess topological properties. We develop an exact mapping from our one-dimensional disorder model into a two-dimensional color Harper model with a fictitious magnetic flux ratio and a fictional dimension controlled by the phase of the weak laser beam. Using this mapping, we evaluate topological invariants, such as the charge-charge Chern number, associated with edge states that emerge in energy gaps of this system, thus revealing the topological insulating nature of several gapped phases. 
Lastly, we identify that these 
topological color-insulator phases can energetically neighbor three different configurations: two extended phases, two localized phases, or one extended and one localized phase, in sharp contrast to conventional topological insulators, where the topological insulating phase always neighbors two extended (conducting) phases.
\end{abstract}

\maketitle

\section{Introduction}
\label{sec:introduction}

Disorder plays a significant role in
condensed matter and atomic systems.
On the condensed matter side, disorder 
is omnipresent producing 
localization in superconducting qubits~\cite{martinis-2017}, 
in Majorana~\cite{mirlin-2019}
and 
Weyl fermions~\cite{fritz-2019}, 
as well as in hard-sphere bosons~\cite{meng-1992}.
On the atomic physics realm, 
disorder produces
matter-wave localization in optical lattices~\cite{lewenstein-2005, castin-2005}, affects collective 
excitations~\cite{graham-2007},
and reduces the condensation 
temperature of  superfluids~\cite{zobay-2006, 
sa-de-melo-2011}.

In ultracold atoms, the effects of disorder have been experimentally studied in Bose~\cite{aspect-2007, inguscio-2008, widera-2020} and Fermi~\cite{widera-2024a, widera-2024b} systems.
Further experimental exploration of the impact of disorder in ultracold quantum gases was achieved in $^{40}{\rm K}$ (fermion)~\cite{aidelsburger-2021a, aidelsburger-2021b}, 
$^{87}{\rm Rb}$ (boson)~\cite{spielman-2020},
Feshbach molecules of $^{6}{\rm Li}_2$ (boson)~\cite{widera-2022},
and very recently in 
$^{84}{\rm Sr}$ (boson)~\cite{weld-2026}.

Topology is also fundamental in describing quantum phases such as the quantum Hall effect~\cite{hatsugai-1997, den-nijs-1982, kohmoto-1985, yasuhiro-1993} and topological insulators~\cite{mele-2005a, mele-2005b, shou-cheng-2006},
where gauge (magnetic) fields and spin-orbit coupling play important roles, respectively. 
The effects of artificial Abelian gauge (magnetic) fields were explored experimentally in ultracold bosonic atoms~\cite{spielman-2009, bloch-2013, goldman-2015}, while the effects of non-Abelian gauge (spin-orbit coupling and Rabi) fields were investigated for bosons~\cite{spielman-2011} and fermions~\cite{zwierlein-2012} using Raman beams. Available fermionic systems, such as $^6{\rm Li}$ and $^{40} {\rm K}$, are more prone to heating, and thus more difficult to study. 

The combined effect of artificial Abelian gauge (magnetic) fields and non-Abelian gauge (spin-orbit coupling and Rabi) fields in SU(2) Fermi systems was only studied theoretically~\cite{sa-de-melo-2019} due to experimental constraints. 
However, more recently, synthetic non-Abelian gauge fields were realized in ultracold $^{87}{\rm Sr}$, where SU(2) dynamics was studied at finite temperatures~\cite{wilkowski-2022}, and the simultaneous effects of disorder and non-Abelian (spin-orbit coupling and Rabi) fields were explored for SU(2) fermions in one-dimensional bichromatic lattices~\cite{sa-de-melo-2025}. These new experimental and theoretical developments allow for the exploration of topological phenomena in SU(2) ultracold fermions. Furthermore, the complex relationship between disorder and non-Abelian gauge fields has been studied in bosonic environments~\cite{sherman-2015, spielman-2020, sa-de-melo-2023}.

Although much is now known about disorder, topology, as well as Abelian and non-Abelian gauge fields in SU(2) fermions, the interplay of these quantities in fermions with three internal states, such as SU(2) pseudospin-1 or SU(3) systems, has been much less explored. 
Early work on SU(N) ultracold atoms of 
$^{173}{\rm Yb}$~\cite{takahashi-2007, ueda-2009, takahashi-2010, takahashi-2012, fallani-2014, folling-2016} and 
$^{87}{\rm Sr}$~\cite{killian-2010, schreck-2010, schreck-2011, killian-2014} opened the door for new developments, where three or more 
internal states play an important role in antiferromagnetic correlations~\cite{takahashi-2022}
and flavor-selective localization~\cite{fallani-2022}, 
while some theory work explored color superfluidity in the continuuum~\cite{sa-de-melo-2019b} and the quantum 
Hall effect in lattices~\cite{sa-de-melo-2022}.

Furthermore, very recently, the stability of $^{6}{\rm Li}$ ultracold fermions with three internal states was investigated~\cite{navon-2026}, quantum gas microscopy for three-color Hubbard systems was developed~\cite{schauss-2025},
non-Abelian gauge fields (color-orbit coupling) were implemented in $^{87}{\rm Sr}$, an SU(3) ultracold Fermi system~\cite{wilkowski-2025}, 
and collective excitations in SU(3) fermions with color-orbit and color-flip fields were investigated~\cite{sa-de-melo-2025b}.
Such achievements opened new avenues for theoretical and experimental investigations of three-internal-state and SU(3) systems. 

In this manuscript, we develop a theory and propose experiments to investigate the interplay of disorder and non-Abelian gauge fields in SU(3) colored fermions.
We investigate their localization and topological properties in one-dimensional bichromatic optical lattices consisting of strong and weak laser beams.
We describe non-Abelian gauge fields in terms of color-orbit coupling and color-flip (Rabi) terms, and obtain a tight-binding description of trapped SU(3) colored fermions corresponding to a generalized three-color Aubry-Andr\'e model.
We show that the presence of color-orbit coupling and color-flip (Rabi) fields explicitly breaks the conventional self-duality of a simple three-color Aubry-Andr\'e system.
This duality-breaking generates mobility regions across the energy spectrum, demonstrating that non-Abelian fields 
can either enhance or hinder color localization. We employ exact diagonalization techniques to obtain the energy spectrum, the eigenfunctions, and the density of states, and use finite-size scaling of the inverse participation ratio to obtain phase diagrams identifying regions of extended or localized bulk states.

In addition, we observe the existence of edge-states which possess topological properties,
due to the presence of color-orbit coupling and color-flip (Rabi) fields.
We identify an exact mapping between our one-dimensional disorder model and a two-dimensional color Harper model with a 
fictitious magnetic flux ratio and a fictional dimension controlled by the phase of the weak laser beam. Using this mapping, we obtain topological invariants, such as the charge-charge Chern number, associated with edge states that may emerge in energy gaps of this system, thus revealing the topological insulating nature of several gapped phases. 
Finally, we identify that these 
topological color-insulator phases can energetically neighbor three different configurations: two extended phases, two localized phases, or one extended and one localized phase, in sharp contrast to conventional topological insulators, where the topological insulating phase always neighbors two extended (conducting) phases.

The remainder of this manuscript is organized as follows.
In Sec.~\ref{sec:continuum-hamiltonian}, we discuss a continuum Hamiltonian describing proposed experiments, involving a strong and a weak laser beams, that produce a one-dimensional optical lattice with bichromatic disorder for SU(3) colored fermions. In Sec.~\ref{sec:lattice-hamiltonian}, we derive a tight-binding lattice Hamiltonian that describes both disorder and non-Abelian gauge (color-orbit coupling and color-flip) fields. In Sec.~\ref{sec:symmetries-of-the-three-color-hamiltonian}, we discuss the symmetries of our three-color lattice Hamiltonian, including a color-gauge symmetry. In Sec.~\ref{sec:energy-spectra-and-inverse-partition-ratio}, we analyze the energy spectrum, eigenstates, and the inverse participation ratio (IPR) to identify states that are localized or extended. 
In Sec.~\ref{sec:finite-size-scaling}, we perform a finite-size scaling analysis of the inverse participation ratio to verify the behavior of localized and extended states in 
the thermodynamic limit. In Sec.~\ref{sec:phase-diagrams}, we obtain 
phase diagrams for fixed disorder and varying 
color-orbit coupling and color-flip (Rabi) fields to show that non-Abelian gauge fields can either enhance or hinder localization.
In Sec.~\ref{sec:color-density-of-states}, 
we analyze the color density of states,
identify localized and extended wavefunctions via the 
IPR, and plot the filling factor as a function of the chemical potential.
In Sec.~\ref{sec:color-harper-model}, 
we map our one-dimensional disorder model into the two-dimensional color Harper model, using
the phase of the weaker laser beam to characterize a fictitious dimension. In Sec.~\ref{sec:topological-considerations},
we show that the aforementioned mapping allows
for identification of topological invariants associated with bulk color insulating phases and color edge states.
In Sec.~\ref{sec:conclusions}, we discuss our main findings.

\section{Continuum Hamiltonian}
\label{sec:continuum-hamiltonian}

\begin{figure}[tb]
\centering
\includegraphics[width=0.4\textwidth]{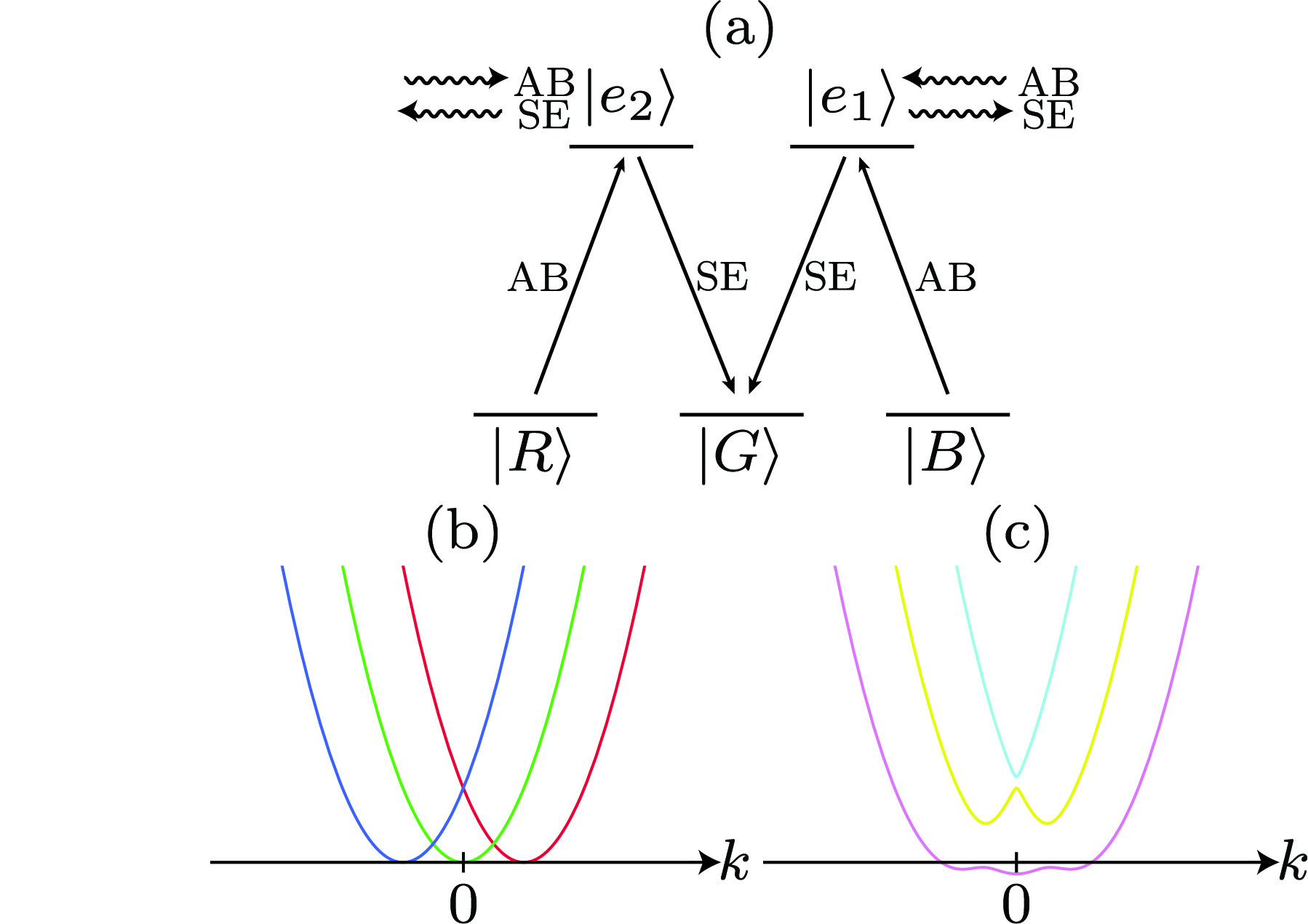}
\caption{(a) Raman coupling scheme for the three-color manyfold $\{|R\rangle,|G\rangle,|B\rangle\}$ of atoms in the continuum. 
Two excited states $|e_{1}\rangle$ and 
$|e_{2}\rangle$ mediate absorption (AB) and stimulated emission (SE) 
processes that generate color-orbit coupling 
and color-flip (Rabi) fields. The color-flip (Rabi) fields couple $|R\rangle \leftrightarrow |G\rangle$ and $|G\rangle \leftrightarrow |B\rangle$, 
but $|R\rangle \leftrightarrow |B\rangle$ transitions are absent. 
(b) Kinetic energy dispersions of the $RGB$ colors in the presence of 
color-orbit coupling but no color-flip fields: the $|R\rangle$ ($|B\rangle$) state is shifted by $k_T$ to the right (left), while $|G\rangle$ remains centered at $k=0$. 
(c) Eigenergies of kinetic energy operator (color-mixed band structure) in the presence of color-orbit coupling and color-flip fields.}
\label{fig:Raman-coupling}
\end{figure}

We consider the non-interacting one-dimensional Hamiltonian in real space 
\begin{equation}
    \label{Eq1}
    \mathcal H = \sum_{\alpha\beta} \int dx\, \psi^\dagger_{\alpha}(x) \left[ K^{\alpha \beta}(\hat k) + V^{\alpha \beta}(x) \right] \psi_\beta(x),
\end{equation}
where $\psi^\dag_{\alpha} (x)$ and $\psi_{\alpha} (x)$ are the creation and annihilation field operators, at position $x$,  for colors 
$\alpha,\beta \in \{R, G, B\}$.
In the presence of counter-propagating Raman beams illustrated in Fig.~\ref{fig:Raman-coupling}(a), 
the kinetic energy operator acquires a color-dependent momentum kick and is given by
\begin{equation}
    \label{eqn:Kinetic-Energy-Matrix}
    K^{\alpha\beta} ({\hat k}) = 
    \begin{pmatrix} 
    \varepsilon({\hat k} - k_T)  &          -h_x/\sqrt{2}         &             0              \\
            -h_x /\sqrt{2}           &   \varepsilon({\hat k})    &           -h_x /\sqrt{2}            \\
              0             &          -h_x /\sqrt{2}        &   \varepsilon({\hat k} + k_T)   \\
    \end{pmatrix},
\end{equation}
where $\varepsilon(\hat k) = \hbar^2 \hat{k}^2 / 2m$ is 
the kinetic energy, with $\hat k = -i \,d/dx$, $\hbar \hat k$
being the momentum operator and $h_x$ is the color-flip (Rabi) field. The R (B) states have a momentum kick to the right (left), while the G state has no momentum kick, see Fig.~\ref{fig:Raman-coupling}(b). We can also write the kinetic energy in compact form 
\begin{equation}
    \label{Eq2}
    K^{\alpha \beta} ({\hat k}) = \left[ \varepsilon(\hat k) \mathbf{I} -h_x \mathbf{S}_x - h_z(\hat k) \mathbf{S}_z + E_T \mathbf{S}_z^2 \right]^{\alpha \beta},
\end{equation}
where $h_z(\hat k) = \hbar^2 \hat k k_T / m$ is the coefficient describing color-orbit coupling, that is, the coupling between the momentum operator 
$\hbar \hat k$ and  $\mathbf{S}_z$. Here, $\mathbf{S}_j$ are the spin-1 matrices with $j= \{ x, y, z \}$ and
$E_T = \hbar^2 k_T^2 / 2m$ is the kinetic energy transferred by the momentum kick $k_T$ generated by the Raman beams.

The second term in Eq.~(\ref{Eq1}) is the bichromatic lattice potential 
\begin{equation}
V^{\alpha\beta}(x) = \left[ V_1(x) + V_{2}^{\alpha} (x) \right] \delta^{\alpha\beta}.
\end{equation}
The strong lattice is represented by the optical potential
\begin{equation}
\label{eqn:strong-optical-potential}
V_1(x) = -c_1 E_{R_1} \cos^2(k_1 x),
\end{equation}
which is taken to be state (color) independent.
In the energy scale $E_{R_1} = \hbar^2 k_1^2 / 2m$, $k_1 = 2\pi/\lambda_1$ is the wavenumber of the strong laser,  with $\lambda_1 = 813 \rm{nm}$ for $^{87}\text{Sr}$.
The weak lattice, described by the potential
\begin{equation}
\label{eqn:weak-optical-potential}
V_{2}^{\alpha}(x) = -c_{2}^{\alpha} E_{R_1}\cos^2(k_2 x + \phi_{2}^{\alpha}),
\end{equation}
is responsible for the bichromatic disorder and may be state (color) dependent through the coefficient $c_{2}^{\alpha}$ and phase $\phi_{2}^{\alpha}$.
The wavenumber of the weak lattice is $k_2 = R_\lambda k_1$, where 
$R_\lambda =  \lambda_1 / \lambda_2$ is the ratio between the wavelengths of the strong $(\lambda_1)$ and weak $(\lambda_2)$ optical lattices.
We choose standard wavelengths used for $^{87} \text{Sr}$, 
that is $\lambda_1 = 813{\rm nm}$ of the strong lattice and and $\lambda_2 = 1064{\rm nm}$ for the weak lattice,  
leading to $R_\lambda= 813/1064 \approx 0.764$.
We choose the coefficient $c_1$ to be positive and a free sign for $c_{2}^{\alpha}$, such that 
$\vert c_{2}^{\alpha} \vert/c_1 \ll 1$. 

Having discussed the continuum Hamiltonian, we present next an analysis of the tight-binding lattice Hamiltonian that emerges in the presence of color-orbit coupling and color-flip (Rabi) fields.

\section{Lattice Hamiltonian}
\label{sec:lattice-hamiltonian}

We study the tight-binding regime established by the strong lattice and use the lowest-band Wannier function $w(x-x_n) = w_n(x)$, centered around the site $x_n = n a$, where $a = \lambda_{1}/2$ is the lattice spacing. 
We write the field operators as 
$\psi^{\dagger}_\alpha(x) = \sum_n w_n^*(x) f^{\dagger}_{n\alpha}$
and
$\psi_\alpha(x) = \sum_n w_n(x) f_{n\alpha} $, 
where $f^{\dagger}_{n\alpha}$ and $f_{n\alpha}$ are fermionic creation and annihilation operators at lattice site $n$ with color $\alpha$.  

The resulting Hamiltonian in second-quantized
notation is
\begin{equation}
\label{eqn:hamiltonian}
\mathcal H = -\sum_{nm\alpha\beta} J_{nm}^{\alpha\beta} f^{\dagger}_{n\alpha} f_{m\beta} + \sum_{nm\alpha\beta} \Delta_{nm}^{\alpha\beta} f^{\dagger}_{n\alpha} f_{m\beta},
\end{equation}
corresponding to the generalization of the Aubry-Andr\'e Hamiltonian~\cite{andre-1980} for colored fermions (three internal states) with color-orbit coupling and 
color-flip fields. The indices $\{n, m\}$ and $\{\alpha \beta\}$ represent lattice sites and color states, respectively. 

The first term in Eq.~(\ref{eqn:hamiltonian}) 
has matrix elements
\begin{equation}
\label{eqs:Jnmalphabeta}
    J_{nm}^{\alpha\beta} = - \int dx\, w_n^*(x) \left[ K^{\alpha\beta}(\hat k) + V_1(x) \delta^{\alpha\beta}\right] w_m(x),
\end{equation}
containing the kinetic energy term $K^{\alpha\beta}(\hat k)$ defined in Eq.~(\ref{eqn:Kinetic-Energy-Matrix}) and the strong lattice potential $V_1 (x)$. Note that $K^{\alpha\beta}(\hat k)$ contains both diagonal and off-diagonal terms in the color indices $\alpha$
and $\beta$.
We briefly note that the color-diagonal terms 
$K^{\alpha \alpha} ({\hat k})$
in Eq.~(\ref{eqn:Kinetic-Energy-Matrix}), have the 
simple form 
\begin{equation}
J_{nm}^{RR} = e^{i \theta^{RR}_{nm}} J_{nm}(k_T = 0)
\end{equation} 
for the Red states with phase $\theta^{RR}_{nm} = k_T (x_n - x_m)$, 
\begin{equation}
J_{nm}^{GG} = e^{ i \theta^{GG}_{nm}} J_{nm}(k_T = 0)
\end{equation}
for the Green states with phase $\theta^{GG}_{nm} = 0$,
and
\begin{equation}
J_{nm}^{BB} = e^{ i \theta^{BB}_{nm}} J_{nm}(k_T = 0)
\end{equation}
for the Blue states with phase $\theta_{BB}^{nm} = - k_T (x_n - x_m)$.
In the expressions above, 
\begin{equation}
\label{eqs:Jnm-kT=0}
    J_{nm}(k_T = 0) = -\int dx\, w_n^*(x) \left[\frac{\hbar^2 \hat k^2}{2m} + V_1(x) \right] w_m(x).
\end{equation}
is a color-independent matrix element for zero color-orbit coupling ($k_T = 0$).
These three diagonal terms can be brought into a single expression
\begin{equation}
\label{eqn:jnm-alpha-alpha}
J_{nm}^{\alpha \alpha} = e^{i\theta^{\alpha \alpha}_{nm}} J_{nm}(k_T = 0),
\end{equation}
which differ from each other only by their phases 
\begin{equation}
\label{eqn:phasenm-alpha-alpha}
\theta^{\alpha \alpha}_{nm} =  k_T \gamma_{\alpha \alpha} \delta x_{nm} = k_T a \gamma_{\alpha \alpha}(n-m),
\end{equation}
with $\delta x_{nm} = x_n - x_m = (n-m)a$, and $\gamma_{R R}  = +1$, 
$\gamma_{GG} = 0$ and $\gamma_{BB} = -1$.

The second term in Eq.~(\ref{eqn:hamiltonian}) is 
associated with the disorder and is controlled 
by the weak-lattice potential $V_2^\alpha (x)$ leading to 
\begin{equation}
\label{eqn:DeltaabnmV2}
\Delta_{nm}^{\alpha\beta} = \int dx\, w_n^*(x) V_2^{\alpha}(x) w_m(x) \delta^{\alpha\beta}.
\end{equation}

We begin our analysis of the lattice Hamiltonian given in 
Eq.~(\ref{eqn:hamiltonian}) by investigating the deep-lattice regime, where a tight-binding approximation is reasonable. 
For a deep lattice, in the vicinity of the minimum located at $x_n$, the strong lattice potential is approximated by the harmonic potential
\begin{equation}
V_1(x) \approx -c_1 E_{R_1} + \frac{m\omega^2}{2}(x-x_n)^2,
\end{equation}
and the Wannier function is approximated by 
\begin{equation}
w_n(x) = \left[\frac{2}{\pi\xi^2}\right]^{1/4}
\exp{\left[ -\frac{(x-x_n)^2}{\xi^2} \right]},
\end{equation}
with $\xi = \sqrt{2\hbar/m\omega} \ll a$. 

Using these approximations, we can explicitly calculate the
matrix elements $J_{nm}^{\alpha \beta}$ and $\Delta_{nm}^{\alpha \beta}$ appearing in Eq.~(\ref{eqn:hamiltonian}). We begin our investigation by calculating  $J_{nm}^{\alpha \beta}$ first and
$\Delta_{nm}^{\alpha \beta}$ second.

In the deep-lattice regime, we calculate the 
diagonal matrix elements $J_{nm}^{\alpha \alpha}$
in Eq.~(\ref{eqn:jnm-alpha-alpha}) by computing
the color-independent contribution shown in Eq.~(\ref{eqs:Jnm-kT=0}), that is, 
\begin{equation}
\label{eqn:jnm-kt0}
J_{nm}(k_T = 0) \approx \left[c_1 E_{R_1} -\frac{\hbar\omega}{2} \right] \delta_{nm} + \left[ c_1 E_{R_1} -\frac{\hbar \omega}{2} \right] B_{nm},
\end{equation}
where $\hbar\omega = \sqrt{2c_1}E_{R_1}$ is the energy associated the harmonic trap frequency $\omega$ of the deep lattice. Here,  
\begin{equation}
\label{eqn:Bnmint}
B_{nm} = \int dx\, w^*_n(x) w_m(x) = \exp{\left[-\frac{(x_n-x_m)^2}{2\xi^2}\right]} 
\end{equation}
is the overlap integral between two 
Wannier functions, which can be also written as
\begin{equation}
\label{eqn:Bnmapp}
B_{nm} = \exp{\left[ - (n - m)^2 
\frac{\hbar \omega}{8 E_a} \right]}, 
\end{equation}
where $E_a = \hbar^2/(2m a^2)$ is a characteristic lattice energy for particle with mass $m$. We introduce $E_a$ to avoid confusion between the fermion mass and the site position labeled by the same symbol $m$.
The exponential dependence shows that onsite and
nearest-neighbor terms are the most important, since 
\begin{equation}
\frac{E_{R_1}} {E_a} = \frac{\frac{\hbar^2 k_1^2}{2m}} {\frac{\hbar^2}{2m a^2}} = {k_1^2 a^2} = \pi^2,
\end{equation}
where we used $k_1 = 2\pi/\lambda_1$ with $\lambda_1 = 2a$, leading to the ratio $\hbar\omega/E_{a} = \pi^2 \sqrt{2c_1}$. For a deep lattice,
with $c_1 = 10$, corresponding to $10 E_{R_1}$, gives
$\hbar\omega/E_{a} \approx 44.14 $, which is relatively large in comparison to one.

Using Eqs.~(\ref{eqn:jnm-alpha-alpha}), ~(\ref{eqn:phasenm-alpha-alpha}) and~(\ref{eqn:jnm-kt0}) leads to local $(n = m)$ and color-diagonal $(\alpha = \beta)$ matrix elements
\begin{equation}
J_{nn}^{RR} = J_{nn}^{GG} = J_{nn}^{BB} = 
J_{nn}(k_T = 0) = -\varepsilon_0,
\end{equation}
since the phases $\varphi_{nn}^{\alpha \alpha} = 0$.
Here, $\varepsilon_0$ is our energy reference, which  
in the deep lattice regime, becomes
\begin{equation}
\varepsilon_0 \approx -c_1 E_{R_1} + \frac{\hbar\omega}{2}
= - E_{R_1} \left( \frac{2 c_1 - \sqrt{2c_1}}{2} \right).
\end{equation}
The expression above is always negative for $c_1 \ge 1$ and naturally does not depend on $k_T$.

In addition to local color-diagonal terms, we have also local color-off-diagonal contributions $J_{nn}^{\alpha \beta}$ $(\alpha \ne \beta)$ arising from color-flip elements of $K^{\alpha \beta} ({\hat k})$ in Eq.~(\ref{eqn:Kinetic-Energy-Matrix}); we denote them as 
\begin{equation}
\label{eqs:J-hopping}
J^{\alpha\beta}_{nn} =
- \int dx\, w_n^*(x) \left\{ - h_x \left[ {\bf S}_x  \right]^{\alpha \beta}  \right\} w_n(x) = 
h_x \left[ {\bf S}_x \right]^{\alpha \beta},
\end{equation}
where ${\bf S}_x$ is a standard spin-1 matrix.
Note that local color flips only occur between neighboring states in the color ladder due to the matrix elements of ${\bf S}_x$, that is, $RG$ and $GB$ are coupled, but $RB$ are not.
As a result, the non-vanishing color-off-diagonal terms become
\begin{equation}
J^{RG}_{nn} = J^{GB}_{nn} = J^{GR}_{nn} = J^{BG}_{nn} 
= \frac{h_x}{\sqrt{2}}
\end{equation}
and are controlled by the color-flip field $h_x$, while the 
vanishing color-off-diagonal terms are
\begin{equation}
J^{RB}_{nn} = J^{BR}_{nn} = 0.
\end{equation}

Having discussed the local matrix elements $J_{nn}^{\alpha \beta}$, we analyze next the non-local terms $J_{nm}^{\alpha \beta}$ with 
$n \ne m$ in Eq.~(\ref{eqs:Jnmalphabeta}). The general tensor structure of $J_{nm}^{\alpha \beta}$ for $n \ne m$ is 
\begin{equation}
J_{nm}^{\alpha \beta} = J_{nm}^{\alpha \alpha} \delta_{\alpha \beta},
\end{equation}
where $J_{nm}^{\alpha \alpha}$ is inferred from Eq.~(\ref{eqn:jnm-alpha-alpha}) and (\ref{eqn:phasenm-alpha-alpha})
and corresponds physically to color-diagonal hopping energies between sites $n \ne m$.

We conclude our discussion of the matrix elements 
$J_{nm}^{\alpha \beta}$ by noting that higher order neighbors give exponentially small contributions in comparison to nearest neighbors. This implies that 
it is sufficient to consider $m = n \pm 1$ for 
the matrix elements $J_{nm}^{\alpha \beta}$ in the lattice
Hamiltonian of Eq.~(\ref{eqn:hamiltonian}).

For later use in Sec.~\ref{sec:energy-spectra-and-inverse-partition-ratio} and beyond, we define the nearest-neighbor hopping amplitude $J$ in the absence of color–orbit coupling as our energy unit. From Eqs.~(\ref{eqn:jnm-kt0}), (\ref{eqn:Bnmint}), and (\ref{eqn:Bnmapp}), the nearest-neighbor hopping term is
\begin{equation}
\label{eqs:Jnaturalenergy}
    J \equiv J_{n,n\pm 1}(k_T = 0) = \left[ c_1 E_{R_1} -\frac{\hbar\omega}{2} \right] B_{n, {n\pm 1}},
\end{equation}
where the dimensionless matrix element is 
\begin{equation}
\label{eqn:Bnnpm1}
B_{n, {n\pm 1}} = \exp{\left[ - 
\frac{\hbar \omega}{8 E_a} \right]}
= \exp{\left[ -\frac{\pi^2 \sqrt{2c_1}}{8} \right]},
\end{equation}
where $\hbar \omega = \sqrt{2c_1}E_{R_1}$. 
Thus, in terms of the recoil energy $E_{R_1}$ and the dimensionless parameter $c_1$ of the strong lattice potential $V_1 (x)$ shown 
in Eq.~(\ref{eqn:strong-optical-potential}), our energy scale becomes 
\begin{equation}
\label{eqn:hopping-unit}
J = \left(c_1 - \sqrt{\frac{c_1}{2}}\right) E_{R_1}
\exp{\left[ -\frac{\pi^2 \sqrt{2c_1}}{8} \right]}.
\end{equation}
In the deep lattice regime where $c_1 \gg 1$ or $c_1 E_{R_1} \gg \hbar \omega / 2$, the term $\sqrt{c_1/2}$ can be dropped.
However, for the $c_1$ in the realistic 
range $[10, 30]$, the term $\sqrt{c_1/2}$ varies from $22\%$ to $13\%$ of $c_1$. 
Thus, we keep this term in the expression of $J$ for realistic parameter ranges.
Note that the prefactor in the expression for $J$ increases with growing $c_1$, but the exponential decreases with increasing $\sqrt{c_1}$.

To finalize the description of
the lattice Hamiltonian in Eq. (\ref{eqn:hamiltonian}), we discuss next the matrix elements 
$\Delta_{nm}^{\alpha \beta}$ shown in Eq.~(\ref{eqn:DeltaabnmV2}) representing the bichromatic disorder produced by the weak optical
lattice potential $V_2^{\alpha} (x)$ defined
in Eq.~(\ref{eqn:weak-optical-potential}).
We begin by investigating the local terms $\Delta_{nn}^{\alpha\beta}$. Using the 
deep lattice approximation, the Wannier 
functions $w_n (x)$ discussed above, 
the trigonometric identity
$\cos^2(k_2x + \phi_2^{\alpha}) = 
\left[ 1 + \cos(2 k_2x + 2\phi_2^{\alpha}) \right]/2$,
and the relation $k_2 = R_{\lambda} k_1$,
the local disorder terms become
\begin{equation}
\Delta_{nn}^{\alpha\beta} = 
\chi_{nn}^{\alpha\beta}
+ \eta_{nn}^{\alpha \beta}.
\end{equation}
Here, the periodic contribution of the weak lattice potential $V_2 (x)$ is described by the first term
\begin{equation}
\chi_{nn}^{\alpha\beta} =
-\frac{c_2^{\alpha} E_{R_1}}{2}
\int dx\, w_n^*(x)
\cos(2 R_\lambda k_1x + 2\phi_2^{\alpha})
w_n (x) \delta^{\alpha\beta},
\end{equation}
while a local energy reference is identified with
the second term
\begin{equation}
\eta_{nn}^{\alpha\beta} =
-\frac{c_2^{\alpha} E_{R_1}}{2}
\int dx\, w_n^*(x)
w_n (x) \delta^{\alpha\beta}.
\end{equation}
Making the variable transformation 
$\widetilde x = x - x_n$, writing the Wannier functions as
$w_n(x) = w(x - x_n) = w(\widetilde x)$, applying the 
trigonometric identity
$\cos (\theta_1 + \theta_2) = 
\cos (\theta_1) \cos(\theta_2) 
-\sin(\theta_1) \sin(\theta_2) 
$
with $\theta_1 = 2 R_\lambda k_1 {\widetilde
x}$ and 
$\theta_2 =  2 R_\lambda k_1 x_n + 2\phi_2^{\alpha}$,
as well as using $x_n = a$, $k_1 = 2\pi/\lambda_1$ 
and $a = \lambda_1/2$ leads to 
the final expression for the local disorder 
\begin{equation}
\Delta_{nn}^{\alpha\beta} = \Delta^{\alpha\beta} \cos(2\pi R_\lambda n + \varphi^\alpha) + \eta^{\alpha\beta}.
\end{equation}
Here, we chose 
$\varphi^\alpha = 2\phi_2^\alpha \pm \pi$ to produce 
a positive amplitude of the position-dependent cosinusoidal modulation
\begin{equation}
\Delta^{\alpha\beta} = 
\frac{c^\alpha_2 E_{R_1}}{2}\int d\widetilde x \cos(2 R_\lambda k_1 \widetilde x) \vert w(\widetilde x) \vert^2 \delta^{\alpha\beta}
\end{equation}
coming from $\chi_{nn}^{\alpha \beta}$.
The position-independent term, coming from $\eta_{nn}^{\alpha \beta}$, describes the
color-dependent disorder reference energy
\begin{equation}
\eta^{\alpha\beta} = -\frac{c_2^{\alpha} E_{R_1}}{2}\int d\widetilde x \, \vert w(\widetilde x) \vert^2 \delta^{\alpha\beta} = -\frac{c_2^{\alpha} E_{R_1}}{2} \delta^{\alpha\beta}.
\end{equation}
Note that the color-diagonal element is simply 
\begin{equation}
\eta^{\alpha\alpha} = - \frac{c_2^{\alpha}E_{R_1}}{2},
\end{equation}
and that $\Delta_{nn}^{\alpha \beta}$ has no sinusoidal dependence on position because of parity, that is, the sinusoidal contributions are identically zero.
Furthermore, the local disorder $\Delta_{nn}^{\alpha\beta}$ and its energy reference $\eta^{\alpha\beta}$ are color-diagonal.

Using the Gaussian approximation for the Wannier functions
$w_n (x)$, we obtain
\begin{equation}
\Delta^{\alpha\beta} = 
\left(\frac{c_2^\alpha E_{R_1}}{2} \right)
\exp \left({-\frac{1}{2} R_\lambda^2 k_1^2 \xi^2} \right) \delta^{\alpha\beta},
\end{equation}
which is finally written as 
\begin{equation}
\Delta^{\alpha\beta}
= \left(\frac{c_2^\alpha E_{R_1}}{2} \right)
\exp\left( - \frac{2R_\lambda^2 E_{R_1}}{\hbar\omega} \right) \delta^{\alpha\beta}.
\end{equation}
Using the deep lattice relation $\hbar \omega = \sqrt{2c_1} E_{R_1}$, the exponent simplifies to 
\begin{equation}
\label{eqn:DeltaAlphaAlpha}
\Delta^{\alpha \alpha} = 
\left( \frac{c_2^\alpha E_{R_1}}{2} \right)
\exp\left(- \frac{2 R_\lambda^2}{\sqrt{2c_1}} \right).
\end{equation}
elucidating the dependence on the dimensionless depth $c_1$ and the ratio 
$R_{\lambda} = \lambda_1/\lambda_2$ between the strong ($\lambda_1$) and
the weak ($\lambda_2$) lattices. 
Notice that while the prefactor of 
$\vert \Delta^{\alpha \alpha} \vert$ 
is proportional to the dimensionless depth $\vert c_2^\alpha \vert$ of the weak lattice, 
the exponent is negative and inversely proportional to $\sqrt{c_1}$.
This dependence produces a smaller exponential factor for larger $\sqrt{c_1}$, thus
increasing 
$\vert \Delta^{\alpha \alpha} \vert$ 
as $c_1$ increases. This is the opposite behavior of the hopping $J$ in Eq.~(\ref{eqn:hopping-unit}), where the exponential factor decreases with increasing $\sqrt {c_1}$.

\begin{figure}
\centering
\includegraphics[width=0.9\linewidth]{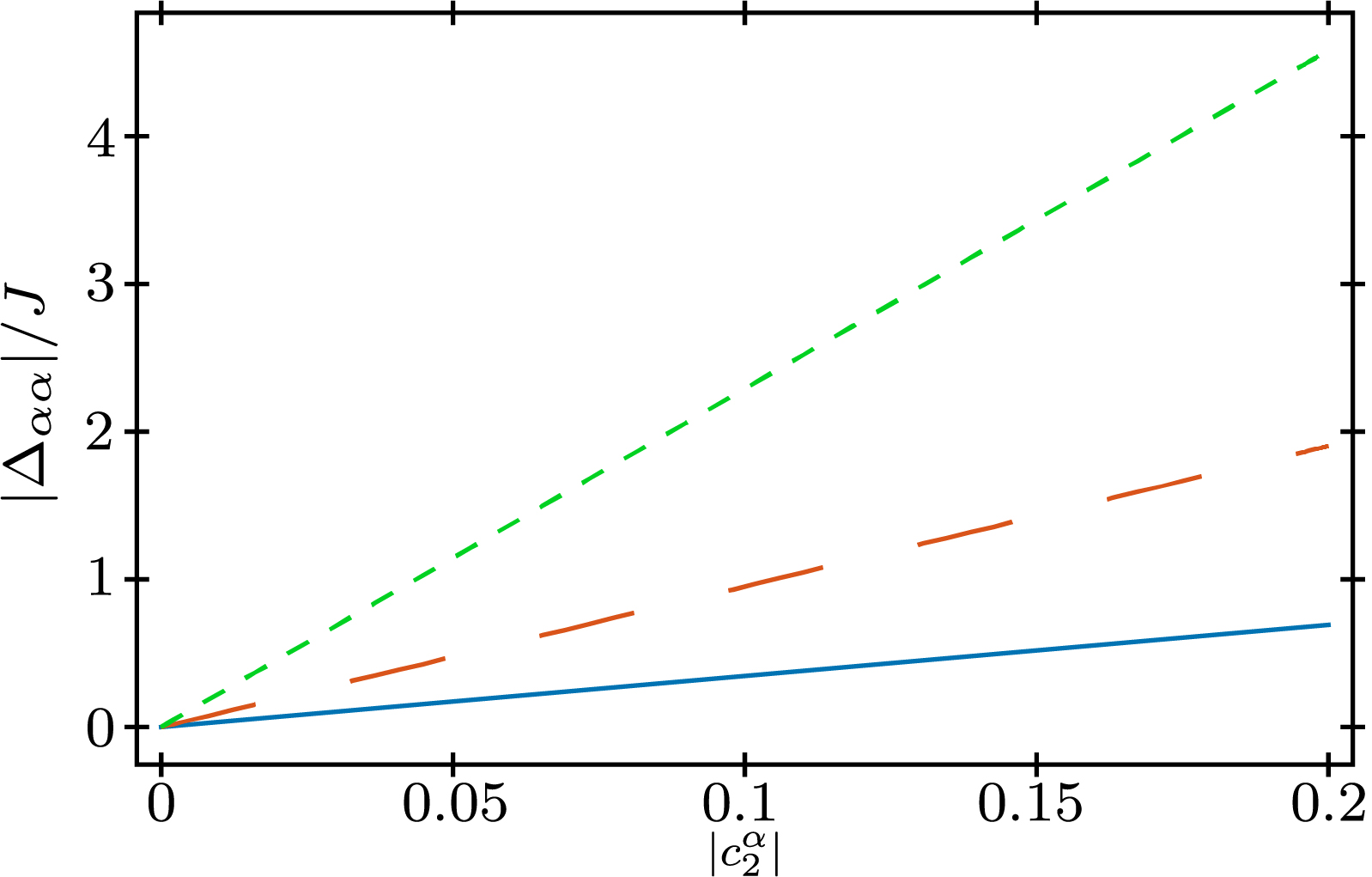}
\caption{Plot of ratio between bichromatic disorder $\vert \Delta^{\alpha\alpha} \vert$ and nearest-neighbor hopping $J$ versus weak lattice parameter
$\vert c_2^\alpha \vert$ for a bichromatic optical lattice in the tight-binding regime. The wavelength ratio used is $R_{\lambda} = 813/1064$, which is compatible with $^{87}{\rm Sr}$.
The solid blue, long-dashed red, short-dashed green lines correspond to strong lattice depth parameters of $c_{1} = 5$, $c_{1} = 10$, and $c_1 = 15$, that is, energy depths of $5E_{R_1}$, $10E_{R_1}$, and $15E_{R_1}$,
respectively.}
\label{fig:delta-over-hopping}
\end{figure}

For later reference, 
we plot in Fig.~\ref{fig:delta-over-hopping} the ratio between 
$\Delta^{\alpha \alpha}$
and $J$ versus the $c_2^{\alpha}$, which 
controls the depth of the weak optical lattice. The parameter
\begin{equation}
\frac{\Delta^{\alpha \alpha} }{J}
= 
\frac{c_2^\alpha}
{2 ( c_1  - \sqrt{\frac{c_1}{2}} )}
\exp\left(- \frac{2R_\lambda^2}{\sqrt{2c_1}} 
+\frac{\pi^2 \sqrt{2c_1}}{8}
\right),
\end{equation}
depends on the lattice-depth parameters
$c_1$ and $c_2^{\alpha}$ and the wavelength ratio $R_{\lambda}$, and is always positive for $c_1 > 1$.

Finally,  we analyze the non-local terms of $\Delta_{nm}^{\alpha\beta}$ shown in Eq.~(\ref{eqn:DeltaabnmV2}).
While it is crucial to retain non-local $J_{nm}^{\alpha\beta}$ since they describe the kinetic energy of particles or holes, the non-local disorder 
$\Delta_{n m}^{\alpha\beta}$ $(n \ne m)$ is exponentially small in comparison to the local disorder $\Delta_{nn}^{\alpha\beta}$, that is, $\vert\Delta^{\alpha\beta}_{n\ne m} \vert \ll \vert \Delta_{nn}^{\alpha\beta} \vert$. 

Having discussed the matrix elements $J_{nm}^{\alpha \beta}$
and $\Delta_{nm}^{\alpha \beta}$ in the Hamiltonian of 
Eq.~(\ref{eqn:hamiltonian}), we can write the final form of the lattice Hamiltonian as 
\begin{equation}
\label{eqn:hamiltonian-nearest-neighbor}
\mathcal H = -\sum_{\braket{nm}\alpha\beta} J_{nm}^{\alpha\beta} f_{n\alpha}^{\dagger} f_{m\beta}
+\sum_{n\alpha\beta} \Gamma_{nn}^{\alpha\beta} f_{n\alpha}^{\dagger} f_{n\beta}.
\end{equation}
The first term in Eq.~(\ref{eqn:hamiltonian-nearest-neighbor}) contains the non-local term $J_{nm}^{\alpha \beta}$ representing color-dependent 
nearest-neighbor hopping $\braket{nm} \to m = n \pm 1$ given by
\begin{equation}
\label{eqn:nearest-neighbor-hopping-alpha-beta}
J_{n, n \pm 1}^{\alpha \beta} = 
J e^{i\theta^{\alpha \alpha}_{n, n\pm 1}} \delta^{\alpha \beta},
\end{equation}
where the color-dependent phase factors are 
\begin{equation}
\label{eqn:theta-alpha-alpha}
\theta^{\alpha \alpha}_{n, n\pm 1} = 
\mp \gamma_{\alpha} k_T a.
\end{equation}
with 
$\gamma_R=+1, \gamma_G=0, \gamma_B=-1$.

The second term in in Eq.~(\ref{eqn:hamiltonian-nearest-neighbor}) describes the local contribution
\begin{equation}
\label{eqn:local-terms-hamiltonian}
\Gamma_{nn}^{\alpha\beta} = 
\varepsilon_0 \delta^{\alpha \beta} 
-h_x \left[\mathbf{S}_x\right]^{\alpha \beta}
+ \eta^{\alpha\beta} + 
\Delta^{\alpha\beta}\cos(2\pi R_{\lambda} n +\varphi^\alpha),   
\end{equation}
containing an overall color-independent energy reference
$\varepsilon_0$, a color-flip term controlled by the
color-flip (Rabi) field $h_x$, a color-dependent disorder energy 
$\eta^{\alpha\beta}$ and a disorder potential with 
color-dependent amplitude $\Delta^{\alpha\beta}$.

The presence of color-orbit coupling ($k_T \ne 0$) and color-flip (Rabi) field ($h_x \ne 0$) reveals that the Hamiltonian in Eq.~(\ref{eqn:hamiltonian-nearest-neighbor}) is a generalization of the Aubry-Andr\'e model (AAM)~\cite{andre-1980} used to describe bichromatic disorder in the absence of color-orbit coupling and Rabi fields~\cite{inguscio-2008}. 
For $k_T = 0$ and $h_x = 0$, the Hamiltonian becomes fully diagonal in the color indices with 
\begin{equation}
J_{n, n \pm 1}^{\alpha \beta} = J
\delta^{\alpha \beta}
\end{equation}
being a color-independent hopping matrix element
and 
\begin{equation}
\Gamma_{nn}^{\alpha \beta} = 
\Gamma_{nn}^{\alpha \alpha} 
\delta^{\alpha \beta},
\end{equation}
being a local matrix element containing color-dependent
disorder
\begin{equation}
\Gamma_{nn}^{\alpha \alpha} = 
\varepsilon_0 + \eta^{\alpha \alpha} 
+ \Delta^{\alpha\alpha}
\cos(2\pi R_{\lambda} n + \varphi^\alpha).    
\end{equation}
Thus, for $k_T = 0$ and $h_x = 0$, the Hamiltonian
in Eq.~(\ref{eqn:hamiltonian-nearest-neighbor}) 
reduces to three color copies of 
the Aubry-Andr\'e model.

The Hamiltonian matrix corresponding to the full Hamiltonian displayed in Eq.~(\ref{eqn:hamiltonian-nearest-neighbor}) is
\begin{equation}
    \mathbf{H} = 
    \begin{pmatrix}
        \ddots & -\mathbf{J}_{\bar2\bar1} & 0 & 0 & 0 \\
        -\mathbf{J}_{\bar1\bar2} & \boldsymbol{\Gamma}_{\bar1\bar1} &
        -\mathbf{J}_{\bar1 0} &  0 & 0 \\
        0 & -\mathbf{J}_{0\bar1} & \boldsymbol{\Gamma}_{00} &
        -\mathbf{J}_{01} & 0 \\
        0 & 0 & -\mathbf{J}_{10} &
        \boldsymbol{\Gamma}_{11} & -\mathbf{J}_{12} \\
        0 & 0 & 0 & -\mathbf{J}_{21} & \ddots \\
    \end{pmatrix}
    \label{eqn:hamiltonian-matrix}
\end{equation}
where $\{\bar n, \bar m\} = \{-n,-m\}$, $\boldsymbol{\Gamma}_{nn}$ are $3\times3$ on-site matrices with elements $\Gamma_{nn}^{\alpha\beta}$ and the off-diagonal block matrices $\mathbf{J}_{n, n \pm 1}$ describe nearest-neighbor hopping with 
matrix elements $J_{n, n\pm 1}^{\alpha \beta}$ given in 
Eq.~(\ref{eqn:nearest-neighbor-hopping-alpha-beta}).
The hopping matrices are Hermitian in color-space and satisfy the relation $(\mathbf{J}_{nm})^{\dagger} = (\mathbf{J}^{T}_{nm})^*$.

Now that the three-color Hamiltonian matrix is established, we discuss next its symmetries.

\section{Symmetries of the three-color Hamiltonian}
\label{sec:symmetries-of-the-three-color-hamiltonian}

To explore the symmetries of the three-color Hamiltonian matrix ${\bf H}$ shown in Eq.~(\ref{eqn:hamiltonian-matrix}),
we analyze the block matrices $\boldsymbol{\Gamma}_{nn}$
representing the local energy contributions, and 
$\mathbf{J}_{n, n \pm 1}$ describing the nearest-neighbor hopping energies. We begin our discussion with the local block matrices $\boldsymbol{\Gamma}_{nn}$.

\subsection{Local terms}
\label{sec:local-terms}

We explore first the on-site matrix 
\begin{equation}
\boldsymbol{\Gamma}_{nn} = \begin{pmatrix}
\varepsilon_R(\varphi^R) & -h_x/\sqrt{2} & 0 \\
-h_x/\sqrt{2} & \varepsilon_G(\varphi^G) & -h_x/\sqrt{2} \\
0 & -h_x/\sqrt{2} & \varepsilon_B(\varphi^B)
\end{pmatrix}
\label{eqn:hamiltonian-local-block}
\end{equation}
where the diagonal elements are simply given by 
\begin{equation}
\Gamma_{nn}^{\alpha \alpha}
= 
\varepsilon_{\alpha}(\varphi^\alpha) = 
\varepsilon_0 + \eta^{\alpha \alpha} + \Delta^{\alpha\alpha}\cos\left(2\pi R_\lambda n + \varphi^\alpha\right),
\label{eqn:hamiltonian-local-block-diagonal-elements}
\end{equation}
where $\alpha \in \{ R,G,B \}$.
This $3\times3$ matrix is written in a basis of pseudospin-1 states described by the mapping 
$\ket{R} \to \ket{1, 1}$, $\ket {G} \to \ket{1, 0}$, and 
$\ket {B} \to \ket{1, -1}$, corresponding to the $\ket{s,m_s}$
states.
Thus, we express the Hamiltonian in terms of spin-1 matrices
as follows
\begin{equation} 
\label{eqs:Gamma}
\mathbf{\Gamma}_{nn} = \varepsilon_G(\varphi^G) \mathbf I -h_x \mathbf{S}_x - h_z \mathbf{S}_z + g_z \mathbf{S}_z^2
\end{equation}
where $h_x$ plays the role of a Zeeman field along the $x$ axis in pseudo-spin space, 
\begin{equation}
h_z = \frac{1}{2}\left[(\varepsilon_B(\varphi^B)-\varepsilon_R(\varphi^R)\right]
\end{equation}
represents a Zeeman field along the $z$ axis in pseudo-spin space, and 
\begin{equation}
g_z = \frac{1}{2}\left[ \varepsilon_R(\varphi^R) + \varepsilon_B(\varphi^B)\right]- \varepsilon_G(\varphi^G)
\end{equation}
describes a quadratic Zeeman shift along the $z$ axis in pseudo-spin space, and thus can be viewed as a pseudo-spin quadrupolar effect.
The color subspace spanned by the matrices ${\bf I}, {\bf S}_x, {\bf S}_z, {\bf S}_z^2$ is a subset of SU(3), and thus are related to the Gellmann matrices $\lambda_i$ covering SU(3) as follows:
${\bf S}_x = (\lambda_1 + \lambda_6) /\sqrt{2} $, 
${\bf S}_y = (\lambda_2 + \lambda_7) /\sqrt{2} $,
${\bf S}_z = (\lambda_3 + \sqrt{3}\lambda_8) /2 $, 
and ${\bf S}_z^2 = (4 {\bf I} + 3\lambda_3 - \sqrt{3}\lambda_8) /6 $.

The presence of the fields \(h_x, h_z,\) and \(g_z\) explicitly lifts the SU(3) degeneracy of the local term $\boldsymbol{\Gamma}_{nn}$, introducing energy splitting among the three color states (Red, Green, and Blue).
These fields originate from experimentally tunable controls and system-specific parameters, each playing a distinct physical role. For fixed $E_{R_1}$, the color-dependent
parameters in ${\bf \Gamma}_{nn}$ are controlled by $c_2^{\alpha}$ and
$\varphi^{\alpha} = 2\phi_2^{\alpha} \pm \pi$, which are directly related  to the color-dependent dimensionless amplitude $(c_2^{\alpha})$ and phase $(\phi_2^{\alpha})$ of the weak optical lattice potential described
in Eq.~(\ref{eqn:weak-optical-potential}).

Below, we explicitly present and describe all the terms appearing in Eq.~(\ref{eqs:Gamma})
in terms of the matrix elements defined in 
Eq.~(\ref{eqn:local-terms-hamiltonian}).
The first term, multiplying the identity matrix ${\bf I}$, is 
\begin{equation}
\varepsilon_G(\varphi^G) = \varepsilon_0 + \eta^{GG} + 
\Delta^{GG} \cos(2\pi R_{\lambda} n + \varphi^G),
\end{equation}
representing the energy of the G state.
The contribution $\varepsilon_0$ represents a color-independent energy offset, the energy shift $\eta^{GG}$ is due to the state-dependent amplitude $c_2^{G}$ of the weak lattice, and the last factor is a spatially modulated potential with amplitude 
$\Delta^{GG}$, also controlled by $c_2^{G}$, wavenumber $k^{G} = 2\pi R_{\lambda}/a$ and phase $\varphi^G$.
The lattice position $x_n = na$ was used to define $k^{G}$ via the relation $k^{G} x_n = 2\pi R_{\lambda} n$. The corresponding wavelength of the modulation is $\lambda^{G} = 2\pi/k^{G} = a/ R_{\lambda}$, which is independent of color (internal state).

The second term $h_x$, multiplying ${\bf S}_x$, represents the color-flip (Rabi) field that induces transitions between adjacent color states, specifically between R and G or B and G.  
These off-diagonal terms appear symmetrically in the matrix and encode the strength of this mixing.
The absence of direct coupling between the R and B states, as indicated by the zero entries in the corners of the matrix in Eq.~(\ref{eqs:Gamma-Explicit-Matrix}), reflects the ladder-like structure of the Raman transitions 
shown in Fig.~\ref{fig:Raman-coupling}.
Moreover, the presence of $h_x$ breaks the SU(3) symmetry by introducing color-flipping and lifts any residual energy degeneracies among color states.

The third term, multiplying ${\bf S}_z$, is a spatially-dependent color Zeeman shift
\begin{align}
h_z  =  \frac12\big[\eta^{BB} + \Delta^{BB} \cos(2\pi R_{\lambda} n + \varphi^B) \nonumber \\ - \eta^{RR} - \Delta^{RR} \cos(2\pi R_{\lambda} n +\varphi^R) \big],
\end{align}
which moves the energies of the R and B states in opposite directions, 
leaving the energy of the G state unaffected. 
The spatially-independent part $(\eta^{BB} - \eta^{RR})/2$ and the
spatially modulated component $\Delta^{BB}\cos(2\pi R_{\lambda} n + \varphi^B) - \Delta^{RR}\cos(2\pi R_{\lambda} n + \varphi^R)$
capture the color-dependent energy shifts involving the R and B states.

Lastly, the fourth contribution, multiplying ${\bf S}_z^2$, is 
\begin{align}
g_z 
 = 
- \eta^{GG} 
- \Delta^{GG} \cos(2\pi R_{\lambda} n + \varphi^G)
+ \frac12 \left[ \eta^{RR} + \eta^{BB} \right]
\nonumber
\\
+ \frac12 \left[ \Delta^{RR} \cos(2\pi R_{\lambda} n + \varphi^R) + \Delta^{BB} \cos(2\pi R_{\lambda} n + \varphi^B) \right]
\end{align}
describing the strength of the quadratic Zeeman shift, that is, a change in 
the energies of R and B states by the same amount, 
without affecting the energy of the 
G state. The term 
$\left[ \eta^{RR} + \eta^{BB} \right]/2 - \eta^{GG}$
represents the spatially-independent contribution, while 
$ \left[ \Delta^{RR} \cos(2\pi R_{\lambda} n + \varphi^R) + 
\Delta^{BB} \cos(2\pi R_{\lambda} n + \varphi^B) \right]/2 
- \Delta^{GG} \cos(2\pi R_{\lambda} n + \varphi^G) $  
describes the spatially-dependent part.
In passing, we note that the term containing $g_z$ does not appear for SU(2) fermions in the presence of spin-orbit coupling and Rabi 
fields~\cite{sa-de-melo-2021, sa-de-melo-2023}.

\subsection{Color Gauge Transformation: Local Terms}
\label{sec:color-gauge-transformation-locat-terms}

To better understand the properties of the local matrix ${\bf \Gamma}_{nn}$ in Eq.~(\ref{eqs:Gamma}), we perform a local color gauge transformation (CGT) via the unitary operator
$\mathbf{U}_n = e^{i k_T x_n \mathbf{S}_z}$ acting into the color RGB basis
states. Since $\mathbf{U}_n$ represents 
a local rotation about the z-axis, it commutes with any function of $\mathbf{S}_z$
and with the identity ${\bf I}$. Therefore, we have
\begin{equation}
\left[ \mathbf{U}_n, \mathbf{S}_z \right] = 0 \quad \text{and} \quad \left[ \mathbf{U}_n, \mathbf{S}_z^2 \right] = 0.
\end{equation}
As a result, the only term in 
$\boldsymbol{\Gamma}_{nn}$ that is affected by the local CGT is the color-flip (Rabi) field appearing with spin operator $\mathbf{S}_x$. To explicitly show the transformation of $\mathbf{S}_x$ under the local CGT, we write the rotated operator as
\begin{equation}
\widetilde{\mathbf{S}}_x = \mathbf{U}_n^\dagger \mathbf{S}_x \mathbf{U}_n = e^{-i k_T x_n \mathbf{S}_z} \mathbf{S}_x e^{+i k_T x_n \mathbf{S}_z}.
\end{equation}
Using the commutation relations 
$\left[ \mathbf{S}_z, \mathbf{S}_x \right] = i \mathbf{S}_y$, $\left[ \mathbf{S}_z, \mathbf{S}_y \right] = -i \mathbf{S}_x$
we obtain
\begin{equation}
    \widetilde{\mathbf{S}}_x = \mathbf{S}_x \cos(k_T x_n) + \mathbf{S}_y \sin(k_T x_n).
\end{equation}
This result shows that the term $-h_x {\bf S}_x $ transforms into 
\begin{equation}
-h_x \widetilde {\bf S}_x
= -h_x \cos(k_T x_n) \mathbf{S}_x - h_x \sin(k_T x_n)\mathbf{S}_y,
\end{equation} 
From this relation, we can see that the local 
color-flip (Rabi) field 
${\widetilde {\bf h}} (x_n) = 
\left( {\widetilde h}_x (x_n), {\widetilde h}_y (x_n) \right)$ has two position-dependent components; the first one is  
\begin{equation}
\label{eqn:hx-CGT}
{\widetilde h}_x (x_n)  = h_x \cos(k_T x_n),
\end{equation}
while the second component is 
\begin{equation}
\label{eqn:hy-CGT}
{\widetilde h}_y (x_n) = h_x \sin(k_T x_n).
\end{equation}
Thus, the local on-site matrix $\boldsymbol{\Gamma}_{nn}$ in Eq.~(\ref{eqn:hamiltonian-local-block}), subjected to the local CGT,  becomes
$\widetilde{\boldsymbol{\Gamma}}_{nn} = \mathbf{U}_n^\dagger \boldsymbol{\Gamma}_{nn} \mathbf{U}_n$ leading to the matrix
\begin{multline} 
\widetilde{\boldsymbol{\Gamma}}_{nn} =
\\
\label{eqs:tildegamma}
\begin{pmatrix}
\varepsilon_R(\varphi^R) & - \left(\frac{h_x}{\sqrt{2}}\right) e^{-i k_T x_n} & 0 \\
- \left(\frac{h_x}{\sqrt{2}}\right) e^{i k_T x_n} & \varepsilon_G(\varphi^G) & - \left(\frac{h_x}{\sqrt{2}}\right) e^{-i k_T x_n} \\
0 & - \left(\frac{h_x}{\sqrt{2}}\right) e^{i k_T x_n} & \varepsilon_B(\varphi^B)
\end{pmatrix}.
\end{multline}
The color-diagonal elements $\varepsilon_\alpha( \varphi^\alpha)$, representing the local energy of each color state including disorder effects, remain unchanged as in Eq.~(\ref{eqn:hamiltonian-local-block}). However, the off-diagonal elements, representing the color-flip (Rabi) field $h_x$,
acquire a spatially-dependent complex phase $k_T x_n$ modulated by the momentum transfer $k_T$.

Having discussed the local terms of the three-color Hamiltonian in the presence of color-orbit
and color-flip fields, we analyze next the hopping matrix elements.  

\subsection{Hopping terms}
\label{sec:hopping-terms}

In the presence of color-orbit coupling characterized by $k_T$, the hopping matrix between sites $n$ and $m$ $(n \ne m)$ is 
\begin{equation}
    \mathbf{J}_{nm} = J_{nm} (0)
    \begin{pmatrix}
         e^{i k_T \delta x_{nm}} & 0 & 0 \\
         0 & 1 & 0 \\
         0 & 0 & e^{-i k_T \delta x_{nm}}       
    \end{pmatrix},
\label{eqn:hamiltonian-hopping-block}
\end{equation}
where $J_{nm} (0) = J_{nm} (k_T = 0)$, defined in Eq.~(\ref{eqs:Jnm-kT=0}), 
sets the energy scale for hopping
and $\delta x_{nm} = x_n - x_m = (n-m)a$ describes the separation between sites $n$ and $m$.
The hopping matrix shows color-dependent terms, controlled by phase factors 
$ e^{\pm i k_T \delta x_{nm}}$ that
carry the net momentum transfer  
$k_T$, representing color-orbit coupling.
These terms preserve a given color state during hopping but modulate its phase, thus affecting the energy level structure of the three-color system, when $h_x \ne 0$.

The phase factors in the hopping matrix emerge because the momentum transfer is color-dependent, as inferred by the Raman processes
shown in Fig.~\ref{fig:Raman-coupling}. 
The R states acquire a momentum shift $k_T$ to the right, the G states have no momentum shift, and the B states acquire a momentum shift $k_T$ to the left.

Using the $3 \times 3$ identity matrix $\mathbf I$, as well as the pseudospin-1 matrices ${\mathbf S}_z$ and 
${\mathbf S}^2_z$, the hopping matrix in Eq.~(\ref{eqn:hamiltonian-hopping-block}) becomes  
\begin{equation}
\mathbf{J}_{nm}/J_{nm} (0) = \mathbf I + i F (k_T \delta x_{nm})  
\mathbf{S}_{z} + G (k_T \delta x_{nm}) \mathbf{S}_{z}^2,
\end{equation}
where the functions 
$F (k_T \delta x_{nm}) =  \sin(k_T \delta x_{nm})$
and $G (k_T \delta x_{nm}) = \cos(k_T \delta x_{nm}) - 1$
represent the color-dependent contributions to the hopping matrix $\mathbf{J}_{nm}$. Thus, the hopping matrix $\mathbf{J}_{nm}$ for the SU(3) system is a linear combination of the identity matrix $\mathbf{I}$, the pseudo-spin operator $\mathbf{S}_z$, and its square $\mathbf{S}_z^2$.

\subsection{Color Gauge Transformation: Hopping}
\label{sec:color-gauge-transformation-hopping}

We represent the local color-gauge transformation 
$\mathbf{U}_n = e^{i k_T x_n \mathbf{S}_z}$ in the basis of red ($\ket{R}$), green ($\ket{G}$), and blue ($\ket{B}$) states as the matrix
\begin{equation}
\mathbf{U}_n =
\begin{pmatrix}
e^{i k_T x_n} & 0 & 0 \\
0 & 1 & 0 \\
0 & 0 & e^{-i k_T x_n}
\end{pmatrix},
\end{equation}
and use it to obtain the new hopping matrix as 
$\widetilde{\mathbf{J}}_{nm}$ becomes
$ \widetilde{\mathbf{J}}_{nm} = \mathbf{U}^\dagger_n \mathbf{J}_{nm} \mathbf{U}_m$. Here, $\mathbf{J}_{nm}$ is the original hopping matrix between sites $n$ and $m$ defined in Eq.~(\ref{eqn:hamiltonian-hopping-block}).
As a result of the local CGT, the color-dependent phase factors are completely removed, leading to a transformed hopping matrix
\begin{equation}
\label{eqs:tildeJ}
\widetilde{\mathbf{J}}_{nm} = J_{nm}(0) \mathbf{I}.
\end{equation}
The resulting transformed hopping matrix $\widetilde{\mathbf{J}}_{nm}$ is diagonal and color-independent, where $J_{nm}(0)$ is the hopping amplitude in the absence of color-orbit coupling. Next, we are ready to discuss the full Hamiltonian matrix after the local color gauge transformation (CGT).

\subsection{Fully Transformed Hamiltonian}
\label{sec:fully-transformed-hamiltonian}

After the CGT, the full Hamiltonian in first-quantization is represented by the matrix
\begin{equation}
\label{eqn:hamiltonian-SGT}
\widetilde{\mathbf{H}} =
\begin{pmatrix}
\ddots & -\widetilde{\mathbf{J}}_{\bar2\bar1} & 0 & 0 & 0 \\
-\widetilde{\mathbf{J}}_{\bar1\bar2} & \widetilde{\boldsymbol{\Gamma}}_{\bar1\bar1} &
-\widetilde{\mathbf{J}}_{\bar1 0} & 0 & 0 \\
0 & -\widetilde{\mathbf{J}}_{0\bar1} & \widetilde{\boldsymbol{\Gamma}}_{00} &
-\widetilde{\mathbf{J}}_{01} & 0 \\
0 & 0 & -\widetilde{\mathbf{J}}_{10} &
\widetilde{\boldsymbol{\Gamma}}_{11} & -\widetilde{\mathbf{J}}_{12} \\
0 & 0 & 0 & -\widetilde{\mathbf{J}}_{21} & \ddots \\
\end{pmatrix},
\end{equation}
containing $3 \times 3$ block matrices, where only local and nearest-neighbor blocks are included. The on-site blocks are
given in Eq.~(\ref{eqs:tildegamma}), and the nearest-neighbor hopping blocks $\widetilde{\mathbf{J}}_{n, n\pm 1} = J_{n, n\pm 1}(0)\mathbf{I}$ are defined in Eq.~(\ref{eqs:tildeJ}). 

It is important to emphasize that the CGT transforms $\mathbf{H}$ into $\widetilde{\mathbf{H}}$, via the mapping $\mathbf{J}_{nm} \rightarrow \widetilde{\mathbf{J}}_{nm}$ and $\boldsymbol{\Gamma}_{nn} \rightarrow \widetilde{\boldsymbol{\Gamma}}_{nn}$.
The new tunneling matrix $\widetilde{\mathbf{J}}_{nm} = J_{nm}(0)\mathbf{I}$,
where $\mathbf{I}$ is the identity, does not contain color-dependent phases.
The local color-diagonal elements $\widetilde{\Gamma}_{nn}^{\alpha\alpha} = \Gamma_{nn}^{\alpha\alpha}$ remain invariant, however the color-off-diagonal elements become $\widetilde{\Gamma}_{nn}^{RG} = \widetilde{\Gamma}_{nn}^{GB} = -h_x e^{-i k_T x_n}/\sqrt{2}$, and $\widetilde{\Gamma}_{n}^{GR} = \widetilde{\Gamma}_{nn}^{BG} = 
-h_x e^{+i k_T x_n}/\sqrt{2}$, as can be seen in 
Eq.~(\ref{eqs:tildegamma}).
The CGT is effectively a local rotation in the direction of the color-flip (Rabi) field $h_x$ by a counter-clockwise angle of $k_T x_n$
leading to the complex field $\widetilde{h}_{\perp} (x_n) = h_x e^{-i k_T x_n} = \widetilde h_x -i \widetilde h_y$, where $\tilde{h}_x = h_x \cos(k_T x_n)$ and $\widetilde{h}_y = h_x\sin(k_T x_n)$.

The results above demonstrate that the local CGT can remove the spatially dependent phases of the hopping matrix,
as seen in Eq.~(\ref{eqs:tildeJ}), and transfer them into a spatially oscillating color-flip (Rabi) field with a local phase controlled by $k_T x_n$, as displayed in Eq.~(\ref{eqs:tildegamma}). This simplification is a key advantage of the CGT, because it reveals two important symmetries of the system. First, the CGT shows that the Hamiltonian $\widetilde {\mathbf H}$ and the field
$\widetilde{h}_{\perp} (x_n)$ are periodic in the color-orbit coupling parameter $k_T a$, with period $2\pi$, as revealed in ${\widetilde{\boldsymbol\Gamma}}_{nn}$ displayed in Eq.~(\ref{eqs:tildegamma}). Second, the CGT also demonstrates a color-gauge symmetry, that is, when $h_x = 0$ there is no dependence of 
$\widetilde {\mathbf H}$ on the color-orbit coupling parameter $k_T a$, which is completely gauged away, as seen again in Eq.~(\ref{eqs:tildegamma}).

These general symmetry properties serve as references
in analyzing mobility regions via the energy spectra and inverse participation ratio (IPR) to be discussed next.  

\section{Energy Spectra and Inverse Participation Ratio}
\label{sec:energy-spectra-and-inverse-partition-ratio}

To obtain the energy spectra and inverse participation ratios associated with the Hamiltonian matrix in Eq.~(\ref{eqn:hamiltonian-matrix}), we analyze the matrix elements of $\boldsymbol{\Gamma}_{nn}$ in 
Eq.~(\ref{eqn:hamiltonian-local-block}) and of 
$\mathbf{J}_{nm}$ in Eq.~(\ref{eqn:hamiltonian-hopping-block}).

Throughout the rest of this paper, we focus on the non-trivial effects of color-orbit coupling and color-flip (Rabi) fields for the simpler case of color-independent disorder, 
In this case, the matrix elements of 
$\boldsymbol{\Gamma}_{nn}$ simplify since $c_2^\alpha$ and $\phi_2^\alpha$ in Eq.~(\ref{eqn:weak-optical-potential})
do not depend on the color state.
This leads to $c_2^\alpha = c_2$ or $c_2^R = c_2^G = c_2^B = c_2$, and to $\varphi^{\alpha} = 
 2\phi_2^\alpha  \pm \pi =  2 \phi_2 \pm \pi = \varphi $ or, equivalently, 
$\varphi^R = \varphi^G = \varphi^B  = -2 \phi_2 = \varphi$, meaning that the weak optical potential affects all color states in the same way.

For color-independent disorder, the matrix elements $\Gamma_{nn}^{\alpha \beta}$ contain disorder amplitudes $\Delta^{\alpha\beta} = \Delta \delta^{\alpha\beta}$ and reference energies $\eta^{\alpha\beta} = \eta \delta^{\alpha\beta}$ which 
are color-independent. Thus, we can set our energy reference to be $\varepsilon_0 + \eta$. 
Using the Gaussian approximation for the Wannier function, we obtain the color-independent disorder amplitude to be 
\begin{equation}
\label{eqn:delta-color-independent}
\Delta = 
\left( \frac{c_2 E_{R_1}}{2} \right)
\exp\left(- \frac{2R_\lambda^2}{\sqrt{2c_1}} \right).
\end{equation}
As a result of color-independent disorder, 
the energies $\varepsilon_R$, $\varepsilon_G$, 
$\varepsilon_B$ are degenerate, that is, 
$\varepsilon_R (\varphi) = \varepsilon_G (\varphi) = 
\varepsilon_B (\varphi) = \varepsilon (\varphi)$, where
\begin{equation}
\label{eqn:energy-disorder-color-diagonal}
\varepsilon (\varphi) 
= 
\varepsilon_0 + \eta + 
\Delta \cos( 2\pi R_\lambda n + \varphi )
\end{equation}
and the non-trivial color-dependent effects in $\boldsymbol{\Gamma}_{nn}$ arise only from the color-flip terms controlled by the Rabi field $h_x$. Notice, however,
that with varying $\varphi$, the local disorder modulation
changes as $\cos(2\pi R_{\lambda} n + \varphi)$.

The nearest-neighbor hopping elements,
seen in Eq.~(\ref{eqn:nearest-neighbor-hopping-alpha-beta}),
are color-diagonal 
\begin{equation}
J_{n, n \pm 1}^{\alpha \alpha} = 
J e^{i\theta^{\alpha \alpha}_{n, n\pm 1}},
\end{equation}
where 
$\theta^{\alpha \alpha}_{n, n\pm 1}$
is given in Eq.~(\ref{eqn:theta-alpha-alpha}),
and $J$ is the hopping parameter defined in 
Eq.~(\ref{eqn:hopping-unit}), which we take from 
now as our energy unit. 

Notice that for $h_x/J = 0$, and color-independent disorder, the Hamiltonian matrix ${\bf H}$ in Eq.~(\ref{eqn:hamiltonian-matrix}) or the 
Hamiltonian operator $\mathcal{H}$ in Eq.~(\ref{eqn:hamiltonian-nearest-neighbor}) 
reduces to three identical copies of the 
Aubry–Andr\'e model, due to the color-gauge symmetry. 
For the phase choice 
$\varphi = 2\phi_2 \pm \pi = 0$, that is, 
$\phi_2 = \mp \pi/2$, 
the cosinoidal dependence in 
Eq.~(\ref{eqn:energy-disorder-color-diagonal})
simplifies to 
\begin{equation}
\label{eqn:energy-disorder-color-diagonal-pi-shift}
\varepsilon (\varphi) 
= 
\varepsilon_0 + \eta + 
\Delta \cos( 2\pi R_\lambda n).
\end{equation}

The experimental range of the ratio $\Delta/J$
between $0$ and $4$ is easily achievable experimentally for primary optical-lattice depth parameters  
in the range of $5 < c_1 < 15$ for the ratio 
$R_{\lambda} = 813/1064$ compatible with optical lattice wavelengths for $^{87}{\rm Sr}$.
Using the harmonic approximation for color-independent disorder, the ratio is 
\begin{equation}
\frac{\Delta}{J} = 
\frac{c_2}
{2 ( c_1  - \sqrt{\frac{c_1}{2}} )}
\exp\left(- \frac{2R_\lambda^2}{\sqrt{2c_1}} 
+\frac{\pi^2 \sqrt{2c_1}}{8}
\right),
\end{equation}
for $c_1 > 1$.

In Fig.~\ref{fig:delta-over-hopping}, the dependence
of $\vert \Delta \vert/J$ versus $\vert c_2 \vert$, the weak optical lattice depth parameter, 
is shown for $c_1 = 5$
in solid blue, $c_1 = 10$ in long-dashed red, and
$c_1 = 15$ in short-dashed green
for the ratio $R_{\lambda} = 813/1064$ compatible with 
lasers used for $^{87}{\rm Sr}$.

In analyzing the energy spectrum of SU(3) fermions in the presence of bichromatic disorder, our first example corresponds to choosing the relative phase of the weak and strong optical lattice lasers to be 
color-independent and equal to
$\pi/2$, that is,
$\phi_2 = \pi/2$ or 
$\varphi = 0$.
In this case, when color-orbit coupling and color-flip (Rabi) fields are zero,  
the Hamiltonian in Eq.~(\ref{eqn:hamiltonian-matrix}) reduces to three 
identical color-independent Aubry-Andr{\'e} models (AAM)~\cite{andre-1980}.
As a consequence, in the limit of $h_x = 0$, Aubry-Andr{\'e} physics is recovered: 
all energy eigenstates are localized for $\vert \Delta \vert/J > 2$ ,
while all energy eigenstates 
are extended for 
$\vert \Delta\vert/J < 2$,
implying that there are no mobility regions in energy.

\begin{figure}[tb]
\centering
\includegraphics[width=0.48\textwidth]{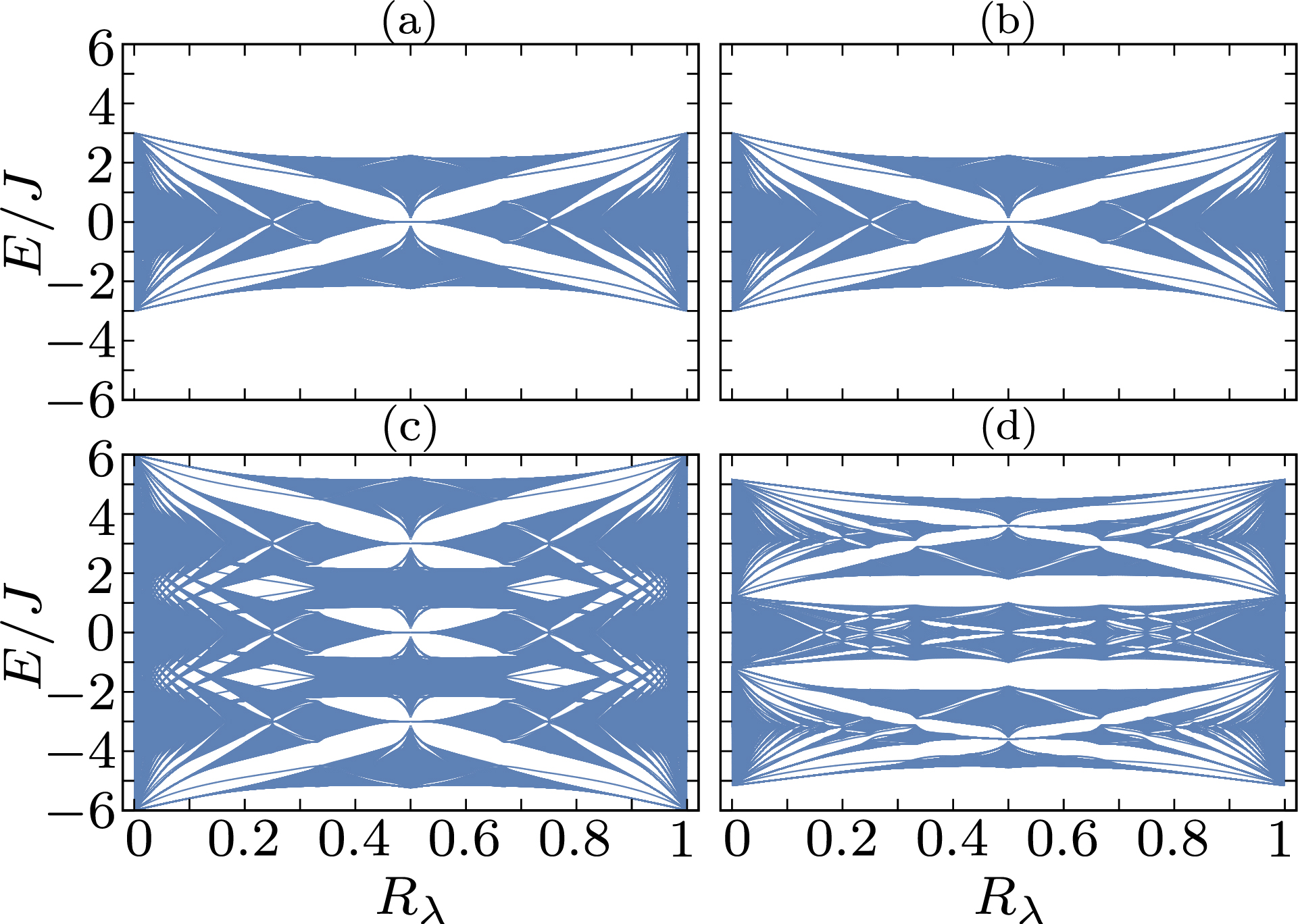}
\caption{Energy spectra versus incommensurability $R_\lambda$ at fixed disorder
$\Delta/J = 1$ for $N=500$ sites, $\varphi = 0$: (a) $h_x/J = 0$, $k_T a = 0$;
(b) $h_x/J = 0$, $k_T a = \pi/2$; (c) $h_x/J = 3$, $k_T a = 0$;
(d) $h_x/J = 3$, $k_T a = \pi/2$. In the absence of color-flip (Rabi) fields, panels (a,b), the energy spectrum is identical, since for any value of $k_T a$ there is color-gauge symmetry when $h_x= 0$. 
Non-zero $h_x$, with $k_T a = 0$, splits the color-degeneracy of the states (panel c), while non-zero 
$h_x$, with $k_T a \neq 0$, 
opens additional gaps by lifting residual
degeneracies (panel d). The spectrum is periodic in $R_{\lambda}$ (mod~1) and symmetric about
$R_{\lambda}= 1/2$.
}
\label{fig:energy-versus-Rlambda}
\end{figure}

In Fig~\ref{fig:energy-versus-Rlambda}, we show the energy eigenvalues $E/J$ versus $R_{\lambda}= \lambda_1/\lambda_2$, using $J$ defined in Eq.~(\ref{eqn:hopping-unit}) as our energy unit, and seting the energy reference 
$\varepsilon_{\rm ref} = \varepsilon_0 + \eta = 0$.
The parameters used are 
$\Delta /J = 1$ for disorder, $N=500$ for number of sites and 
$\phi_2 = \pi/2$ $(\varphi = 0)$ for relative phase. In panel (a), where $h_x$ and $k_T a$ are zero, we recover 
the results of the three-fold degenerate AAM. In panel (b), where $h_x = 0$ and $k_T a = \pi/2$, we obtain the same result as in (a) due to the color-gauge symmetry, which holds for any value of color-orbit coupling, when $h_x = 0$. In panel (c), where $h_x/J = 3$ and $k_T a = 0$, the effect of $h_x$ is to mix R-G and G-B colors and split the three-fold degeneracy of the RGB states that occurs for 
$h_x = 0$.
In panel (d), where $h_x/J = 3$ and $k_T a= \pi/2$, we see additional energy gaps in comparison to panel (c), due to the presence of color-orbit coupling $k_T a$ that lifts residual degeneracies.
The spectrum is periodic in $R_{\lambda}$ (mod 1) and is symmetric about $R_{\lambda} = 1/2$. 

We use exact diagonalization to examine localization properties as a function of energy and calculate the inverse participation ratio (IPR) given by
\begin{equation} 
\label{eqn:IPR} 
{\rm IPR}
= \frac{\sum_n \chi_n^2}
{\left[\sum_n \chi_n \right]^2}
= 
\frac{\sum_{n}\left(\vert\psi_{nR} \vert^2 + \vert \psi_{nG} \vert^2 + \vert \psi_{nB} \vert^2 \right)^2}
{\left[ 
\sum_n \left(\vert\psi_{nR} \vert^2 + \vert \psi_{nG} \vert^2 + \vert \psi_{nB} \vert^2 \right)
\right]^2},
\end{equation}
for an eigenstate $\vert \psi \rangle$ with eigenenergy $E$, where its local projection into the bra $\bra{n,\alpha}$ 
is $\psi_{n \alpha} = {\langle{n, \alpha} \vert \psi\rangle}$.
Here, $n$ represents the site, $\alpha$ the color, and
we used the relation $\chi_n = (\vert \psi_{n_R}\vert^2 + \vert \psi_{n_G} \vert^2 + \vert \psi_{n_B}\vert^2)$. For a fully normalized state the relation $\sum_n \chi_n = 1$ holds. 

For a fully extended eigenstate $\vert \psi \rangle$ with eigenenergy $E$, the spinor components scale 
as $\vert \psi_{n \alpha} \vert^2 = p_\alpha/N$, with 
$\sum_{\alpha} p_{\alpha} = 1$. Thus, the denominator tends to $1$, and the numerator tends to $1/N$, leading to an IPR that scales with 
$1/N$ and approaches zero when the system size $N \to \infty$. However, when $\vert \psi \rangle$ is fully localized at site $n = n_0$, the spinor components 
behave as $\vert \psi_{n \alpha}\vert^2 = p_{\alpha} \delta_{n n_0}$.
Thus, both the denominator and numerator tend to $1$, leading to an IPR that converges to 1. The IPR is bounded from below by $0$ and from above by $1$, that is, $0 \le {\rm IPR} \le 1$.

Given the recent experimental creation of color-orbit coupling in $^{87}{\rm Sr}$~\cite{wilkowski-2025}, we discuss from now on, only the case of 
$R_{\lambda} = 813/1064$, which is compatible with the lasers used for $^{87}{\rm Sr}$.

\begin{figure}[t]
\centering
\includegraphics[width=0.48\textwidth]{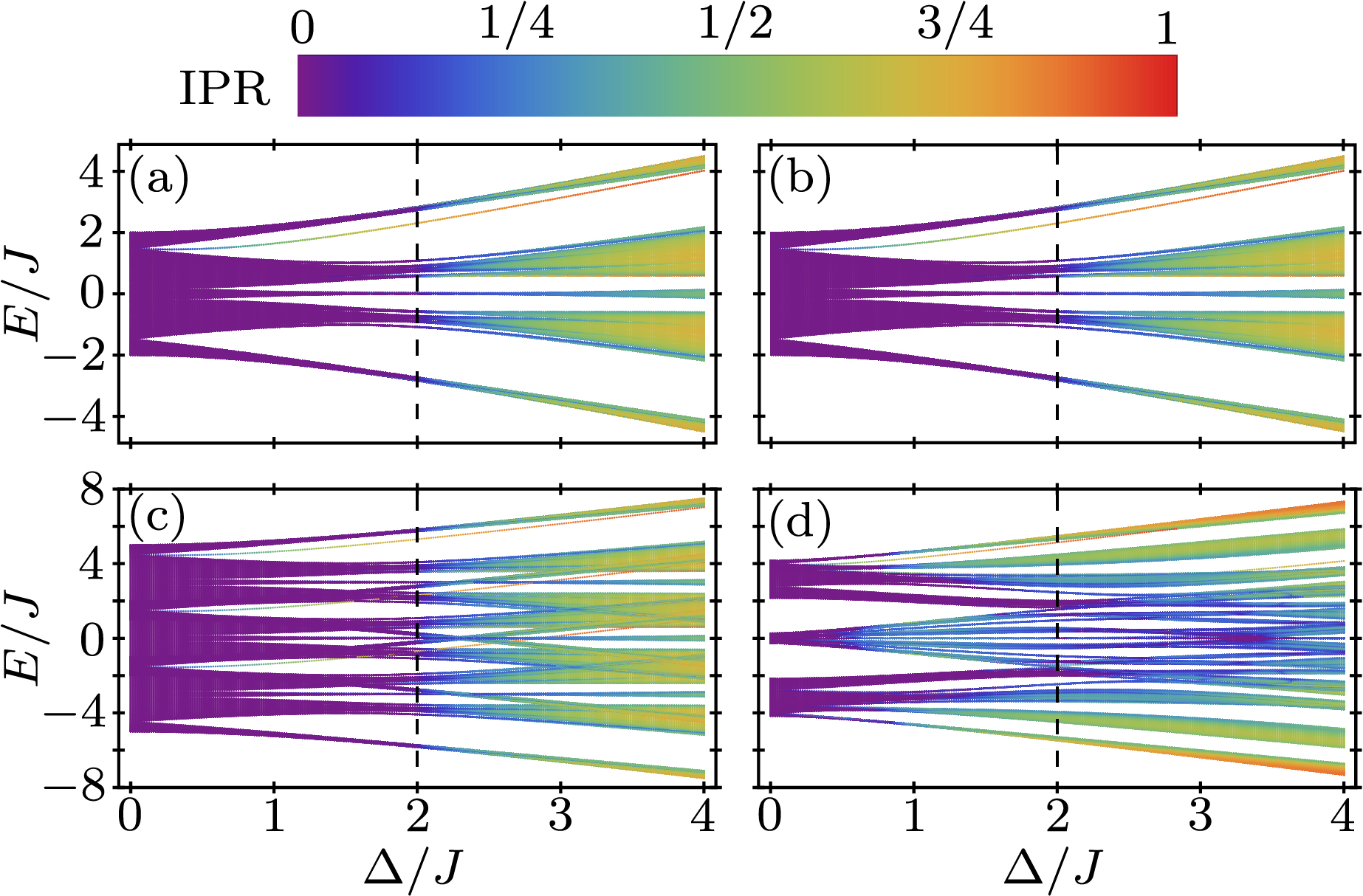}
\caption{Plots of $E/J$ versus $\Delta/J$ showing mobility regions at various color-orbit coupling and color-flip (Rabi) fields
with $R_{\lambda} = 813/1064$, $N = 500$, 
$\varphi = 0$: (a) $h_x/J = 0, k_T a =0$; (b) $h_x/J = 0, k_T a =\pi/2$; (c) $h_x/J = 3, k_T a =0$; (d) $h_x/J = 3, k_T a =\pi/2$. 
The continuous IPR color scheme varies from violet for extended states to non-violet (blue to red) for localized states. 
There is a sharp transition in (a), (b), and (c) at $(\Delta/J)^{AA}_{c} = 2$. For (a) and (b), this occurs because of the color-gauge symmetry, while in (c), this occurs because we have color-independent disorder and three copies of energy-shifted AAM. 
In (d), the locally inhomogeneous color-flip (Rabi) fields lead to mobility regions both below and above 
$(\Delta /J)^{AA}_{c} = 2$.}
\label{fig:energy-versus-delta}
\end{figure}

In Fig.~\ref{fig:energy-versus-delta}, we show the eigenenergy spectrum $E/J$ as a function of 
$\Delta/J$ for various values of color-orbit-coupling and color-flip (Rabi) fields.
The IPR is represented by the color scale, where violet indicates extended states and non-violet (blue to red) denotes localized states.
This enables a clear visualization of localization transitions and the emergence of mobility regions.

In Fig.~\ref{fig:energy-versus-delta}(a), both the color-flip (Rabi) fields and color-orbit-coupling 
are absent ($h_x/J = 0$, $k_T a = 0$).
In this case, the system reduces to the standard Aubry-Andr\'e model and exhibits a sharp localization transition at the critical value $(\Delta /J)_{c}^{\text{AA}} = 2$. For $\Delta/J < (\Delta /J)_{c}^{\text{AA}}$, the states remain fully extended, while for $\Delta/J > 
(\Delta /J)_{c}^{\text{AA}}$ all states become localized.

In Fig.~\ref{fig:energy-versus-delta}(b), 
color-orbit coupling is introduced while the color-flip (Rabi) field remains zero  
($h_x/J = 0$, $k_T a = \pi/2$), 
the localization transition still occurs at 
$(\Delta/J)_{c}^{\text{AA}}$ due the color-gauge symmetry.

In Fig.~\ref{fig:energy-versus-delta}(c), where a finite color-flip (Rabi) field is present ($h_x/J = 3$, $k_T a = 0$), but it is spatially uniform, so the overall transition from extended to localized states remains consistent with the behavior observed in panels (a) and (b). This means that states are extended for 
$\Delta/J < (\Delta/J)_{c}^{\text{AA}} = 2$ and become localized beyond this threshold, because there are three copies of AAM, with different energy-reference shifts caused by $h_x/J$.

In Fig.~\ref{fig:energy-versus-delta}(d), where both color-flip (Rabi) fields and color-orbit coupling are present ($h_x/J = 3$, $k_T a = \pi/2$), a very rich structure emerges, 
because the color-flip (Rabi) fields are spatially non-uniform. 
Multiple mobility regions emerge across the energy spectrum. For fixed values of $\Delta /J < 2$, such as $\Delta /J = 1$, there are regions where the states are localized and where the states are extended. Similarly, for fixed values of 
$\Delta/J < 2$, such as $\Delta/J = 3$, 
similar mobility regions emerge. This is in sharp contrast with the standard results of AAM, where there are no mobility regions for
fixed $\Delta/J.$ Thus, Fig.~\ref{fig:energy-versus-delta}(d), reveals an important feature of the simultaneous existence of color-orbit coupling $(k_T a \ne 0)$ and color-flip (Rabi) fields $(h_x \ne 0)$: the existence of mobility regions.

When both $h_x \ne 0$ and $k_T a \ne 0$, the use of the CGT leads to spatially dependent 
color-flip (Rabi) fields with period $\ell_{h_x} = 2\pi/k_T a$ in the site index $n$, as seen 
in Eqs.~(\ref{eqn:hx-CGT}) and~(\ref{eqn:hy-CGT}). In contrast, the spatial period of the
bichromatic disorder in the site index $n$ is $ \ell_{\Delta} = 1/R_{\lambda} \approx 1.31$. The interplay of the periods $\ell_{h_x}$ and $\ell_{\Delta}$, as well the amplitudes $h_x $ and 
$\Delta $ affect the IPR, and thus the localization 
properties of the states.

\begin{figure}[t]
\centering
\includegraphics[width=0.48\textwidth]{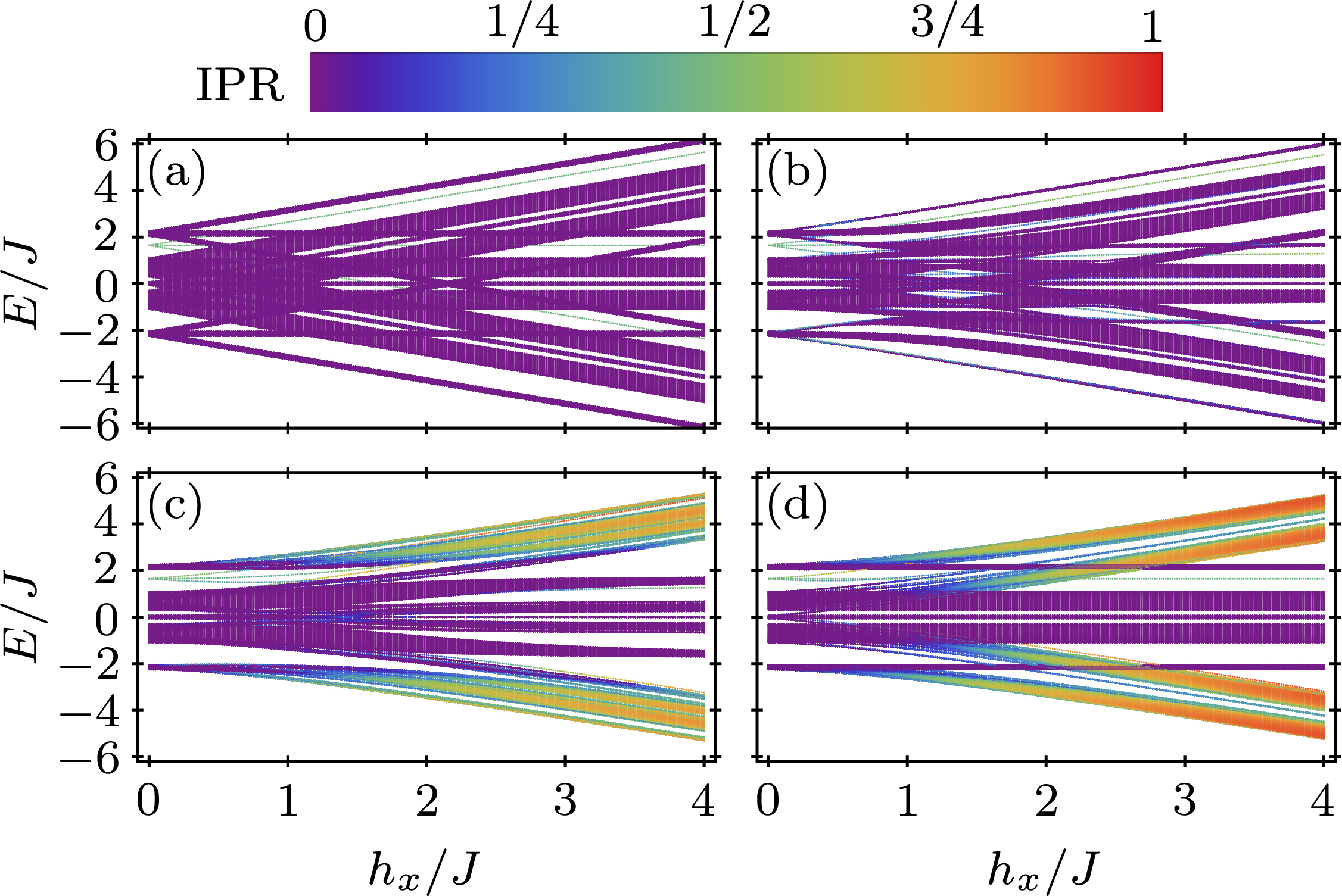}
\caption{Plots of $E/J$ versus $h_x/J$ showing mobility regions (violet) for different values of color-orbit parameter $k_T a$ with $\Delta/ J = 1$, $R_\lambda = 813 / 1064$, $N = 500$, $\varphi = 0$: (a) $k_T a = 0$; (b) $k_T a = \pi/4$; (c) $k_T a=3\pi/4$; (d) $k_T a=\pi$. 
The continuous IPR color scheme varies from violet for extended states to non-violet (blue to red)
for localized states. 
The emergence of mobility regions shows 
the interplay between the magnitude of the
color-flip (Rabi) $h_x$ and disorder fields $\Delta$ and their
spatial variation controlled by 
$k_T a$ and $2\pi R_{\lambda}$, respectively, within the color-gauge transformed Hamiltonian.}
\label{fig:energy-versus-hx}
\end{figure}

In Fig.~\ref{fig:energy-versus-hx}, we show the energy 
$E/J$ versus color-flip (Rabi) field $h_x/J$, for different values of the color-orbit parameter $k_Ta$ at fixed disorder $\Delta/J = 1$. The IPR is also shown using the same color code as in Fig.~\ref{fig:energy-versus-delta}.

In Fig.~\ref{fig:energy-versus-hx}(a), where $k_T a = 0$, the color-flip (Rabi) field is position independent, and we have three color copies of the AAM, where $E/J$ varies with $h_x/J$.
Since the disorder is color-independent, all bulk states are extended since 
$\Delta/ J = 1$
is below the critical value 
$(\Delta/J)^{AA}_{c} = 2$.
The only localized states are edge states that may appear in energy gaps.

In Figs.~\ref{fig:energy-versus-hx}(b),~\ref{fig:energy-versus-hx}(c) and~\ref{fig:energy-versus-hx}(d), where
$k_T a = \pi/4$, $k_T a = 3\pi/4$ and $k_T a = \pi$,
respectively, the color-flip (Rabi) field ($h_x \ne 0$) 
is spatially non-uniform, so some states with low effective kinetic energy, that is, low-energy (particles) and high-energy (holes) states, start to localize even for $\Delta/J = 1$, because the effective tunneling probability from site-to-site is reduced. For fixed $h_x/J$, we note that mobility (violet) regions of the states are reduced as $k_T a$ changes from $0$ to $\pi$, since the local color-flip (Rabi) field period $\ell_{h_x}$ changes from uniform (infinite period in $n$) to staggered (period 2 in $n$). 

\begin{figure}[t]
\centering
\includegraphics[width=0.48\textwidth]{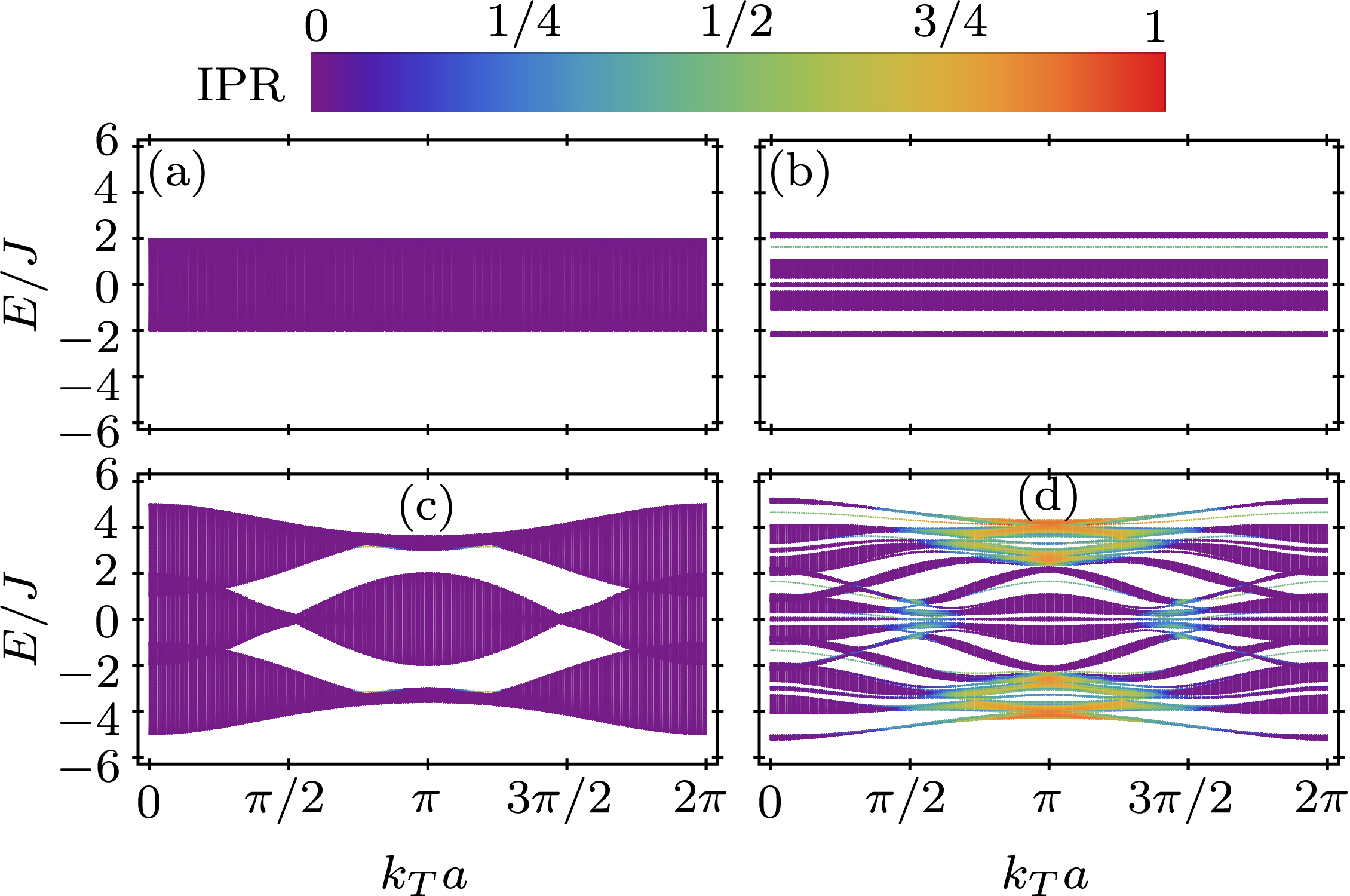}
\caption{Plots of $E/J$ versus $k_T a$ showing mobility regions (violet)  for different values of 
$h_x/J$ and $\Delta/J$ with $R_{\lambda} = 813 / 1064$, $N = 500$, $\varphi = 0$: 
(a) $h_x/J = 0$, $\Delta / J = 0$; 
(b) $h_x/J = 0$, $\Delta / J = 1$; 
(c) $ h_x / J = 3$, $\Delta / J = 0$; 
(d) $h_x /J = 3$, $\Delta/ J = 1$.
The continuous IPR color scheme varies from violet for extended states to non-violet (blue to red) for localized states. 
In panels (a) and (b), where $h_x/J = 0$, the energies $E/J$ are independent of $k_T a$ due to the color-gauge symmetry. In panels (c) and (d), where $h_x/ J \ne 0$, the energies $E/J$ change as a function of $k_T a$, due to the broken color-gauge symmetry. Degeneracies are lifted for $\Delta/J \ne 0$ when comparing panels (a) and (b), as well as panels (c) and (d).}
\label{fig:energy-versus-kTa}
\end{figure}

In Fig.~\ref{fig:energy-versus-kTa}, we show the energy spectrum $E/J$ versus color-orbit coupling parameter $k_Ta$ for selected values of $h_x/J$ and 
$\Delta/J$. The displayed IPR uses the same color code as Fig.~\ref{fig:energy-versus-hx}.
In Fig.~\ref{fig:energy-versus-kTa}, we highlight that $E/J$ and the IPR are symmetric about $k_T a = \pi$. This symmetry is a reflection of the 
chirality of the color-flip (Rabi) field ${\widetilde {\bf h}} (x_n)$ described by Eqs.~(\ref{eqn:hx-CGT}) and~(\ref{eqn:hy-CGT}): for $0 < k_T a < \pi$, ${\widetilde {\bf h}} (x_n)$ has a
right-handed polarization; for $k_T a = \pi$, ${\widetilde {\bf h}} (x_n)$ is staggered and linearly polarized;
for $\pi < k_T a < 2\pi$, ${\widetilde {\bf h}} (x_n)$ has a left-handed
polarization; and 
for $k_T a = \{ 0, 2\pi\}$, ${\widetilde {\bf h}} (x_n)$ is uniform.

In Fig~\ref{fig:energy-versus-kTa}(a), 
where $h_x/J = 0$ and $\Delta/ J = 0$, all the bulk states are delocalized (violet) for any $k_Ta$ due to the color-gauge symmetry and the absence of disorder. 

In Fig~\ref{fig:energy-versus-kTa}(b), 
where $h_x/J = 0$ and  $\Delta/ J = 1$, all the bulk states are delocalized (violet) for any $k_Ta$ due to the color-gauge symmetry and to weak disorder.
In this case, the system is described by three identical copies of the AAM, which 
are below the critical disorder 
$\Delta /J < (\Delta /J)_{c}^{\text{AA}} = 2$, necessary for localization. Thus, all bulk states are extended, and the mid-gap states are localized at one of the edges.

In Fig~\ref{fig:energy-versus-kTa}(c), 
where $h_x/J = 3$ and  $\Delta / J = 0$, all the bulk states are delocalized (violet) for any $k_Ta$ due to the absence of disorder. The emergence of gaps in the energy spectrum is due to the interplay between the strength of the color-flip (Rabi) field 
$h_x/J$ and its spatial modulation 
$\ell_{h_x} = 2\pi/k_T a$, controlled
by $k_T a$. 

In Fig~\ref{fig:energy-versus-kTa}(d), 
where $h_x/J = 3$ and  $\Delta / J = 1$, we can see clearly the existence of mobility regions (violet) separated by localized states, and the presence of energy gaps due to the interplay between $h_x/J$ and $\Delta / J$ and the respective periods $\ell_{h_x}$
and $\ell_{\Delta}$ of their spatial modulation. 
The strongest localization of bulk states occurs in the neighborhood of $k_T a = \pi$, where the spatial profile of the color-flip (Rabi) field is staggered, that is, $\ell_{h_x} = 2$. Here, the localization is strongest at lower energies (particles) and higher
energies (holes). Again, mid-gap states are localized at one of the edges.

Taken together, Figs.~\ref{fig:energy-versus-delta}, \ref{fig:energy-versus-hx} and \ref{fig:energy-versus-kTa} demonstrate that the interplay of disorder, colof-flip (Rabi) fields, and color-orbit coupling 
generates a rich mobility-region structure beyond that of the conventional Aubry-Andr\'e model, where mobility regions are absent. In addition, they show that 
localized bulk states can occur
for $\Delta/J < (\Delta/J)_{c}^{\text{AA}} = 2$. 
Lastly, not shown in Fig.~(\ref{fig:energy-versus-kTa}), extended bulk states can 
occur for $\Delta/J > (\Delta/J)_{c}^{\text{AA}} = 2$, a phenomenon that is not possible within standard Aubry-Andr\'e physics.

Although disorder alone produces a sharp transition at $(\Delta/J)^{AA}_c = 2$ independent of the energy $E/J$, 
finite color-flip (Rabi) fields and color-orbit coupling create intermediate regimes with mobility regions in the energy spectrum. The above results provide clear evidence that both color-orbit coupling and color-flip (Rabi) fields can either hinder or enhance localization
for fixed values of disorder. Furthermore, the mid-gap states that appear in Figs.~\ref{fig:energy-versus-delta}, \ref{fig:energy-versus-hx}, and \ref{fig:energy-versus-kTa} are localized at the edges and have a topological origin that further enriches the physics of this interplay, as discussed below in 
Sec.~\ref{sec:topological-considerations}.

Having established the localization properties and mobility regions that emerge in the energy spectra, we turn 
our attention next to a finite-size scaling analysis of the IPR. This investigation allows us to quantify the evolution of localization properties with system size $N$ and to construct phase diagrams covering extended and localized regions across the full parameter space.

\section{Finite Size Scaling} 
\label{sec:finite-size-scaling}

To characterize the localization properties of eigenstates in finite systems, it is necessary to distinguish between states that are extended
or localized in the thermodynamic limit, where the number of sites 
$N \to \infty$. 
A natural diagnostic is provided by the IPR, 
which scales differently for extended or localized
states as the number of sites $N$ increases.

For an extended state in a system of size $L = (N-1)a$, where $a$ is the lattice spacing, the IPR decreases 
as $1/N \sim 1/L$ for large $N$ or $L$. However, for a localized state, the IPR saturates to a finite value that is independent of $N$ and is in the 
interval $(0, 1]$ as $N \to \infty$.
This behavior is captured by the finite-size scaling 
relation~\cite{sanchez-palencia-2019}
\begin{equation}
    \mathrm{IPR} \;\sim\; L^{-\tau},    
\end{equation}
where the scaling exponent $\tau$ provides a quantitative measure of the 
localization properties of the state. Explicitly, $\tau$ is defined as
\begin{equation}
\label{eqs:tau}
    \tau = \frac{d \, \log_{10}(\mathrm{IPR})}{d \, \log_{10}(1/N)}.
\end{equation}
In this framework, $\tau = 1$ corresponds to a fully extended state, 
$\tau = 0$ corresponds to a fully localized state, and intermediate values 
$0 < \tau < 1$ may arise near transition points, reflecting the existence of critical states and the emergence of multifractal wavefunctions~\cite{sanchez-palencia-2019}. Thus, $\tau$ serves as a convenient order parameter for identifying localization-delocalization transitions and mapping mobility regions.

For the conventional Aubry-Andr\'e 
model, where there are no mobility regions, 
the finite-size scaling relation in Eq.~(\ref{eqs:tau}), the scaling exponent $\tau$ gives $\tau = 1$ $(\Delta/J < 2)$ for extended and $\tau = 0$ $(\Delta/J > 2)$ 
for localized states. 
At the critical point of the Aubry–André transition, wave functions are neither fully extended nor fully localized, but instead exhibit multifractal scaling. In this case, the scaling exponent takes an intermediate value
$\tau = 0.5$ at the critical value 
$\Delta/J = 2$
indicating the presence of critical states. 

\begin{figure}[tb]
\centering
\includegraphics[width=0.48\textwidth]{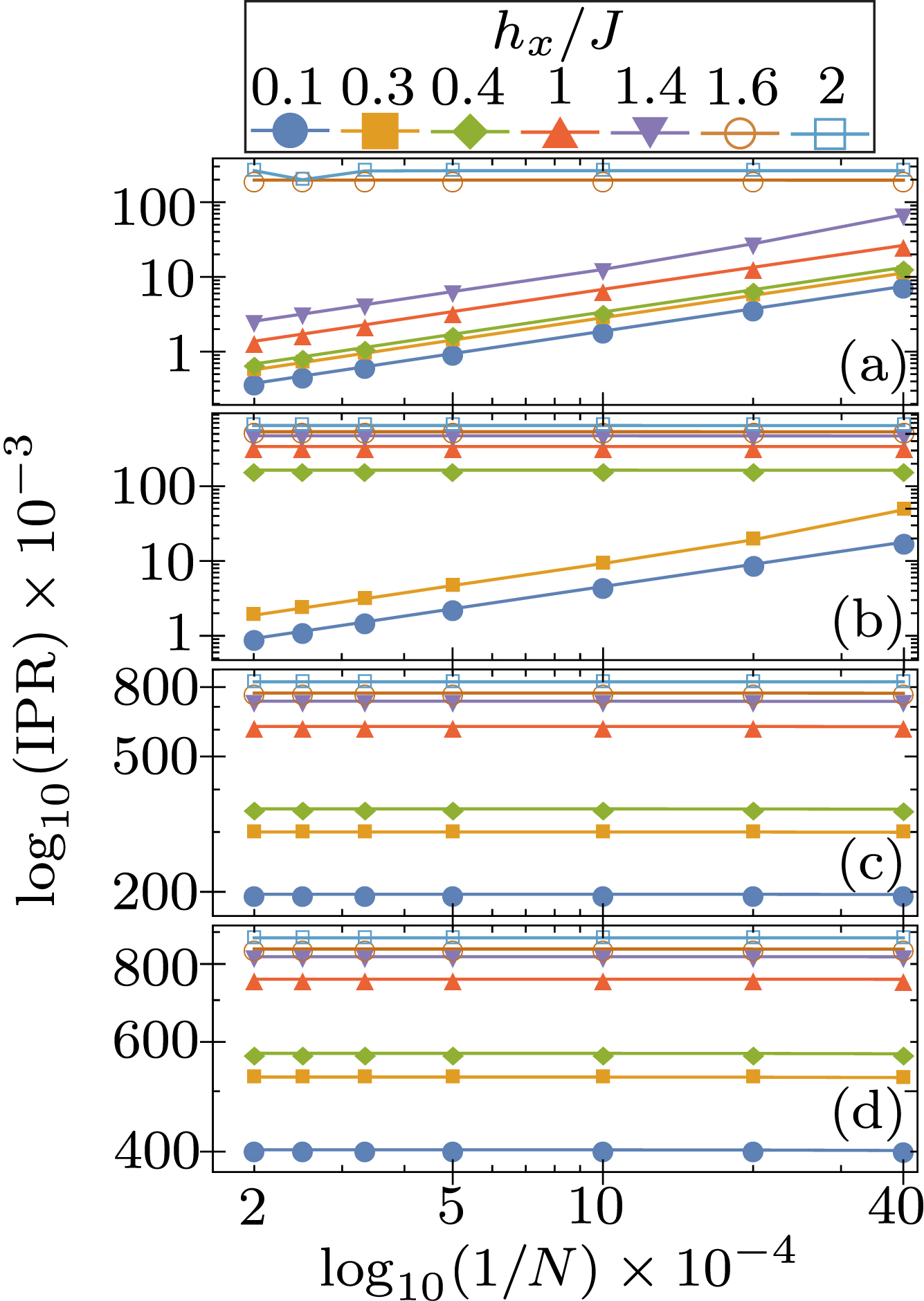}
\caption{Plots of IPR versus inverse system size $1/N$ for the ground state $(\nu \approx 0)$ in $\log_{10}$-$\log_{10}$ scale, 
with $k_T a =\pi$, and $\varphi = 0$. Each of the panels corresponds to a different disorder: 
(a) $\Delta/J = 0.5$; (b) $\Delta/J = 1$; 
(c) $\Delta/J = 1.5$; (d) $\Delta/J = 2$. 
The different lines in each panel correspond to $h_x/J = \{0.1, 0.3, 0.4, 1.0, 1.4, 1.6, 2.0\}$ for IPRs calculated with $N = \{250, 500, 1000, 2000, 3000, 4000, 5000\}$.  
}
\label{fig:IPR-scaling}
\end{figure}

\begin{figure}[t]
\centering
\includegraphics[width=0.4\textwidth]{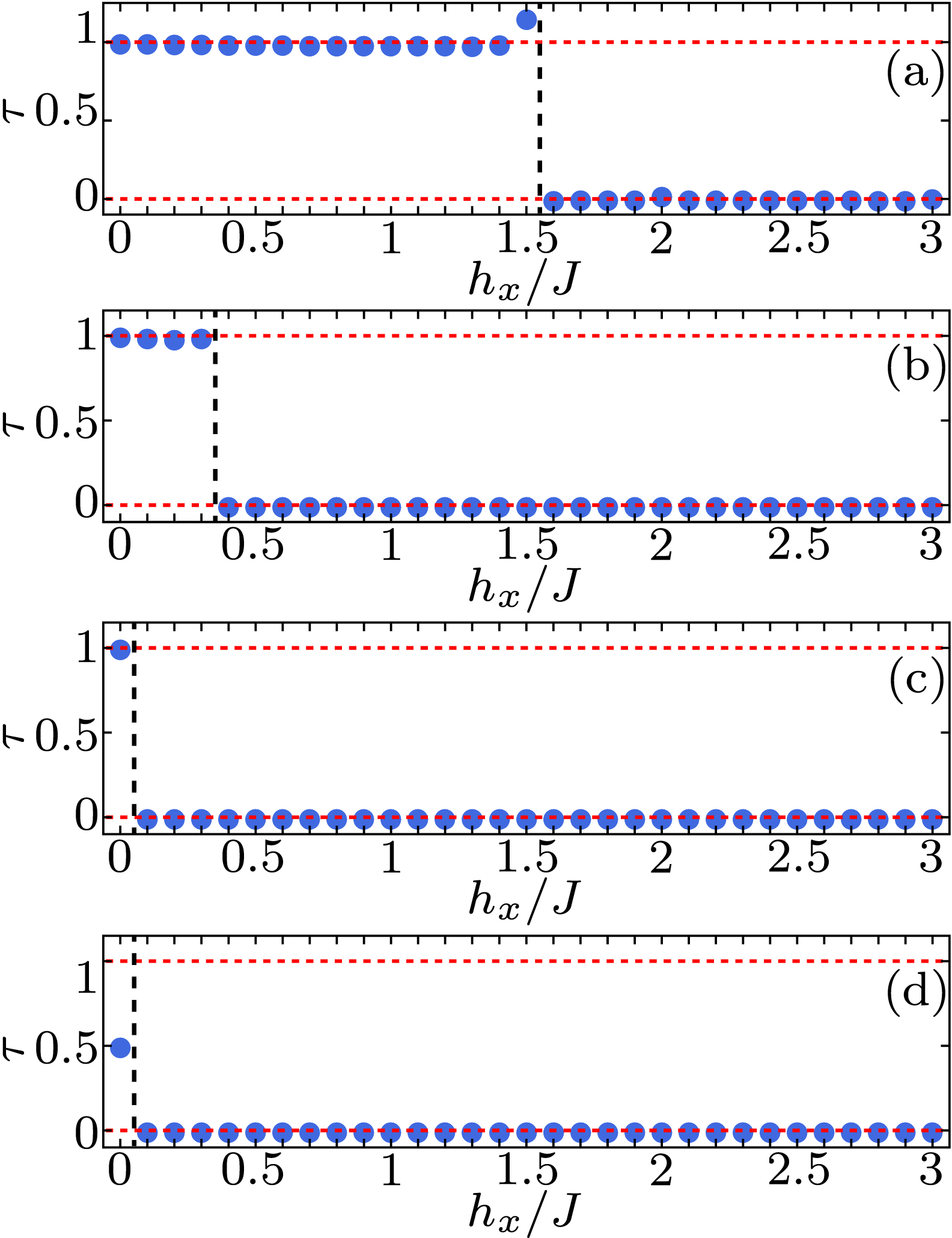}
\caption{Plots of $\tau$ as a function of $h_x/J$ for the lowest energy state $(\nu \approx 0)$ for $R_{\lambda} = 813/1064$, $k_T a = \pi$, and 
$\varphi = 0$. The values of disorder are: 
(a) $\Delta/ J = 0.5$; 
(b) $\Delta/ J = 1$; 
(c) $\Delta/ J = 1.5$; 
(d) $\Delta/ J = 2$.
The exponent $\tau$ distinguishes between extended ($\tau \approx 1$) and localized ($\tau \approx 0$) bulk states.
The special case of $\tau = 0.5$ at $h_x/J = 0$ for panel (d) 
is the Aubry-Andr\'e limit.}
\label{fig:tau-versus-hxdelta}
\end{figure}

For our model with color-orbit coupling and color-flip (Rabi) fields, we did not investigate in detail the possibility of critical states with multifractal structure at the boundaries of the mobility regions; instead, we focused on the exploration of jumps from $\tau = 1$ to $\tau = 0$, indicating the existence of a transition from extended to localized states.
To obtain $\tau$, we performed simulations with system sizes up to $N = 5000$ sites for various filling factors
\begin{equation}
\nu = N_p/N,
\label{eqn:filling-factor}
\end{equation}
where $N_p$ is the number of particles. The filling factor $\nu$ lies in the range $\left[0, 3\right]$, reflecting the number of fermions per site due to their three-color states.
The scaling exponent $\tau$ is obtained from the slope of 
$\log_{10}(\mathrm{IPR})$ versus $\log_{10}(1/N)$ at large $N$, 
where finite-size corrections are negligible. 
This procedure guarantees that $\tau$ reflects the 
thermodynamic limit $N \to \infty$.
In Fig.~\ref{fig:IPR-scaling}, we show an example of how
to obtain $\tau$ numerically, and plot $\tau$ explicitly in Fig.~\ref{fig:tau-versus-hxdelta}.

In Fig.~\ref{fig:IPR-scaling}, we show the scaling behavior of the IPR, that is, IPR versus inverse system size $1/N$ for the lowest energy state $(\nu \approx 0)$ in $\log_{10}$-$\log_{10}$ scale, 
with $k_T a =\pi$, and $\varphi = 0$. Each of the panels corresponds to a different disorder: 
(a) $\Delta/J = 0.5$; (b) $\Delta/J = 1$; 
(c) $\Delta/J = 1.5$; (d) $\Delta/J = 2$. The different lines in each panel correspond to $h_x/J = \{0.1, 0.3, 0.4, 1.0, 1.4, 1.6, 2.0\}$ for IPRs calculated with $N = \{250, 500, 1000, 2000, 3000, 4000, 5000\}$. The legend shows different values of 
$h_x/J$. These results are used to determine the scaling exponent $\tau$ in the thermodynamic limit $N \to \infty$. The transition from extended to localized states occurs when $\tau$ changes from 
$\tau =  1$ to $\tau = 0$.

In Fig.~\ref{fig:tau-versus-hxdelta}, we show the scaling exponent 
$\tau$ versus $h_x/J$ for the lowest energy state $(\nu \approx 0)$ at fixed values of $\Delta/J$ inferred from the
method outlined in Fig.~\ref{fig:IPR-scaling}.
The change from $\tau \approx 1$ for extended bulk states to $\tau \approx 0$ for localized bulk states serves as a useful diagnostic for localization. In all panels, the parameters used are $R_{\lambda} = 813/1064$, $k_T a = \pi$, 
$\varphi = 0$.
Panel (a) shows the results for   
$\Delta/ J = 0.5$, where 
$\tau$ changes from $1$ to $0$ at approximately 
$h_x/J = 1.5$. 
Panel (b) shows results for
$\Delta/ J = 1.0$,
where 
$\tau$ changes from $1$ to $0$ at approximately 
$h_x/J = 0.35$.
Panel (c) shows results for 
$\Delta/J = 1.5$,
where $\tau$ changes from $1$ to $0$ for $h_x/J$ between 
$0$ and $0.1$.
Panel (d) shows results for
$\Delta/ J = 2.0$,
where $\tau$ changes from $0.5$ to $0$ at 
arbitrarily small $h_x/J$.
The special case of $\tau = 0.5$ at $h_x/J = 0$ in panel (d) 
is the Aubry-Andr\'e limit for $\Delta/J = 2$
corresponding to a critical bulk state.

The examples illustrated in Figs.~\ref{fig:IPR-scaling} and~\ref{fig:tau-versus-hxdelta} demonstrate that the finite-size scaling exponent 
$\tau$ is a robust indicator of localization properties for bulk states. 
A deeper analysis of Fig.~\ref{fig:tau-versus-hxdelta}, where
$k_T a = \pi$ and $\varphi = 0$, shows
that transitions between extended $(\tau = 1)$ to localized 
$(\tau = 0)$ states occur as a function of $h_x/J$ even 
for values of $\Delta/J < 2 $, where the standard Aubry-Andr\'e model has only extended
states. The interplay between 
$\Delta/J$, $h_x/J$, and $k_T a$ on the IPR was already seen in the energy diagrams of Figs.~\ref{fig:energy-versus-delta}, \ref{fig:energy-versus-hx} and \ref{fig:energy-versus-kTa}.
Further analysis of $\tau$ for a large set
$\Delta/J$, $h_x/J$, and $k_T a$ for different
filling factors $\nu$ reveal the transitions from 
$\tau = 1$ to $\tau = 0$ in the thermodynamic limit. 
For all examples that we investigated, the number of 
sites $N = 500$ is generally already quite close to the 
thermodynamic limit, thus we use the IPR and its scaling behavior to explore next phase diagrams (contour plots of the IPR) in the
$h_x / J$ versus $k_T a$ plane, for various situations.

\section{Phase diagrams}
\label{sec:phase-diagrams}

\begin{figure}[t]
\centering
\includegraphics[width=0.48\textwidth]{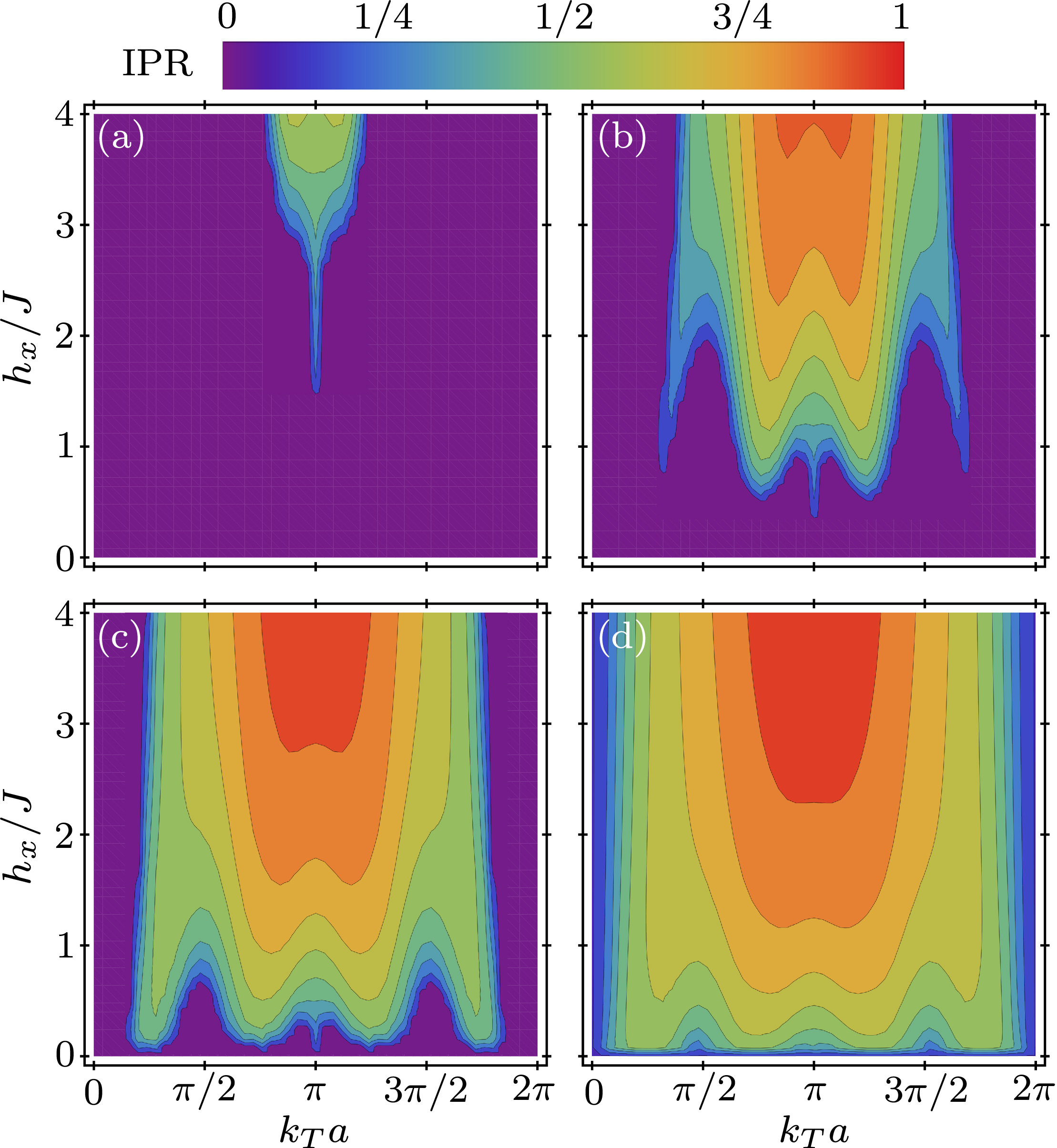}
\caption{Contour plots of the IPR in the $h_x/J$ versus $k_T a$ plane for the lowest energy state $(\nu \approx 0)$ at $R_{\lambda} = 813/1064$, system size $N = 500$, and phase $\varphi = 0$, for several disorder strengths: 
(a) $\Delta/J = 0.5$, (b) $\Delta/J = 1$, 
(c) $\Delta/J = 1.5$, and (d) $\Delta/J = 2$. 
Violet regions correspond to extended states, while non-violet regions indicate localized states. 
}
\label{fig:contourplots-IPR-hx-versus-kTa-fixed-Delta}
\end{figure}

\begin{figure}[t]
\centering
\includegraphics[width=0.48\textwidth]{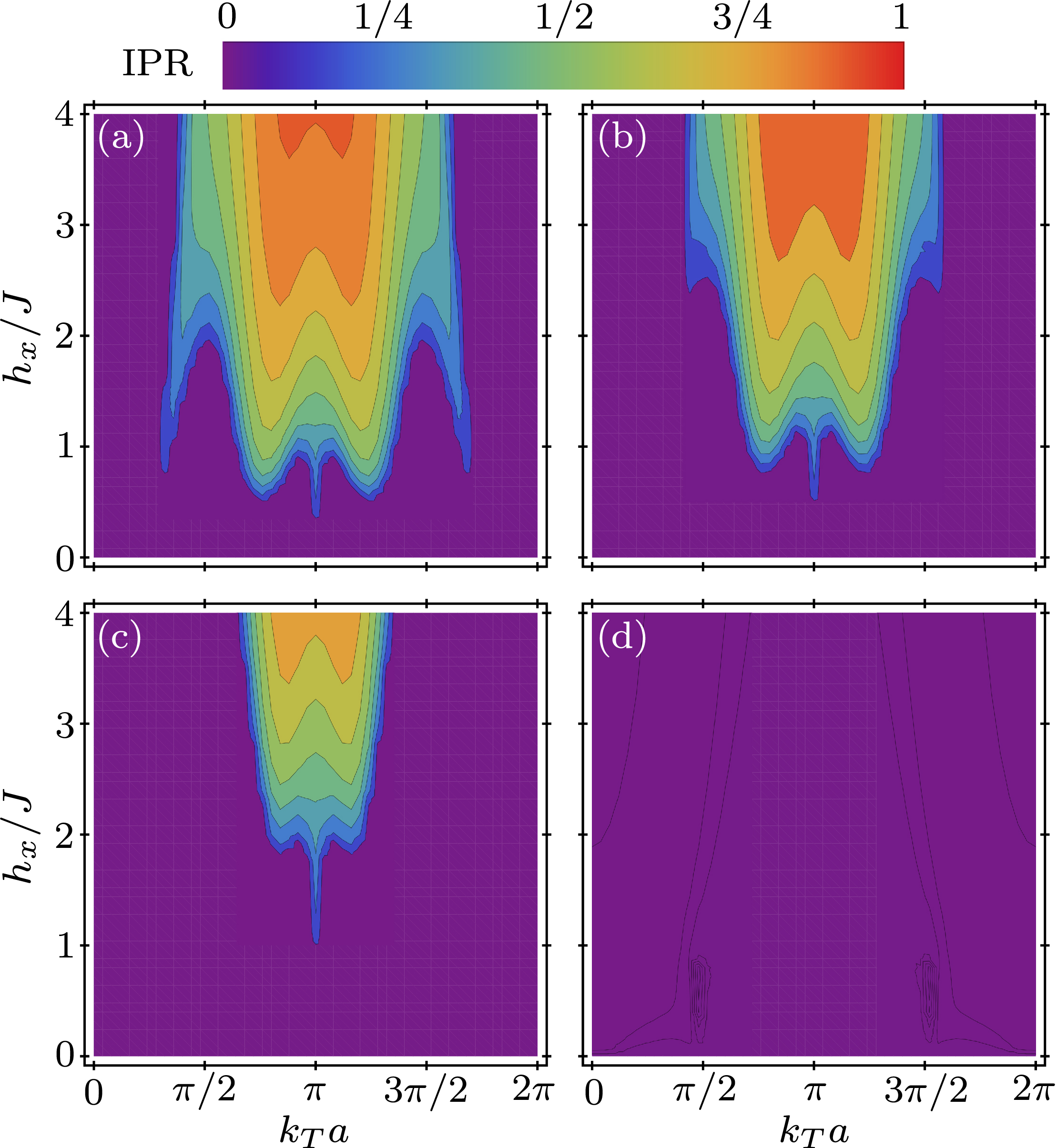}
\caption{Contour plots of the IPR in the $h_x/J$ versus $k_Ta$ plane 
for the lowest energy state $(\nu \approx 0)$ with 
$\Delta/J = 1$, $N=500$,  $R_{\lambda} = 813/1064$, and 
$\varphi^G = 0$, while varying the weak-lattice phases 
$\varphi^R = \varphi^B$ as follows: 
(a) $\varphi^R = \varphi^B = 0$; 
(b) $\varphi^R = \varphi^B = \pi/4$; 
(c) $\varphi^R = \varphi^B = \pi/2$; 
(d) $\varphi^R = \varphi^B = \pi$. Violet regions correspond to extended states, while non-violet regions indicate localized states.}
\label{fig:contourplots-IPR-hx-versus-kTa-fixed-phi}
\end{figure}

\begin{figure}[t]
\centering
\includegraphics[width=0.48\textwidth]{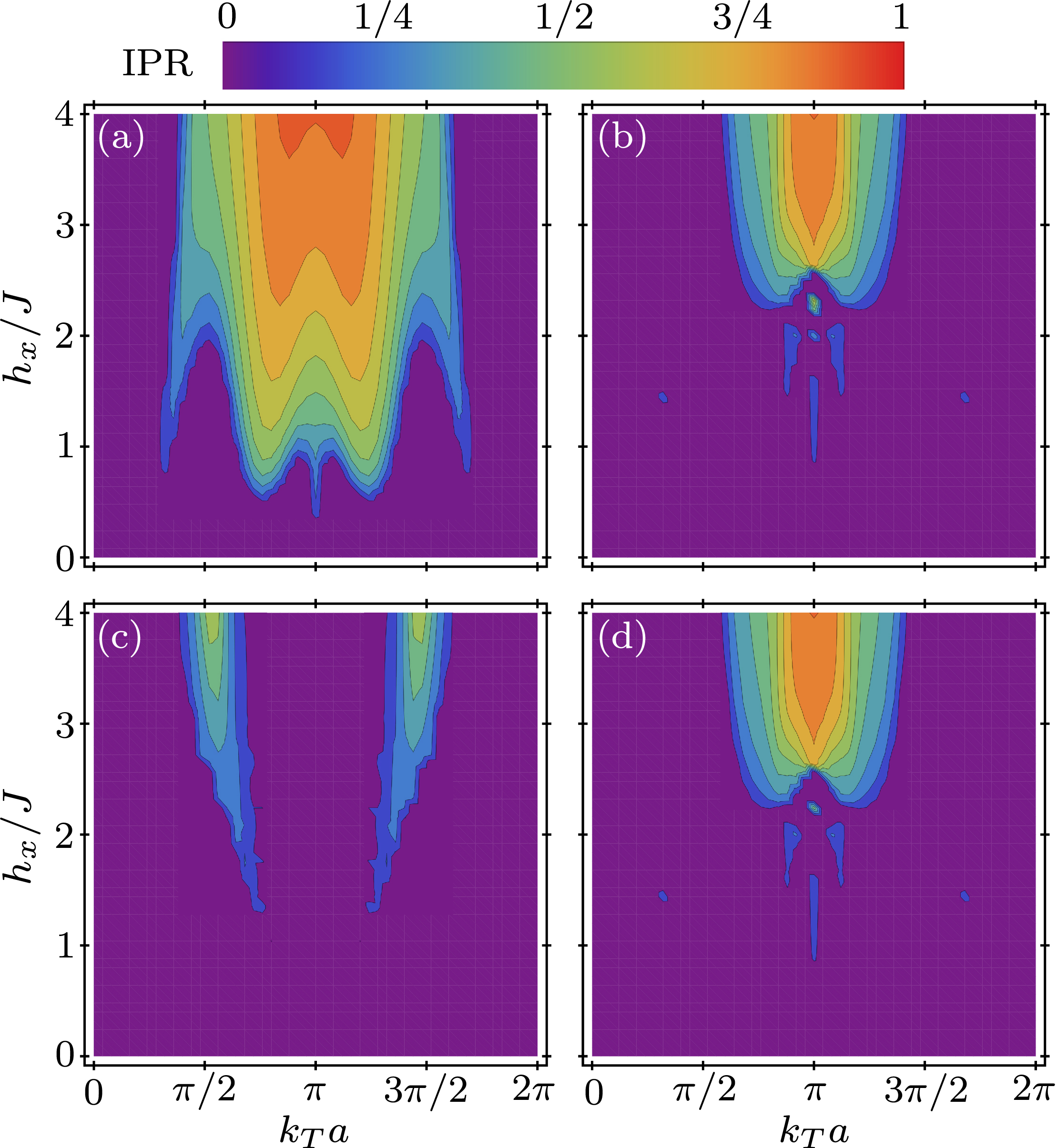}
\caption{Contour plots of the IPR in the $h_x/J$ versus $k_T a$ plane with $\Delta/J = 1$  for $R_{\lambda} = 813/1064$, system size $N = 500$, and phase $\varphi = 0$,
for several filling factors: 
    (a) $\nu \approx 0$, (b) $\nu = 3/4$, (c) $\nu = 3/2$, and (d) $\nu = 9/4$. 
Violet regions correspond to extended states, while non-violet regions indicate localized states.}
\label{fig:contourplots-IPR-hx-versus-kTa-fixed-nu}
\end{figure}

We show phases diagrams (contour plots of the IPR) 
for $h_x/J$ versus $k_T a$ in Figs.~\ref{fig:contourplots-IPR-hx-versus-kTa-fixed-Delta},~\ref{fig:contourplots-IPR-hx-versus-kTa-fixed-phi}, and~\ref{fig:contourplots-IPR-hx-versus-kTa-fixed-nu}, where the violet regions represent extended bulk states, while the other regions (non-violet) represent localized bulk states. 
In all panels of these figures,
the lines $h_x/J = 0$ describe the results of the standard Aubry-Andr\'e model with three degenerate bands and color-independent disorder, while
the lines $k_T a = 0$ and $k_T a = 2\pi$ describe the results of the Aubry-Andr\'e model with three non-degenerate bands ($h_x/J \ne 0)$ and color-independent disorder. Note that all phase diagrams are symmetric about $k_T a = \pi$ due to the chiral symmetry of the color-flip (Rabi) field ${\widetilde {\bf h}} (x_n)$ discussed during the analysis of Fig.~\ref{fig:energy-versus-kTa} in 
Sec.~\ref{sec:energy-spectra-and-inverse-partition-ratio}

In Fig.~\ref{fig:contourplots-IPR-hx-versus-kTa-fixed-Delta}, we show phase diagrams (contour plots of the IPR) of $h_x/J$ versus $k_T a$ for the lowest energy state $(\nu \approx 0)$ at $R_{\lambda} = 813/1064$, system size $N = 500$, and phase 
$\varphi = 0$, at disorder strengths: 
(a) $\Delta/J = 0.5$, (b) $\Delta/J = 1$, 
(c) $\Delta/J = 1.5$, and (d) $\Delta/J = 2$. 
Violet regions correspond to delocalized bulk states, while non-violet regions indicate localized bulk states. This figure demonstrates how tuning 
$\Delta/J$, $h_x/J$, and $k_T a$ controls the localization landscape. For the range of parameters $0 \le h_x/J \le 4$ and $0 \le k_T a \le 2\pi$, we use a grid of $50 \times 50$ points to compute the IPR contour plots. We note that changing the overall phase $\varphi$ to any value between $-\pi$ and $\pi$ has basically no effect on the contour plots and the IPR for bulk states, since $\varphi$ is essentially a reference phase. However, as we shall see in Sec.~\ref{sec:topological-considerations}, $\varphi$ has an important effect 
on the energy and location of edge states.

In Fig.~\ref{fig:contourplots-IPR-hx-versus-kTa-fixed-Delta}(a), 
we show the case for weaker disorder 
$\Delta/J = 0.5$. The violet region, corresponding to extended bulk states, dominates the phase diagram except near the high-symmetry line $k_T a = \pi$, where the spatially dependent color-flip (Rabi) 
field ${\widetilde {\bf h}} (x_n)$ is staggered, as seen 
from Eqs.~(\ref{eqn:hx-CGT}) and~(\ref{eqn:hy-CGT}).
Near $k_T a=\pi$, a sufficiently large $h_x/J$ induces a region of enhanced localization. This result is very different from
that of the standard Aubry-Andr\'e model, since localized bulk states occur for 
$\Delta/J < 2$ due to the spatially dependent color-flip (Rabi) field ${\widetilde {\bf h}} (x_n)$.

In Figs.~\ref{fig:contourplots-IPR-hx-versus-kTa-fixed-Delta}(b), (c), and (d), the disorder parameters are $\Delta/J = 1$, $\Delta/J = 1.5$, and 
$\Delta/J = 2$, respectively. Together, these figures show an expansion 
of the localized regions with increasing disorder. The intricate fingering pattern reflects the interplay between $h_x/J$ and $k_T a$ as disorder increases, due to commensurability effects between 
the spatially modulation of  ${\widetilde {\bf h}} (x_n)$,
in Eqs.~(\ref{eqn:hx-CGT}) and~(\ref{eqn:hy-CGT}),
and $\Delta (x_n) = \Delta \cos( 2\pi R_\lambda n +\varphi )$
in Eq.~(\ref{eqn:energy-disorder-color-diagonal}).
At stronger disorder, $\Delta/J = 2$, the lowest energy state becomes localized for almost all regions of the $h_x/J$ versus $k_T a$ phase space.

In Fig.~\ref{fig:contourplots-IPR-hx-versus-kTa-fixed-phi}, 
we show phase diagrams (contour plots of the IPR) in the 
$h_x/J$ versus $k_T a$ plane for the lowest energy state
$(\nu \approx 0)$ with parameters 
$\Delta/J = 1$, $N=500$, $R_\lambda = 813/1064$, and $\varphi^G = 0$, while varying the relative phases 
$\varphi^R = \varphi^B = \{0, \pi/4, \pi/2, \pi\}$
in panels (a), (b), (c), and (d), respectively. This figure reveals the effects of color-dependent disorder on localization via color-dependent phases $\{\varphi^R,\varphi^G, \varphi^B\}$, as $h_x/J$ and $k_T a$ change. For the range of parameters $0 \le h_x/J \le 4$ and $0 \le k_T a \le 2\pi$, we use a grid of $50 \times 50$ points to compute the IPR contour plots.

Following Figs.~\ref{fig:contourplots-IPR-hx-versus-kTa-fixed-phi}(a), (b), (c) and (d), where $\Delta/J = 1$,
we see a monotonic decrease of the localization region. The phase space for localization is largest when all the phases are the same, that is, 
$\varphi^G = \varphi^R = \varphi^B = 0$, and smallest when 
$\varphi^G = 0$ and $\varphi^R = \varphi^B = \pi$. This can be understood as a constructive interference effect that enhances disorder when all the color phases are the same (in phase), and 
a destructive interference effect that reduces disorder 
when $\varphi^G$ and $\varphi^B = \varphi^R$ differ by 
$\pi$ (out of phase). 
It is also important to mention (not shown) that the phase diagrams are symmetric with respect to the sign of the phases 
$\varphi^R = \varphi^B$, that is, the phase diagrams are symmetric for $\varphi^G = 0$ and 
$\varphi^B = \varphi^R = \{\pm \pi/4, \pm \pi/2, \pm \pi \}$.
Remarkably, in panel (d), where $\varphi^R = \varphi^B = \pi$, there are no localization regions; thus, the system is extended throughout the shown parameter space.
A similar monotonic decrease in the localization region is also observed for larger values of 
$\Delta/J$ 
(for example, $\Delta/J \ge 2$, not shown), however, even in the destructive interference regime 
($\varphi^G = 0$ and $\varphi^B = \varphi^R = \pi$), the effective disorder amplitude is sufficiently large to produce bulk localized states and localization regions in contrast to panel (d), where all bulk states are delocalized for $\Delta/J = 1$. The results shown in Fig.~\ref{fig:contourplots-IPR-hx-versus-kTa-fixed-phi} demonstrate that the phases of the weak-lattice lasers $\phi_2^\alpha$ (or $\varphi^\alpha = 2\phi_2^\alpha \pm \pi$) produce color-dependent 
localization and play an important role in the localization properties of the system, serving also as a control knob similar to $h_x/J$ and $k_T a$.

In Fig.~\ref{fig:contourplots-IPR-hx-versus-kTa-fixed-nu}, 
we show phase diagrams (contour plots of the IPR) in the $h_x/J$ versus $k_T a$ plane at fixed disorder 
$\Delta/J = 1$, for several filling factors $\nu = \{0, 3/4, 3/2, 9/4\}$, but for color-independent disorder, that is, $\varphi = \varphi^G =  \varphi^R = \varphi^B = 0$. For the range of parameters $0 \le h_x/J \le 4$ and $0 \le k_T a \le 2\pi$, we use a $50 \times 50$ grid to compute the IPR contour plots.

As seen in Figs.~\ref{fig:contourplots-IPR-hx-versus-kTa-fixed-nu}(a), (b), (c), and (d), the phase diagrams are remarkably sensitive to the filling 
factor $\nu$ defined in Eq.~(\ref{eqn:filling-factor}) or chemical potential $\mu$, and reflect the mobility regions discussed in Figs.~\ref{fig:energy-versus-delta},~\ref{fig:energy-versus-hx},~\ref{fig:energy-versus-kTa}.
Furthermore, these phase diagrams are essentially particle-hole symmetric, except for slight differences due to edge states, also reflecting the near
particle-hole symmetry already seen in Figs.~\ref{fig:energy-versus-delta},~\ref{fig:energy-versus-hx},~\ref{fig:energy-versus-kTa}.
The contour plots of the IPR are virtually the same for filling factors 
$\nu$ and $(3-\nu)$, with {\it half-filling} corresponding to $\nu = 3/2$. See panels (b) and (d), where $\nu = 3/4$ and $\nu = 9/4$,
respectively. The very small islands (almost point-like) in these panels correspond to edge states, while the larger islands (almost line-like) refer to bulk states. 

Having discussed the phase diagrams in the $h_x/J$ versus $k_T a$ plane for a variety of situations, we discuss next the color density of states and the dependence of the filling factor $\nu$ on the chemical potential $\mu$.
  
\section{Color Density of States }
\label{sec:color-density-of-states}

The density of states (DOS) provides insight into the degeneracy of energy levels and can reveal the existence of
gaps in the energy spectrum. In the absence of disorder, when the chemical potential $\mu$ lies within bulk energy gaps, the system is a band insulator if no edge states cross from one band to another. When edge states connect two bands, displaying a chiral nature, a topological insulator arises when the $\mu$ lies in the bulk gap. However, due to the presence of disorder, $\mu$ can lie in a region with finite DOS, but the states can be localized, thus producing a disorder-induced insulator. In the case of bichromatic disorder, the bulk insulating states are standard Aubrey-Andr\'e insulators when no color-flip fields or color-orbit coupling are present. Since the DOS can be potentially measured in ultracold atoms, we calculate it for our color states, highlight the mobility regions, and investigate the dependence of the filling factor $\nu$ on the chemical potential 
$\mu$.

Although in this section, we calculate only the total color density of states (TCDOS), we obtain its expression in terms of the local and color-dependent density of states, for future use. 

The TCDOS is obtained from the imaginary part of the resolvent operator
\begin{equation}
    \hat {\bf G} (z) = \frac{1}{z \hat{\mathbf{1}} - \hat{\mathbf{H}}}, 
\end{equation}
with $z$ being a complex energy.
Using the states $\vert n, \alpha \rangle$ and 
$\vert m, \beta \rangle$, where 
$n, m$ label sites and $\alpha, \beta$ label colors,
we write the matrix elements
$G_{nm}^{\alpha \beta} (z)  = 
\langle n, \alpha \vert \hat {\bf G} (z) 
\vert m, \beta \rangle$. Considering the energy 
eingenstates $\vert \psi (j) \rangle$, where $j$ labels
the eigenvalue $E_j$ of $\hat {\bf H}$, we use
the closure relation $\hat {\bf 1} = \sum_j \vert \psi (j) \rangle
\langle \psi (j) \vert$ before and after ${\hat {\bf G}} (z)$, to get  
\begin{equation}
G_{nm}^{\alpha \beta} (z) = 
\sum_{ij}
\psi_{n \alpha} (i) 
\langle \psi (i) \vert {\hat {\bf G} (z) \vert \psi (j) \rangle \psi_{m \beta} ^*(j)}.
\end{equation}
Since, $\vert \psi (i) \rangle$ and $\vert \psi (j) \rangle$ are eigensates of $\hat {\bf H}$, the matrix element
\begin{equation}
\langle \psi (i) \vert {\hat {\bf G} (z) \vert \psi (j) \rangle
= 
\frac{\delta_{ij}}{ z - E_j}},
\end{equation}
is used to obtain the result
\begin{equation}
G_{nm}^{\alpha \beta} (z) = 
\sum_{j}
\frac{\psi_{n \alpha} (j) \psi_{m \beta} ^*(j)}
{z - E_j}.
\end{equation}

The total color density of states $\rho (E)$
is obtained by using $z = E + {\rm i} \varepsilon$ from 
the relation
\begin{equation}
\rho(E) = - \frac{1}{\pi} 
\lim_{\varepsilon \to 0} \left\{ \Im  \left[ {\rm Tr}~{\hat {\bf G}} (E+ {\rm i}\varepsilon) \right] \right\}, 
\end{equation}
where the trace of the resolvent operator is 
\begin{equation}
{\rm Tr}~{\hat {\bf G}} (E + {\rm i}\varepsilon)
= 
\sum_{j, n, \alpha} 
\frac{\vert \psi_{n \alpha} (j) \vert^2 }
{E - E_j + {\rm i} \varepsilon}.
\end{equation}
leading to a simple expression for the imaginary part
\begin{equation}
\Im  \left[ {\rm Tr}~{\hat {\bf G}} (E+ {\rm i} \varepsilon) \right]
= 
- \sum_{j, n, \alpha} \vert \psi_{n \alpha} (j) \vert^2
\frac{\varepsilon} {(E - E_j)^2 + \varepsilon^2}.
\end{equation}
The limiting procedure $(\varepsilon \to 0)$ is performed
using the Lorentzian delta sequence 
\begin{equation}
\delta_{\varepsilon} (E - E_j) 
= \frac{1}{\pi} \frac{\varepsilon}{(E - E_j)^2 + \varepsilon^2},
\label{eqn:delta-sequence}
\end{equation}
such that the total color density of states (TCDOS)
becomes 
\begin{equation}
\rho (E) = 
\lim_{\varepsilon \to 0} 
\sum_{j, n, \alpha} \vert \psi_{n \alpha} (j) \vert^2
\delta_{\varepsilon} (E - E_j).
\label{eqn:total-color-density-of-states}
\end{equation}

We also express the TCDOS  as the sum 
\begin{equation}
\rho (E) = 
\sum_{j} \rho_j (E),
\label{eqn:total-color-density-of-states-sum}
\end{equation}
where the spectral weight associated with the eigenvalue $E_j$ is 
\begin{equation}
\rho_j (E) = 
\lim_{\varepsilon \to 0} 
\sum_{n, \alpha} \vert \psi_{n \alpha} (j) \vert^2
\delta_{\varepsilon} (E - E_j).
\label{eqn:spectral-weight-of-Ej}
\end{equation}

We note that $\rho (E)$ has dimensions of inverse energy, that is, $\left[ \rho (E) \right] = \left[ E \right]^{-1}$
and that the integral over energy, the TCDOS has the normalization
\begin{equation}
\int dE\rho (E) = 
\sum_{j, n, \alpha} \vert \psi_{n \alpha} (j) \vert^2
= 3 N
\end{equation}
since the total number of states is $3N$, that is, the product of the number of colors 
$(3)$ and the number of sites $(N)$. In contrast, the integral over energy for individual spectral weights is  
\begin{equation}
\int dE\rho_j (E) = 
\sum_{n, \alpha} \vert \psi_{n \alpha} (j) \vert^2
= 1,
\end{equation}
indicating the normalization of state $\vert \psi (j) \rangle$.

In this section, we do not discuss
the local color density of states (LCDOS) 
$\rho_{n \alpha} (E)$
and the color density of states 
$\rho_{\alpha} (E)$. However, we can easily obtain them from Eq.~(\ref{eqn:total-color-density-of-states}) 
leading to the LCDOS
\begin{equation}
\rho_{n \alpha} (E) = 
\lim_{\varepsilon \to 0} 
\sum_{j} \vert \psi_{n \alpha} (j) \vert^2
\delta_{\varepsilon} (E - E_j).
\label{eqn:local-color-density-of-states}
\end{equation}
and to the CDOS  
\begin{equation}
\rho_{\alpha} (E) = 
\lim_{\varepsilon \to 0} 
\sum_{j,n } \vert \psi_{n \alpha} (j) \vert^2
\delta_{\varepsilon} (E - E_j)
= \sum_n \rho_{n \alpha} (E).
\label{eqn:local-color-density-of-states}
\end{equation}
\begin{figure}[t]
\centering
\includegraphics[width=0.48\textwidth]{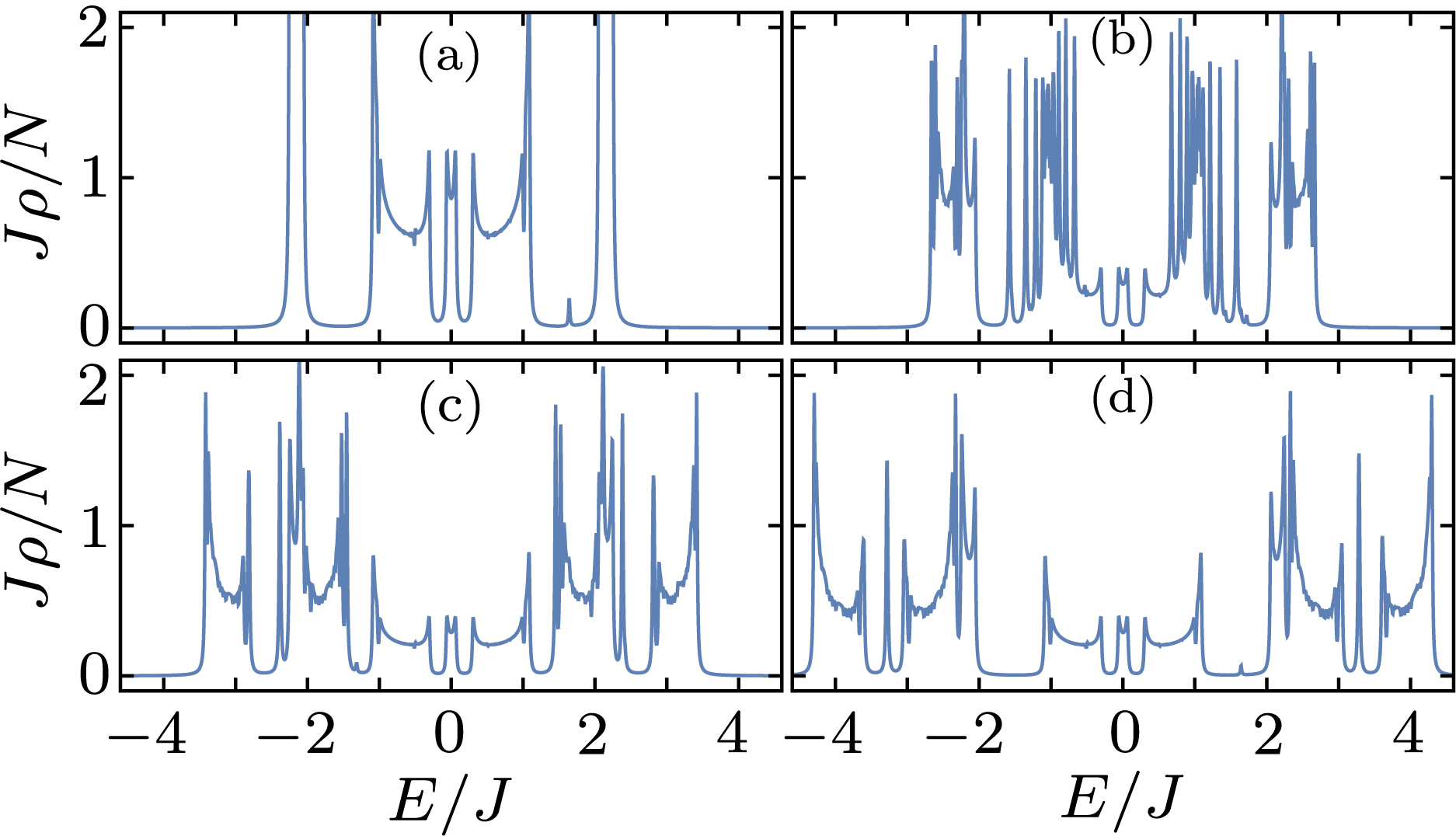}
\caption{Plot of the dimensionless total color density of states (TCDOS) $J \rho/N$ versus the dimensionless energy $E/J$ 
for $\Delta/J = 1$, 
$N = 500$, $R_{\lambda} = 813/1064$, $\varphi = 0$, and $k_T a = \pi$. The panels reflect different color-flip (Rabi) fields:
(a) $h_x/J = 0$, (b) $h_x/J = 1$, 
(c) $h_x/J = 2$, (d) $h_x/J = 3$, corresponding to vertical 
cuts in the energy spectrum $E/J$ vs $h_x/J$ shown in 
Fig.~{\ref{fig:energy-versus-hx}}(d).
}
\label{fig:DOS-versus-energy}
\end{figure}

\begin{figure}[t]
\centering
\includegraphics[width = \columnwidth]{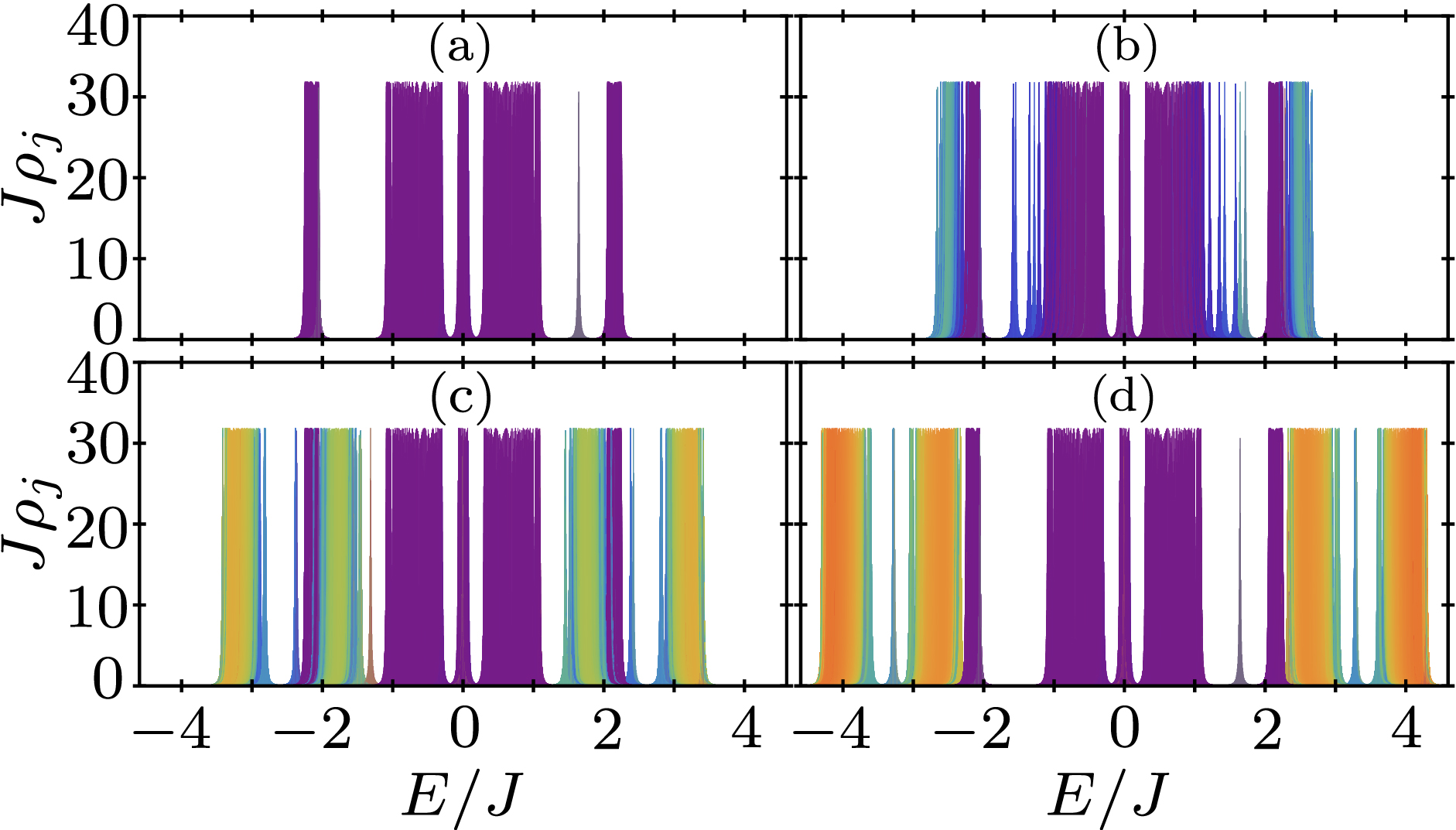}
\caption{
Plot of the dimensionless spectral weight $J \rho_j$ versus the dimensionless energy $E/J$ 
for $\Delta /J = 1$, 
$N = 500$, $R_{\lambda} = 813/1064$, $\varphi = 0$, and $k_T a = \pi$. The panels reflect different color-flip (Rabi) fields:
(a) $h_x/J = 0$, (b) $h_x/J = 1$, 
(c) $h_x/J = 2$, (d) $h_x/J = 3$, corresponding to vertical 
cuts in the energy spectrum $E/J$ vs $h_x/J$ shown in 
Fig.~{\ref{fig:energy-versus-hx}}(d).
The color code reflects IPR values shown in the legend of
Fig.~(\ref{fig:contourplots-IPR-hx-versus-kTa-fixed-nu}),
for example.}
\label{fig:DOS-versus-energy-ipr}
\end{figure}

In Figs.~\ref{fig:DOS-versus-energy} and~\ref{fig:DOS-versus-energy-ipr},
we use the small width $\varepsilon = J/100$ to show the dimensionless total color
density of states $J \rho (E)/N$ and the 
spectral weight $J\rho_j (E)$, respectively.
Here, $\rho (E) = \sum_j \rho_j (E)$ is the total color density of states described in Eqs.~(\ref{eqn:total-color-density-of-states}) and~(\ref{eqn:total-color-density-of-states-sum}), with 
$\rho_j (E) = \delta_{\varepsilon} (E - E_j)$ defined 
in Eq.~(\ref{eqn:spectral-weight-of-Ej}) taking into account the normalization of the 
state $\vert \psi (j) \rangle$.

In Fig.~\ref{fig:DOS-versus-energy}, we show the dimensionless TCDOS $J \rho (E)/N $ versus $E/J$ for 
$\Delta/J = 1$, $N = 500$, $R_{\lambda} = 813/1064$, 
$\varphi = 0$, and $k_T a = \pi$. 
The panels represent different color-flip (Rabi) fields:
(a) $h_x/J = 0$, (b) $h_x/J = 1$, 
(c) $h_x/J = 2$, (d) $h_x/J = 3$, corresponding to vertical 
cuts in the energy spectrum $E/J$ versus $h_x/J$ shown in 
Fig.~{\ref{fig:energy-versus-hx}}(d).
We choose the color-orbit coupling to be $k_T a = \pi$, 
because it corresponds to a staggered spatially dependent 
color-flip field ${\widetilde {\bf h}} (x_n)$,  
where localization is easier to achieve.
The increasing energy range (overall bandwidth) of the TCDOS reflects the growing values of $h_x/J$. The main purpose of this figure is to highlight the total color density of states, energy gaps, and the location of mid-gap edge states. For example, in panel (a), the small 
peak between $E/J = 1$ and $E/J = 2$ is a mid-gap edge state, while, in panel (d) the isolated peak between $E/J = 3$ and $E/J = 3.5$ is also a mid-gap edge state.

In Fig.~\ref{fig:DOS-versus-energy-ipr}, we plot all dimensionless spectral weights $J \rho_j (E)$ (color-coded by their IPR) versus $E/J$ for energy eigenstates $\vert \psi (j) \rangle$. 
The parameters used are the same as
in Fig.~\ref{fig:DOS-versus-energy-ipr}, that is, 
$\Delta/J = 1$, $N = 500$, $R_{\lambda} = 813/1064$, 
$\varphi = 0$, and $k_T a = \pi$. 
The panels represent different color-flip (Rabi) fields:
(a) $h_x/J = 0$, (b) $h_x/J = 1$, 
(c) $h_x/J = 2$, (d) $h_x/J = 3$, corresponding to vertical 
cuts in the energy spectrum $E/J$ vs $h_x/J$ shown in 
Fig.~{\ref{fig:energy-versus-hx}}(d).
The extended (localized) states are indicated by the violet (non-violet) color, matching the IPR values 
displayed in Fig.~{\ref{fig:energy-versus-hx}}(d), and revealing the
mobility regions. Since we use the delta
sequence $\delta_\varepsilon (E)$, defined
in Eq.~(\ref{eqn:delta-sequence}), to plot
$J\rho_j (E)$, all the peaks reach a maximum 
at $100/\pi$. In some cases, the small deviations from the maximum is due to the sampling of $E/J$, which is done in increments of $1/1000$.

From the TCDOS in Eq.~(\ref{eqn:total-color-density-of-states}), we obtain the filling factor 
\begin{equation}
\nu(\mu) = \frac{N_{\rm st}(\mu)}{N}
 = \frac{1}{N} \int_{E_{\rm min}}^{\mu} dE \, \rho(E),
\end{equation}
where $N_{\rm st}(\mu)$ is the number of occupied states up to the chemical potential $\mu$ and $N$ is the total number of sites. 
The values of $\mu$ range from $E_{\rm min}$ (the minimum energy) 
to $E_{\rm max}$ (the maximum energy), while the values of $\nu (\mu)$
range from $\nu (E_{\rm min}) = 0 $ to $\nu (E_{\rm max}) = 3$.

\begin{figure}[t]
\centering
\includegraphics[width = \columnwidth]{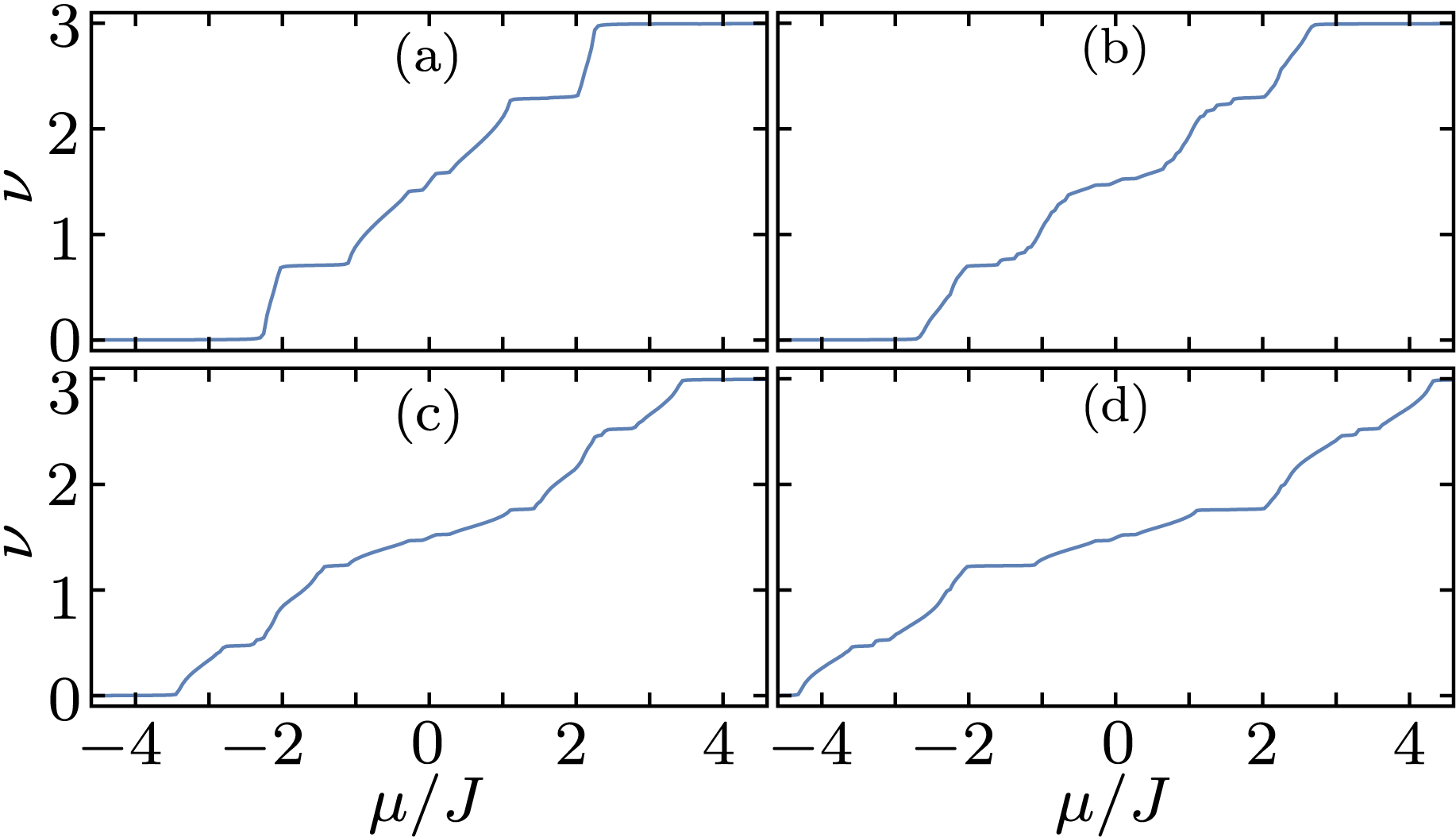}
\caption{Plots of the filling factor $\nu$ versus dimensionless chemical potential $\mu/J$
for $\Delta/J = 1$, 
$N = 500$, $R_{\lambda} = 813/1064$, $\varphi = 0$, and $k_T a = \pi$. The panels reflect different color-flip (Rabi) fields:
(a) $h_x/J = 0$, (b) $h_x/J = 1$, 
(c) $h_x/J = 2$, (d) $h_x/J = 3$, corresponding to vertical 
cuts in the energy spectrum $E/J$ vs $h_x/J$ shown in 
Fig.~{\ref{fig:energy-versus-hx}}(d).
}
\label{fig:filling-factor-versus-mu}
\end{figure}

In Fig.~\ref{fig:filling-factor-versus-mu}, we display $\nu(\mu)/J$ versus $\mu/J$ for the same set of parameters used in Fig.~\ref{fig:DOS-versus-energy} and in Fig.~\ref{fig:DOS-versus-energy-ipr}. For each panel in Fig.~\ref{fig:filling-factor-versus-mu}, the filling factor $\nu$ increases as $\mu/J$ grows, exhibiting plateaus corresponding to energy gaps where the TCDOS vanishes. 
These plateaus indicate incompressible states and describe band insulators. Furthermore, the presence of mid-gap edge states potentially indicates the existence of topological band insulators. 

In Fig.~\ref{fig:filling-factor-versus-mu}(a),
where $h_x/J = 0$, for example, plateaus appear at $\nu \approx \{ 0.72, 1.44, 1.56, 2.34 \}$.
In Fig.~\ref{fig:filling-factor-versus-mu}(b), where $h_x/J = 1$, for example, 
plateaus appear at $\nu \approx \{0.72, 1.48, 1.52, 2.34 \}$.
In Fig.~\ref{fig:filling-factor-versus-mu}(c), where $h_x/J = 2$, and in Fig.~\ref{fig:filling-factor-versus-mu}(d), where $h_x/J = 3$, for example, share a common set of plateaus at
$\nu \approx \{0.48, 1.24, 1.48, 1.52, 1.78, 2.56 \}$, 
but the plateaus' widths are different.
The plateaus in $\nu (\mu)$ characterize the specific densities at which the system is a band insulator, that is, $\mu$ lies in a band gap, as seen in analogous panels of 
Fig.~{\ref{fig:DOS-versus-energy-ipr}}.

When we correlate corresponding panels between Fig.~\ref{fig:filling-factor-versus-mu} and Fig.~\ref{fig:DOS-versus-energy-ipr}, we see that localized states (non-violet) are still compressible, since $d \nu/d{\mu} \ne 0$, but fermions occupying these states are not mobile, that is, they are not within mobility regions (violet). These localized states, outside the mobility regions, correspond to Aubry-Andr\'e insulators, a
terminology that we use due to the bichromatic disorder.

Having discussed the total color density of states, the spectral weights of eigenstates with their IPR and connected mobility regions to the filling factor and compressibility, we are ready to classify the emergent phases during the interplay between 
disorder, color-orbit coupling, and color-flip (Rabi) fields. 
For this purpose, we explore next the connection to the color Harper model.

\section{Color Harper Model}
\label{sec:color-harper-model}

In this section, we map our 1D disorder Hamiltonian into the 2D color Harper model~\cite{sa-de-melo-2021, sa-de-melo-2021b}. For this purpose, we use the first quantization matrix representation of ${\bf H}$, given in 
Eq.~(\ref{eqn:hamiltonian-matrix}), to map it into the color-dependent Harper's matrix (CDHM)~\cite{sa-de-melo-2021b}.
The diagonal (local) block matrices ${\boldsymbol\Gamma}_{nn}$, 
shown in Eqs.~(\ref{eqn:hamiltonian-local-block}) and~(\ref{eqn:hamiltonian-local-block-diagonal-elements}),
are mapped into the ${\bf A}_n$ blocks of the CDHM
and the off-diagonal (hopping) block matrices $-{\bf J}_{n, n-1}$,
shown in Eq.~(\ref{eqn:hamiltonian-hopping-block})
are mapped into the ${\bf B}^*$ blocks of the CDHM~\cite{sa-de-melo-2021b}. This mapping is perfect for a color-independent disorder
$\Delta^{\alpha \alpha} = \Delta$ with color-independent energy references set to zero, that is, $\varepsilon_0 + \eta^{\alpha \alpha} = \varepsilon_0 + \eta = 0$ in Eq.~(\ref{eqn:hamiltonian-local-block-diagonal-elements}).

We consider our matrix Hamiltonian in Eq.~(\ref{eqn:hamiltonian-matrix}), and relate it to a tight-binding model defined in a two-dimensional lattice, where the $x$ direction is in real space and the $y$ direction is a synthetic dimension controlled 
by the phase $\varphi$. The color Harper model has a tri-diagonal block structure that couples nearest-neighbor sites and exhibits discrete translational invariance along the $y$ direction, enabling a momentum-space
description along $y$. This formulation is a generalization of the well-known Harper model, extended to three-internal state (color or pseudospin-1) fermions, that is, the 
2D color Harper model.

The Hamiltonian matrix for our 1D color-disorder model is %
\begin{equation}
    \mathbf{H} = 
    \begin{pmatrix}
        \ddots & -\mathbf{J}_{\bar2\bar1} & 0 & 0 & 0 \\
        -\mathbf{J}_{\bar1\bar2} & \boldsymbol{\Gamma}_{\bar1\bar1} &
        -\mathbf{J}_{\bar1 0} &  0 & 0 \\
        0 & -\mathbf{J}_{0\bar1} & \boldsymbol{\Gamma}_{00} &
        -\mathbf{J}_{01} & 0 \\
        0 & 0 & -\mathbf{J}_{10} &
        \boldsymbol{\Gamma}_{11} & -\mathbf{J}_{12} \\
        0 & 0 & 0 & -\mathbf{J}_{21} & \ddots \\
    \end{pmatrix},
\label{eqn:hamiltonian-matrix-for-mapping}
\end{equation}
where the on-site matrix 
\begin{equation}
\label{eqs:Gamma-Explicit-Matrix}
\boldsymbol{\Gamma}_{nn} = \begin{pmatrix}
\varepsilon_R(\varphi^R) & -h_x/\sqrt{2} & 0 \\
-h_x/\sqrt{2} & \varepsilon_G(\varphi^G) & -h_x/\sqrt{2} \\
0 & -h_x/\sqrt{2} & \varepsilon_B(\varphi^B)
\end{pmatrix}
\end{equation}
has diagonal elements
\begin{equation}
\Gamma_{nn}^{\alpha \alpha}
= 
\varepsilon_{\alpha}(\varphi^\alpha) = 
\varepsilon_0 + \eta^{\alpha \alpha} + \Delta^{\alpha\alpha}\cos\left(2\pi R_\lambda n +\varphi^\alpha\right),
\end{equation}
with $\alpha \in \{ R,G,B \}$.
In the presence of color-orbit coupling characterized by $k_T$, the hopping matrix between sites $n$ and $m$ $(n \ne m)$ is 
\begin{equation}
\label{eqn:Jhopping-matrix}
    \mathbf{J}_{nm} = J_{nm} (0)
    \begin{pmatrix}
         e^{i k_T \delta x_{nm}} & 0 & 0 \\
         0 & 1 & 0 \\
         0 & 0 & e^{-i k_T \delta x_{nm}}       
    \end{pmatrix},
\end{equation}
where $J_{nm} (0) = J_{nm} (k_T = 0)$, defined in Eq.~(\ref{eqs:Jnm-kT=0}), 
sets the energy scale for hopping
and $\delta x_{nm} = x_n - x_m = (n-m)a$ describes the separation between sites $n$ and $m$.

To map the Hamiltonian in Eq.~(\ref{eqn:hamiltonian-matrix-for-mapping})
into the CDHM~\cite{sa-de-melo-2021b}, 
we compare ${\boldsymbol \Gamma}_{nn}$ and 
${\bf J}_{nm}$ with the matrices ${\bf A}$ and ${\bf B}$ of the CDHM. The CDHM has dimensions 
$3N_x \times 3N_y$, and contains the local $3 \times 3$ matrices 
\begin{equation}
{\bf A}_n =
\begin{pmatrix}
A_{nR} & -{h_x} / \sqrt{2} & 0 \\
-{h_x} / \sqrt{2} & A_{n G} & -{h_x} / \sqrt{2} \\
0 & -{h_x} / \sqrt{2} & A_{m B}
\end{pmatrix},
\end{equation}
where the diagonal elements are 
\begin{equation}
A_{nR} = A_{nG} = A_{nB} = -2t_y \cos(k_y a_y + 2\pi \alpha n).
\label{eqn:A-diagonal-matrix-elements}
\end{equation}
Here, $\alpha = \Phi / \Phi_0 \) is the magnetic flux ratio, where $\Phi = B a_x a_y$  is the magnetic flux through a single plaquette of area $a_x a_y$, and $\Phi_0 = hc/e$ is the magnetic flux quantum. Here, we used the Peierls substitution $k_y \to k_y 
-qA_y/\hbar c$, with charge $q = -e$ and and vector potential component $A_y = B x$, leading to $k_y a_y \to k_y a_y + e B x a_y/\hbar c$ in the CGS system. Writing $x = n a_x$ gives 
$k_y a_y \to k_y a_y + e B n a_x a_y/\hbar c$. The last term
is then simply $2\pi \alpha n$, as it appears in 
Eq.~(\ref{eqn:A-diagonal-matrix-elements}).

Also, due to translational invariance along the $y$ direction, $k_y$ represents the momentum along $y$. In comparing the matrices 
${\boldsymbol \Gamma}_{nn}$, for color-independent disorder, with ${\bf A}_{n}$,
we can immediately see the mapping:
$\varepsilon_0 + \eta \to 0$, $\Delta \rightarrow -2t_y$, $2\pi R_{\lambda} n \rightarrow 2\pi n \alpha$, $\varphi \rightarrow k_y a$ and  $h_x \rightarrow h_x$, as shown in 
Tab.~\ref{tab:1D-to-2D-mapping} fully.

\begin{table}[t] 
\centering
\begin{tabular}{| c | c | c |}
\hline
1D disorder model &  & 2D color Harper model \\ 
\hline
$\varepsilon_0 + \eta$ & $\leftrightarrow$  & $0$ \\
\hline
$\Delta$ & $\leftrightarrow$  & $-2t_y$ \\
\hline
$2\pi R_{\lambda} n$ & 
$\leftrightarrow$  & $2\pi \alpha n$ 
\\
\hline
$R_{\lambda} $ & 
$\leftrightarrow$  & $\alpha = \Phi/\Phi_0$ \\
\hline
$\varphi$ & 
$\leftrightarrow$  & $k_y a_y$ 
\\
\hline
$h_x$ & 
$\leftrightarrow$  & $h_x$ 
\\
\hline
$J$ & 
$\leftrightarrow$  & $t_x$ 
\\
\hline
$a$ & 
$\leftrightarrow$  & $a_x$ 
\\
\hline
\end{tabular}
\caption{Mapping of parameters from the 1D disorder model to the 2D color Harper model. This mapping helps clarify topological properties of the edge states in the 1D disorder model, where $\varphi$ plays the role of a 
synthetic dimension.}
\label{tab:1D-to-2D-mapping}
\end{table}

The off-diagonal block matrices ${\bf B}$ in 
the CDHM~\cite{sa-de-melo-2021b} describe 
color-dependent hopping along the $x$ direction
\begin{equation}
{\bf B}^* =
\begin{pmatrix}
-t_x e^{+ik_T a_x} & 0 & 0 \\
0 & -t_x & 0 \\
0 & 0 & -t_x e^{-ik_T a_x}
\end{pmatrix},
\end{equation}
where $k_T a_x$ represent the color-dependent
phase of hopping matrix elements. 
Since, we are including only nearest neighbor
hopping, a comparison between ${\bf B}^*$ and 
\begin{equation}
\label{eqn:Jhopping-matrix}
\mathbf{J}_{10} = J_{10} (0)
\begin{pmatrix}
e^{i k_T a} & 0 & 0 \\
0 & 1 & 0 \\
0 & 0 & e^{-i k_T a}      
\end{pmatrix},
\end{equation}
is sufficient.
The the connection between the 1D color-disorder model and the 2D color Harper model, requires the identification 
${\bf J}_{10} = -{\bf B}^*$, and $J_{10} = J$,
with $J$ real $(J \in \mathcal{R})$,
leading to the mapping:
$J \rightarrow t_x$, and $a \to a_x$ (See 
Tab.~\ref{tab:1D-to-2D-mapping}).

For a color-independent disorder, the Hamiltonian matrix in 
Eq.~(\ref{eqn:hamiltonian-matrix-for-mapping}) becomes
\begin{equation}
\label{eqn:hamiltonian-matrix-for-color-independent-disorder}
\mathbf{H} (\varphi) = 
\begin{pmatrix}
\ddots & -\mathbf{J} & 0 & 0 & 0 \\
-\mathbf{J}^* & \boldsymbol{\Gamma}_{\bar1\bar1} &
-\mathbf{J}  &  0 & 0 \\
0 & -\mathbf{J}^* & \boldsymbol{\Gamma}_{00} &
-\mathbf{J} & 0 \\
0 & 0 & -\mathbf{J}^* &
\boldsymbol{\Gamma}_{11} & -\mathbf{J} \\
0 & 0 & 0 & -\mathbf{J}^* & \ddots \\
\end{pmatrix},
\end{equation}
where the diagonal matrices are 
\begin{equation}
\label{eqs:gamma-nn-color-independent}
\boldsymbol{\Gamma}_{nn} = \begin{pmatrix}
\varepsilon (\varphi) & -h_x/\sqrt{2} & 0 \\
-h_x/\sqrt{2} & \varepsilon (\varphi)  & -h_x/\sqrt{2} \\
0 & -h_x/\sqrt{2} & \varepsilon (\varphi)
\end{pmatrix},
\end{equation}
with $\varepsilon (\varphi) = 
\Delta \cos\left( \varphi + 2\pi R_\lambda n \right)$, 
and the hopping matrix along the x direction is 
\begin{equation}
\label{eqn:hopping-matrix-color-independent-disorder}
\mathbf{J} = J
\begin{pmatrix}
e^{-i k_T a} & 0 & 0 \\
0 & 1 & 0 \\
0 & 0 & e^{+i k_T a}      
\end{pmatrix}.
\end{equation}

With the correspondence between the 1D color-independent disorder model and the 2D color Harper model, 
listed in Tab.~\ref{tab:1D-to-2D-mapping}, we can use our knowledge of its color Hofstadter spectrum~\cite{hofstadter-1976, sa-de-melo-2021b} to understand topological properties of the emergent phases and their edge states. 

\begin{figure}[tb]
\centering
\includegraphics[width=0.47\textwidth]{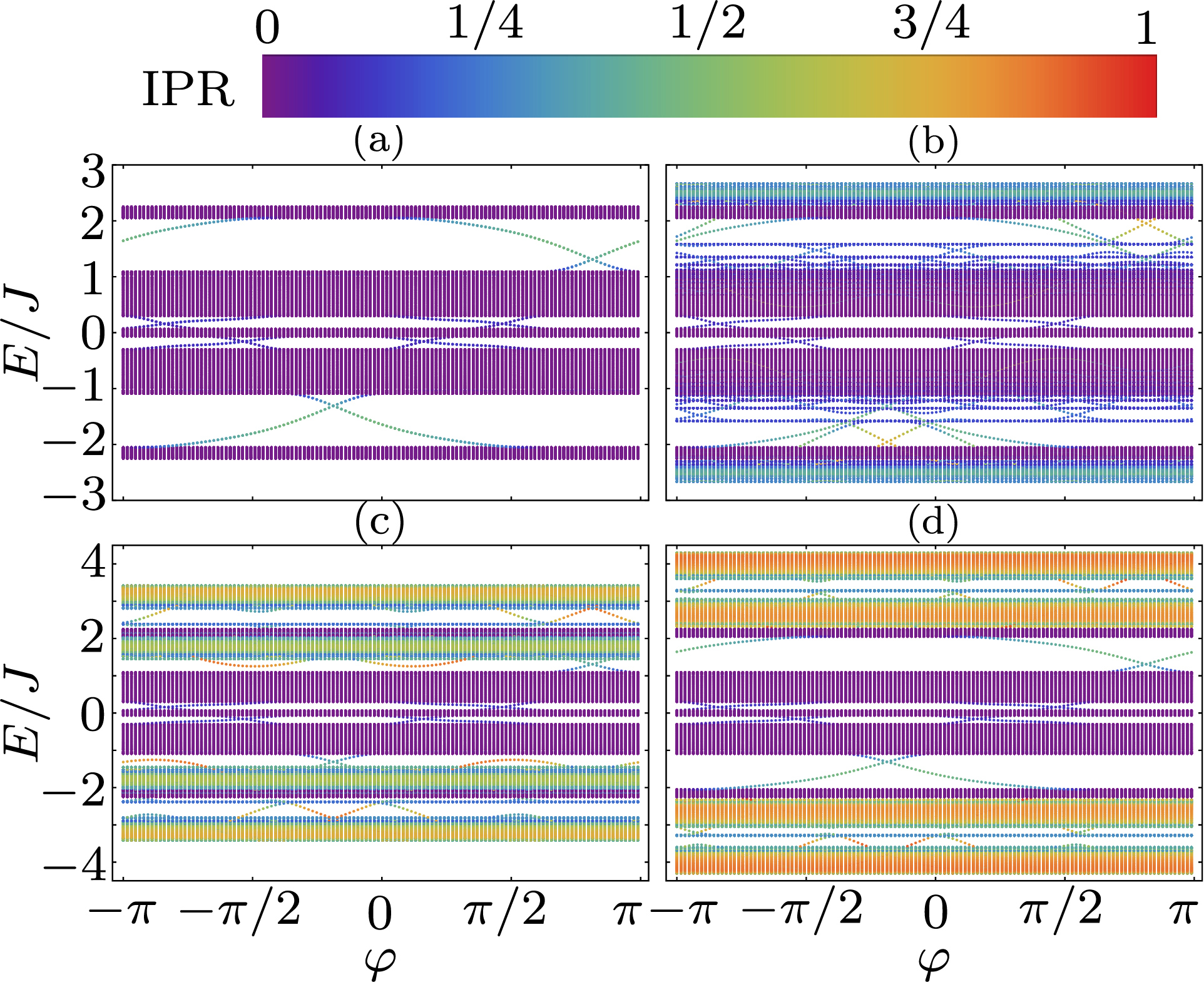}
\caption{
Plots of $E/J$ versus $\varphi$ for $\Delta/J = 1$, $k_T a = \pi$ 
with $R_{\lambda} = 813/1064$ and $N = 500$. The parameters are: 
(a) $h_x/J = 0$; (b) $h_x/J = 1$; (c) $h_x/J = 2$; (d) $h_x/J = 3$. 
The continuous IPR color scheme varies from violet, describing the extended state, to non-violet (blue to red) representing localized states.
The panels shown correspond to
vertical cuts in Fig.~\ref{fig:energy-versus-delta} 
at $\Delta/ J = 1$.
}
\label{fig:energy-spectrum-versus-phase}
\end{figure}
\begin{figure}[tb]
\centering
\includegraphics[width=0.47\textwidth]{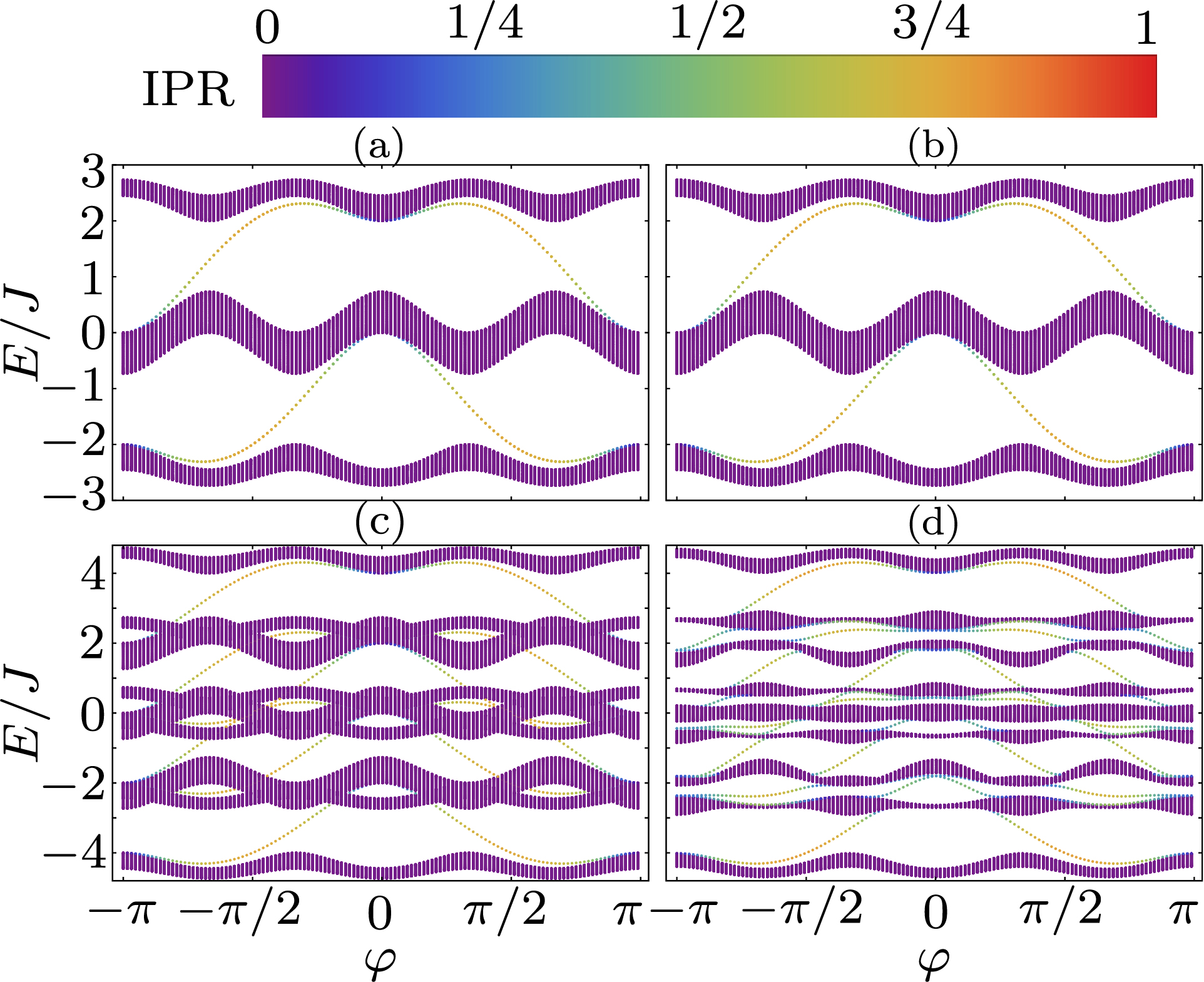}
\caption{
Plots of $E/J$ vs. $\varphi$ for $\Delta/J = -2$ 
with $R_{\lambda} = 1/3$ and $N = 500$.  
The parameters are: 
a) $h_x/J = k_T a = 0$; b) $h_x/J = 0$, $k_T a = \pi/8$; 
c) $h_x/J = 2, k_T a = 0$; d) $h_x/J = 2, k_T a = \pi/8$.
These parameters reproduce the expected results for SU(3) fermions with the corresponding color-orbit coupling, color-flip (Rabi) fields and magnetic flux in a 2D square lattice according to the mapping
in Tab.~\ref{tab:1D-to-2D-mapping}.
}
\label{fig:energy-versus-phase-delta2-r13}
\end{figure}

\section{Topological Considerations}
\label{sec:topological-considerations}

To understand the topological implications of bichromatic disorder in 1D systems, we investigate, in this section, 
the charge-charge Chern numbers that emerge from the mapping of our 1D disorder model into the 2D color Harper model.
We use the emergent band structure 
versus the synthetic dimension $\varphi$ and their eigenstates to extract the Chern indices of the bands, and use the bulk-edge correspondence for 
SU(3) fermions~\cite{sa-de-melo-2022} 
to obtain the Chern numbers associated with edge states. In addition, we monitor the IPR in the band structure to determine if topological insulating phases are neighbored by extended (conducting) and/or localized (insulating) phases. 

In Fig.~\ref{fig:energy-spectrum-versus-phase}, we show $E/J$ versus $\varphi$ for $\Delta /J = 1$, $k_T a = \pi$, $R_{\lambda} = 813/1064$ and $N = 500$. The panels in this figure represent cuts, at $\Delta/ J = 1$, in the energy spectrum shown for each panel of Fig.~\ref{fig:energy-versus-delta}. The color-flip (Rabi) field in each case is:  (a) $h_x/J = 0$; (b) $h_x/J = 1$; (c) $h_x/J = 2$; (d) $h_x/J = 3$.
The continuous IPR color scheme varies from violet, representing extended states, to non-violet (blue to red) describing localized states.
Each panel shows several energy bands and mini-gaps, where edge states emerge.
Some of these edge states cross from one energy band to another, indicating the existence of topological phases.
Because disorder is independent of color, the energy spectrum and the IPR for bulk states 
are particle-hole symmetric, that is, the bulk energy bands are symmetric with respect to $E/J = 0$ with an added $\pi/1064$ phase shift. 
Furthermore, the dispersions of the edge states are symmetric with respect to $E/J = 0$ with an added $\pi$ phase shift.

In Fig.~\ref{fig:energy-versus-phase-delta2-r13}, we show $E/J$ versus $\varphi$ for the disorder strength $\Delta/J = -2$, $R_{\lambda} = 1/3$ and $N = 500$ to 
directly connect the 1D disorder model to the 2D color Harper model for SU(3) fermions, as shown in Tab.~\ref{tab:1D-to-2D-mapping}. In the later case, $R_\lambda = \alpha$ represents a magnetic flux ratio $ \alpha = \Phi/\Phi_0 = 1/3$ in a square lattice, since $\Delta/J = -2$ means that 
$J \leftrightarrow t_x \leftrightarrow t_y$, as 
inferred from Tab.~\ref{tab:1D-to-2D-mapping}.
The parameters distinguishing the panels in this specific case are: (a) $h_x/J = 0$, 
$k_T a = 0$; (b) $h_x/J = 0$, $k_T a = \pi/8$; (c) $h_x/J = 2$, $k_T a = 0$; and (d) $h_x/J = 2$, $k_T a = \pi/8$, 
reproducing the expected results for SU(3) fermions with color-orbit coupling, color-flip (Rabi) fields, and magnetic flux in a 2D square lattice~\cite{sa-de-melo-2021, sa-de-melo-2021b}.
The mid-gap states crossing from one 
band to the other are chiral edge states. Notice that particle-hole symmetry exists in the bulk energy bands with respect to $E/J = 0$, followed by a
$\pi/3$ phase shift, as well as in the dispersions of the edge states with respect to $E/J = 0$, followed by $\pi$ phase shift. Notice that the bulk states in all panels are extended, because $R_\lambda = \alpha = 1/3$,
even though $\Delta/J = -2$, because of the perfect periodicity
of the system reflected in the magnetic Brillouin zone.
 
\begin{figure}[tb]
\centering
\includegraphics[width=0.47\textwidth]{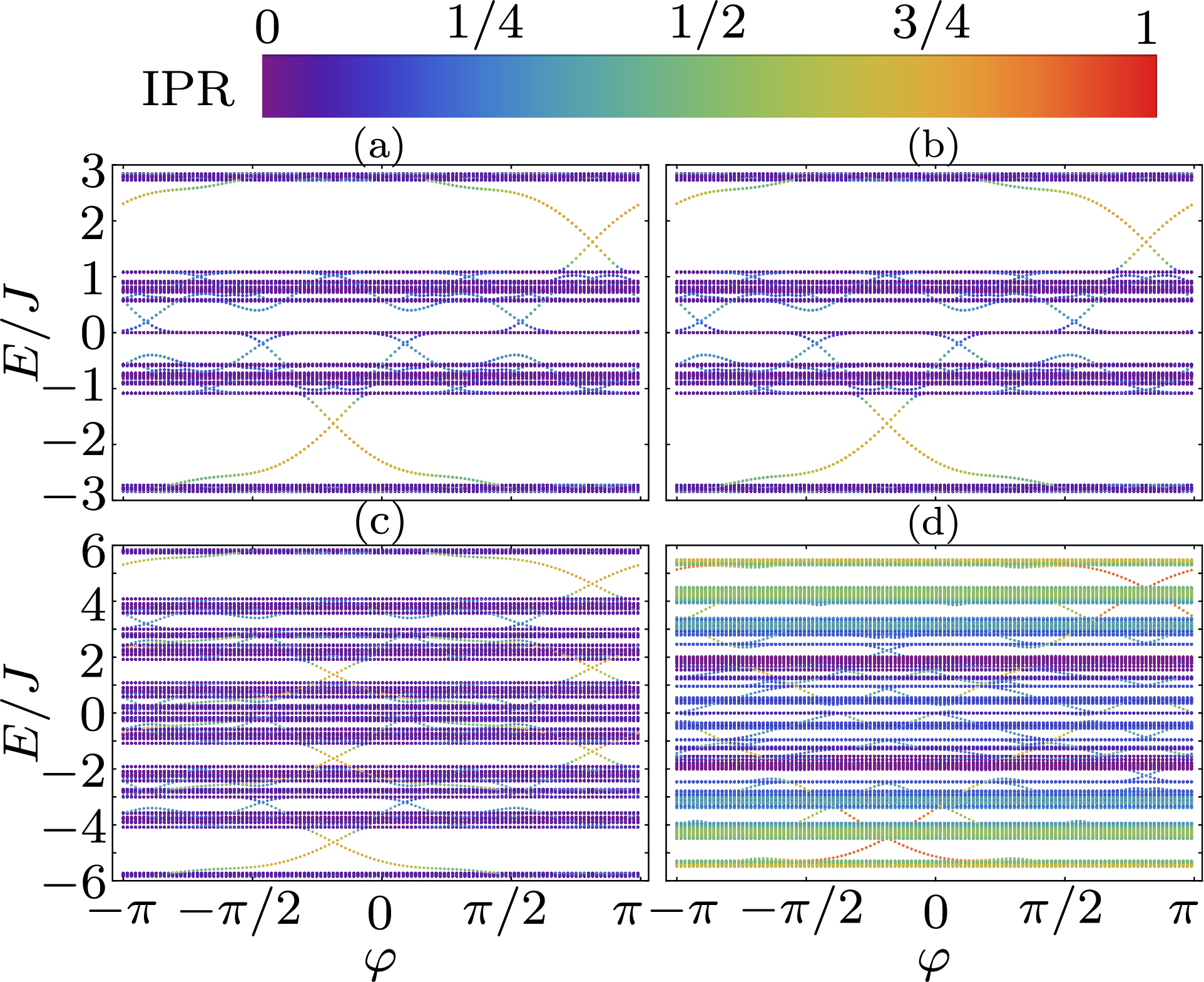}
\caption{
Plots of $E/J$ vs. $\varphi$ for 
$\Delta/J = 2$ 
with $R_{\lambda} = 813/1064$ and $N = 500$.  
The parameters are: 
(a) $h_x/J = 0, k_T a =0$; (b) $h_x/J = 0, k_T a =\pi/2$; (c) $h_x/J = 3, k_T a =0$; (d) $h_x/J = 3, k_T a =\pi/2$.
The continuous IPR color scheme varies from violet for extended states to non-violet (blue to red) for localized states.}
\label{fig:energy-versus-phase-delta2-Rb87}
\end{figure}

In Fig.~\ref{fig:energy-versus-phase-delta2-Rb87}, we return to experimental parameters compatible with $^{87}\text{Rb}$ 
and display the energy spectrum $E/J$ versus $\varphi$ for disorder parameter 
$\Delta/J = 2$ and $R_{\lambda} = 813/1064$, maintaining $N = 500$ sites.
The parameter sets for these panels are (a) $h_x/J = 0$, $k_T a = 0$; (b) $h_x/J = 0$, $k_T a = \pi/2$; (c) $h_x/J = 3$, $k_T a = 0$; and (d) $h_x/J = 3$, $k_T a = \pi/2$. Once again the energy eigenstates states are classified, via their IPR color scheme, as extended (violet) or localized (non-violet). Using Tab.~\ref{tab:1D-to-2D-mapping}, we see that the 1D disorder model
is mapped into a 2D color Harper model for a square lattice with magnetic flux ratio $\alpha = R_{\lambda} = 813/1064$.
As previously discussed, because disorder is independent of color, 
the bulk energy spectrum is particle-hole symmetric, that is, the bulk energy bands are symmetric with respect to $E/J = 0$, followed by a 
$\pi/1064$ phase shift and the energy dispersions of the edge states are symmetric with respect to $E/J = 0$,  followed by a $\pi$ phase shift. 

To rigorously classify these topological phases in the charge sector, we must obtain the topological invariants (Chern numbers)  of each phase. Using the mapping between the 1D disorder model and the 2D color Harper model in Tab.~\ref{tab:1D-to-2D-mapping}, 
we investigate the Chern numbers for different values of 
color-flip (Rabi) fields $h_x$ and color-orbit coupling $k_T a$
for fixed incommensurability ratio $R_\lambda$.

To calculate the Chern numbers as a bulk topological
invariant, we impose periodic boundary conditions along the
real x and the fictitious y directions, and compactify the cylinder into a torus. The compactification along the fictitious $y$ direction is already manifest since the 
phase $\varphi \leftrightarrow k_y a_y$, while the compactification along $x$ is achieved by using the 
ratio $R_{\lambda} = p/q$ to write the color-dependent Harper’s
Hamiltonian as a $3q \times 3q$ matrix
\begin{equation}
\label{eqn:H-kx-varphi-matrix}
{\bf H} (k_x, \varphi) = 
\begin{pmatrix}
{\bf H}_{RR} & {\bf H}_{\rm flip} & 0 \\
{\bf H}_{\rm flip} & {\bf H}_{GG} & {\bf H}_{\rm flip} \\
0 & {\bf H}_{\rm flip} &  {\bf H}_{BB}
\end{pmatrix}.
\end{equation}
This Hamiltonian matrix is the equivalent of 2D color Harper model defined in the magnetic Brilluoin zone determined
by $q$, and represents essentially the Fourier transform of the real space Hamiltonian matrix 
${\bf H} (\varphi)$ shown in Eq.~(\ref{eqn:hamiltonian-matrix-for-color-independent-disorder}).
For $^{87}{\rm Sr}$, $R_{\lambda} = 813/1064$, thus
$q = 1064$.

In Eq.~(\ref{eqn:H-kx-varphi-matrix}), the off-diagonal block matrix is 
${\bf H}_{\rm flip} = -(h_x/\sqrt{2}) {\bf I}_{q \times q}$, where ${\bf I}_{q \times q}$ is the $q \times q$ identity matrix representing the spatially uniform color-flip (Rabi) field.
The color-diagonal block matrices ${\bf H}_{cc}(k_x, \varphi)$, for colors $c \in \{R, G, B\}$, are $q \times q$ matrices describing the hopping $J$ along the $x$ direction and the local disorder parameter landscape controled by $\Delta$, which is independent of color:
\begin{multline}
{\bf H}_{cc}(k_x, \varphi) = \\
\begin{pmatrix}
D_1 & -J e^{i \theta_c} & 0 & \dots & -J e^{-i \theta_c} \\
-J e^{-i \theta_c} & D_2 & -J e^{i \theta_c} & \dots & 0 \\
\vdots & \vdots & \ddots & \ddots & \vdots \\
0 & 0 & \dots & D_{q-1} & -J e^{i \theta_c} \\
-J e^{i \theta_c} & 0 & \dots & -J e^{-i \theta_c} & D_q
\end{pmatrix}.
\end{multline}
The diagonal entries in ${\bf H}_{cc} (k_x \varphi)$ are the energies $D_n = \Delta \cos(\varphi + 2\pi R_\lambda n)$, where the site index $n$ runs from $1$ to $q$. The off-diagonal elements represent the real-space nearest-neighbor hopping amplitude $J$, modified by the color-dependent phase $\theta_c = (k_x - \gamma_c k_T)a$, where $\gamma_R = 1$, $\gamma_G = 0$, and $\gamma_B = -1$. Again, the analysis above maps exactly into the two-dimensional color Harper model via the correspondence listed
in Tab.~\ref{tab:1D-to-2D-mapping}.

The energy spectrum associated with Hamiltonian matrix ${\bf H} (k_x, \varphi)$ has $3q$ color-disorder bands $E_{l_\gamma}(k_x, \varphi)$, which are labeled by a band number $l$ with a generalized color index $\gamma$ corresponding to mixed color states. We identify these mixed color states as cyan (C), magenta (M), and yellow (Y) or via a pseudospin basis $\{C, M, Y\} \leftrightarrow \{\Uparrow, 0, \Downarrow\}$. 

The Chern index for the $l$-th band with generalized color index $\gamma$ is the topological invariant 
\begin{equation}
C_{l_\gamma} = \frac{1}{2\pi i} \int_{-\pi}^{\pi} d\varphi \int_{-\pi/qa}^{\pi/qa} dk_x F_{k_x \varphi}^{(l_\gamma)}(k_x, \varphi),
\end{equation}
where the domain of integration covers the
momentum $k_x$ spanning $2\pi/qa$ of the magnetic Brillouin zone and the synthetic dimension phase 
$\varphi$ spanning a full $2\pi$ cycle. The integrand function is the Berry curvature
\begin{equation}
F_{k_x \varphi}^{(l_\gamma)}(k_x, \varphi) = \partial_{k_x} A_{\varphi}^{(l_\gamma)}(k_x, \varphi) - \partial_{\varphi} A_{k_x}^{(l_\gamma)}(k_x, \varphi),
\end{equation}
which is purely imaginary and is expressed in terms of the Berry connection components
\begin{equation}
A_{j}^{(l_\gamma)}(k_x, \varphi) = \langle u_{l_\gamma}(k_x, \varphi) | \partial_j | u_{l_\gamma}(k_x, \varphi) \rangle,
\end{equation}
where the indices in the partial derivatives are 
$j \in \{k_x, \varphi\}$, and $\vert u_{l_\gamma}(k_x, \varphi)\rangle$ are the eigenstates of the Hamiltonian ${\bf H} (k_x, \varphi)$.

Chern indices are properties of the individual bands $E_{l_\gamma}(k_x, \varphi)$ or band bundles as described above. However, Chern numbers characterizing the macroscopic insulating state are defined within the band gaps and depend on the location of the chemical potential. If the chemical potential $\mu$ is located in a band gap labeled by index $r$ and corresponding to a filling factor $\nu = r/q$, then the Chern number in the gap is
\begin{equation}
C_r = \sum_{E_{l_\gamma} < \mu} C_{l_\gamma},
\end{equation}
representing the sum of the Chern indices of all bands strictly below $\mu$. Using our normalization, the maximum filling factor is $\nu_{\rm max} = 3$, indicating that the maximum number of color states per site is three.

We have also investigated the Chern numbers by analyzing the total chirality of the mid-gap edge states that arise in a cylindrical geometry (compactification along the fictitious $y$ direction) with open boundary conditions along the real spatial direction $x$. We found that they coincide perfectly with the Chern numbers $C_r$ calculated from the toroidal bulk geometry without edges, precisely as expected from the bulk-edge correspondence for SU(3) fermions~\cite{sa-de-melo-2022}. From an analysis of the energy spectrum and the calculated topological invariants, the filling factor 
$\nu_g$ associated with bulk gap $g$
is related to the Chern number $C_g$ and to the
ratio $R_\lambda$ through the Diophantine equation
\begin{equation}
\nu_g = S_g + R_\lambda C_g,
\end{equation}
which is the color generalization~\cite{sa-de-melo-2021b}
of the Wannier-Claro gap labeling theorem~\cite{wannier-1979}.
Here, $S_g$ is the supplementary topological invariant analogous to the quantization of the induced charge density resulting from the lattice potential screening in the corresponding 2D color Harper model, that is, the induced charge density 
$\widetilde{\rho}_{\rm ind}$ at gap $g$
is strictly proportional to this invariant:
\begin{equation}
\widetilde{\rho}_{\rm ind} = \frac{e}{a^2}S_g.
\end{equation}
Therefore, for a filling factor $\nu_g$, where the chemical potential lies inside a gap, we identify the insulating phases in the phase diagrams via the ordered pair of topological numbers $(S_g, C_g)$. We note that additional 
topological invariants (charge-color and color-color Chern numbers) can refine this initial classification~\cite{sa-de-melo-2021, sa-de-melo-2022}, but for our current purpose of 
exploring the charge-charge sector $S_g$ and $C_g$ are sufficient.
\begin{table}[tpb]
\centering
\begin{tabular}{l c c c c c c c c} 
\hline\hline 
Label/gap & $g_1$ & $g_2$ & $g_3$ & $g_4$ & $g_5$ & $g_6$ & $g_7$ & $g_8$ \\ 
\hline 
$\nu_g$ & 0.47 & 0.53 & 1.24 & 1.47 & 1.53 & 1.77 & 2.47 & 2.53 \\ 
$S_g$ & 2 & -1 & 2 & 3 & 0 & 1 & 4 & 1 \\ 
$C_g$ & -2 & 2 & -1 & -2 & 2 & 1 & -2 & 2 \\ 
$N_{\rm st}$ & 236 & 265 & 618 & 736 & 765 & 883 & 1236 & 1265 \\ 
\hline\hline 
\end{tabular}
\caption{List of filling factors $\nu_g$, topological invariants $S_g$ and $C_g$ and number of occupied states $N_{\rm st}$ for the gaps $g_i$ ordered sequentially from the lowest ($i = 1$) to the highest ($i = 8$) values of $\nu_g$.
The parameters used
are $R_\lambda = 813/1064$, $k_T a = \pi$, 
$N = 500$, $h_x/J = 3$ and 
$\Delta/J = 1$,
reflecting the energy bands in Fig.~\ref{fig:energy-spectrum-versus-phase} (d).
The exact filling factor is $\nu = N_{\rm st}/N$ 
matches extremely well to the result from the
gap labelling theorem $\nu_g = S_g + R_{\lambda} C_g$, within the numerical accuracy 
displayed.
}
\label{tab:chern-numbers-exact}
\end{table}

Applying this theoretical framework provides valuable insight into the properties of {\it edge} states of our 1D cisorder model and into the classification of the phases that emerge in 
our phase diagrams. 
In Tab.~\ref{tab:chern-numbers-exact},
we show the filling factor $\nu_g$, and the topological invariants $S_g$ and $C_g$ for a set
of gaps appearing in Fig.~\ref{fig:energy-spectrum-versus-phase}(d) labeled by $\nu_g$ or the number of occupied states 
$N_{\rm st} (\mu) \equiv N_{\rm st}$ up to the chemical potential $\mu$. 
The parameters used
are $R_\lambda = 813/1064$, $k_T a = \pi$, $N = 500$, $h_x/J = 3$ and 
$\Delta/J = 1$. 
In Fig.~\ref{fig:energy-versus-phase-delta2-Rb87}(d), we most prominently identify $8$ gaps, which we label as $g_{i}$, with $i$ varying from $1$ to $8$ corresponding to filling factors $\nu_g$ and
number of occupied sates $N_{\rm st}$. We recall 
that the filling factor $\nu$ varies from $0$ to $3$, as we have a maximum of $3$ colors per site, 
and that $N_{\rm st}$ varies from $1$ to $3N$, that is, 
from $1$ to $1500$ for $N = 500$.
It is important to emphasize that the exact filling factor is $\nu = N_{\rm st}/N$ 
matches extremely well to the result from the
gap labelling theorem $\nu_g = S_g + R_{\lambda} C_g$, within the numerical accuracy used, as shown in Tab.~\ref{tab:chern-numbers-exact}.
Non-zero Chern numbers $C_g$ arise due to the chirality of the {\it edge} states, confirming the topological nature of the insulating phases, 
which emerge due to the interplay of color-orbit coupling, 
color-flip (Rabi) fields and disorder.

We also use this mapping and the analysis of the IPR, to show that topologically non-trivial color insulating phases (with chiral edge states) can be energetically neighbored by: a) two extended bulk phases; b) two localized bulk phases; or c) one extended and one localized bulk phase. Examples of case a) are seen for the phases neighboring gaps $g_3, g_4, g_5, g_6$ from Tab.~\ref{tab:chern-numbers-exact} and Fig.~\ref{fig:energy-spectrum-versus-phase}(d). Examples of case b) are 
seen for the phases neighboring gaps $g_1, g_2, g_7, g_8$ from Tab.~\ref{tab:chern-numbers-exact} and Fig.~\ref{fig:energy-spectrum-versus-phase}(d). Examples of case c) are seen for the phases neighboring gaps between $-2 < E/J < -1$
and between $1 > E/J > 2$ in Fig.~\ref{fig:energy-spectrum-versus-phase}(c). These findings generalize the types of possible phases that can emerge in topological color insulators when disorder is considered.

Having completed our discussion about topological invariants for the 1D disorder model using its connection to the 2D color Harper model, we are ready to state our conclusions.

\section{Conclusions}
\label{sec:conclusions}

We studied localization and topological properties of ultracold fermions with three internal states, such 
as $^{173}{\rm Yb}$ and $^{87}{\rm Sr}$, within a SU(3) manifold. We described the internal states as colors (Red, Green, and Blue) and investigated a one-dimensional Fermi system with bichromatic disorder in the presence of color-orbit and color-flip (Rabi) fields. The bichromatic disorder
emerged from lattice potentials created by strong and weak laser
beams with incommensurate wavelengths.
We considered the general case of color-dependent disorder for SU(3) fermions, but focused on the simpler example of color-independent disorder for parameters compatible with  
$^{87}{\rm Sr}$. We analyzed the symmetries of our three-color Hamiltonian, its energy spectrum, and eigenstates. 
We used exact diagonalization to obtain the inverse participation ratio and identify localized or extended states as a function of bichromatic disorder, color-orbit, and color-flip (Rabi) fields. 

We found that the interplay of bichromatic disorder, color-orbit coupling, and color-flip (Rabi) fields produces mobility regions that are absent in standard bichromatic localization systems; that is, color orbit and color-flip (Rabi) fields can either enhance or hinder localization depending on their magnitudes and length scales in comparison to the strength and spatial modulation of disorder. Due to this interplay, not only localized bulk states appear, but also edge states emerge with interesting topological properties. We explored topological features by mapping our one-dimensional disorder model into a two-dimensional color Harper model, where the phase of the weak laser
beam corresponded to a second fictitious dimension.
Using this mapping, we extracted topological invariants characterizing edge states and topological color insulating phases. We showed that the topological color insulating phases can be energetically neighbored by two extended bulk phases, by two localized bulk phases, or by one extended and one localized bulk phase. These findings generalize the types of possible phases that can emerge in topological color insulators when disorder is considered.
In summary, our results pave the way for the exploration of topological phase transitions in SU(3) fermions via a tunable platform described by the interplay of disorder, color-orbit coupling, and color-flip (Rabi) fields.

\begin{acknowledgments}
C.A.R.S.d.M. acknowledges support
from the Mercator Fellowship of the German Research Foundation (DFG) through the Collaborative Research Center SFB/TR185 (Project No. 277625399).     
\end{acknowledgments}

\bibliography{References}

\providecommand{\noopsort}[1]{}\providecommand{\singleletter}[1]{#1}
\begin{thebibliography}{60}%
\makeatletter
\providecommand \@ifxundefined [1]{%
 \@ifx{#1\undefined}
}%
\providecommand \@ifnum [1]{%
 \ifnum #1\expandafter \@firstoftwo
 \else \expandafter \@secondoftwo
 \fi
}%
\providecommand \@ifx [1]{%
 \ifx #1\expandafter \@firstoftwo
 \else \expandafter \@secondoftwo
 \fi
}%
\providecommand \natexlab [1]{#1}%
\providecommand \enquote  [1]{``#1''}%
\providecommand \bibnamefont  [1]{#1}%
\providecommand \bibfnamefont [1]{#1}%
\providecommand \citenamefont [1]{#1}%
\providecommand \href@noop [0]{\@secondoftwo}%
\providecommand \href [0]{\begingroup \@sanitize@url \@href}%
\providecommand \@href[1]{\@@startlink{#1}\@@href}%
\providecommand \@@href[1]{\endgroup#1\@@endlink}%
\providecommand \@sanitize@url [0]{\catcode `\\12\catcode `\$12\catcode
  `\&12\catcode `\#12\catcode `\^12\catcode `\_12\catcode `\%12\relax}%
\providecommand \@@startlink[1]{}%
\providecommand \@@endlink[0]{}%
\providecommand \url  [0]{\begingroup\@sanitize@url \@url }%
\providecommand \@url [1]{\endgroup\@href {#1}{\urlprefix }}%
\providecommand \urlprefix  [0]{URL }%
\providecommand \Eprint [0]{\href }%
\providecommand \doibase [0]{https://doi.org/}%
\providecommand \selectlanguage [0]{\@gobble}%
\providecommand \bibinfo  [0]{\@secondoftwo}%
\providecommand \bibfield  [0]{\@secondoftwo}%
\providecommand \translation [1]{[#1]}%
\providecommand \BibitemOpen [0]{}%
\providecommand \bibitemStop [0]{}%
\providecommand \bibitemNoStop [0]{.\EOS\space}%
\providecommand \EOS [0]{\spacefactor3000\relax}%
\providecommand \BibitemShut  [1]{\csname bibitem#1\endcsname}%
\let\auto@bib@innerbib\@empty
\bibitem [{\citenamefont {Roushan}\ \emph {et~al.}(2017)\citenamefont
  {Roushan}, \citenamefont {Neill}, \citenamefont {Tangpanitanon},
  \citenamefont {Bastidas}, \citenamefont {Megrant}, \citenamefont {Barends},
  \citenamefont {Chen}, \citenamefont {Chen}, \citenamefont {Chiaro},
  \citenamefont {Dunsworth}, \citenamefont {Fowler}, \citenamefont {Foxen},
  \citenamefont {Giustina}, \citenamefont {Jeffrey}, \citenamefont {Kelly},
  \citenamefont {Lucero}, \citenamefont {Mutus}, \citenamefont {Neeley},
  \citenamefont {Quintana}, \citenamefont {Sank}, \citenamefont {Vainsencher},
  \citenamefont {Wenner}, \citenamefont {White}, \citenamefont {Neven},
  \citenamefont {Angelakis},\ and\ \citenamefont {Martinis}}]{martinis-2017}%
  \BibitemOpen
  \bibfield  {author} {\bibinfo {author} {\bibfnamefont {P.}~\bibnamefont
  {Roushan}}, \bibinfo {author} {\bibfnamefont {C.}~\bibnamefont {Neill}},
  \bibinfo {author} {\bibfnamefont {J.}~\bibnamefont {Tangpanitanon}}, \bibinfo
  {author} {\bibfnamefont {V.~M.}\ \bibnamefont {Bastidas}}, \bibinfo {author}
  {\bibfnamefont {A.}~\bibnamefont {Megrant}}, \bibinfo {author} {\bibfnamefont
  {R.}~\bibnamefont {Barends}}, \bibinfo {author} {\bibfnamefont
  {Y.}~\bibnamefont {Chen}}, \bibinfo {author} {\bibfnamefont {Z.}~\bibnamefont
  {Chen}}, \bibinfo {author} {\bibfnamefont {B.}~\bibnamefont {Chiaro}},
  \bibinfo {author} {\bibfnamefont {A.}~\bibnamefont {Dunsworth}}, \bibinfo
  {author} {\bibfnamefont {A.}~\bibnamefont {Fowler}}, \bibinfo {author}
  {\bibfnamefont {B.}~\bibnamefont {Foxen}}, \bibinfo {author} {\bibfnamefont
  {M.}~\bibnamefont {Giustina}}, \bibinfo {author} {\bibfnamefont
  {E.}~\bibnamefont {Jeffrey}}, \bibinfo {author} {\bibfnamefont
  {J.}~\bibnamefont {Kelly}}, \bibinfo {author} {\bibfnamefont
  {E.}~\bibnamefont {Lucero}}, \bibinfo {author} {\bibfnamefont
  {J.}~\bibnamefont {Mutus}}, \bibinfo {author} {\bibfnamefont
  {M.}~\bibnamefont {Neeley}}, \bibinfo {author} {\bibfnamefont
  {C.}~\bibnamefont {Quintana}}, \bibinfo {author} {\bibfnamefont
  {D.}~\bibnamefont {Sank}}, \bibinfo {author} {\bibfnamefont {A.}~\bibnamefont
  {Vainsencher}}, \bibinfo {author} {\bibfnamefont {J.}~\bibnamefont {Wenner}},
  \bibinfo {author} {\bibfnamefont {T.}~\bibnamefont {White}}, \bibinfo
  {author} {\bibfnamefont {H.}~\bibnamefont {Neven}}, \bibinfo {author}
  {\bibfnamefont {D.~G.}\ \bibnamefont {Angelakis}},\ and\ \bibinfo {author}
  {\bibfnamefont {J.}~\bibnamefont {Martinis}},\ }\bibfield  {title} {\bibinfo
  {title} {{Spectroscopic signatures of localization with interacting photons
  in superconducting qubits}},\ }\href
  {https://doi.org/10.1126/science.aao1401} {\bibfield  {journal} {\bibinfo
  {journal} {Science}\ }\textbf {\bibinfo {volume} {358}},\ \bibinfo {pages}
  {1175} (\bibinfo {year} {2017})}\BibitemShut {NoStop}%
\bibitem [{\citenamefont {Karcher}\ \emph {et~al.}(2019)\citenamefont
  {Karcher}, \citenamefont {Sonner},\ and\ \citenamefont
  {Mirlin}}]{mirlin-2019}%
  \BibitemOpen
  \bibfield  {author} {\bibinfo {author} {\bibfnamefont {J.~F.}\ \bibnamefont
  {Karcher}}, \bibinfo {author} {\bibfnamefont {M.}~\bibnamefont {Sonner}},\
  and\ \bibinfo {author} {\bibfnamefont {A.~D.}\ \bibnamefont {Mirlin}},\
  }\bibfield  {title} {\bibinfo {title} {{Disorder and interaction in chiral
  chains: Majoranas versus complex fermions}},\ }\href
  {https://doi.org/10.1103/PhysRevB.100.134207} {\bibfield  {journal} {\bibinfo
   {journal} {Phys. Rev. B}\ }\textbf {\bibinfo {volume} {100}},\ \bibinfo
  {pages} {134207} (\bibinfo {year} {2019})}\BibitemShut {NoStop}%
\bibitem [{\citenamefont {Sikkenk}\ and\ \citenamefont
  {Fritz}(2019)}]{fritz-2019}%
  \BibitemOpen
  \bibfield  {author} {\bibinfo {author} {\bibfnamefont {T.~S.}\ \bibnamefont
  {Sikkenk}}\ and\ \bibinfo {author} {\bibfnamefont {L.}~\bibnamefont
  {Fritz}},\ }\bibfield  {title} {\bibinfo {title} {{Interplay of disorder and
  interactions in a system of subcritically tilted and anisotropic
  three-dimensional Weyl fermions}},\ }\href
  {https://doi.org/10.1103/PhysRevB.100.085121} {\bibfield  {journal} {\bibinfo
   {journal} {Phys. Rev. B}\ }\textbf {\bibinfo {volume} {100}},\ \bibinfo
  {pages} {085121} (\bibinfo {year} {2019})}\BibitemShut {NoStop}%
\bibitem [{\citenamefont {Huang}\ and\ \citenamefont {Meng}(1992)}]{meng-1992}%
  \BibitemOpen
  \bibfield  {author} {\bibinfo {author} {\bibfnamefont {K.}~\bibnamefont
  {Huang}}\ and\ \bibinfo {author} {\bibfnamefont {H.-F.}\ \bibnamefont
  {Meng}},\ }\bibfield  {title} {\bibinfo {title} {{Hard-sphere Bose gas in
  random external potentials}},\ }\href
  {https://doi.org/10.1103/PhysRevLett.69.644} {\bibfield  {journal} {\bibinfo
  {journal} {Phys. Rev. Lett.}\ }\textbf {\bibinfo {volume} {69}},\ \bibinfo
  {pages} {644} (\bibinfo {year} {1992})}\BibitemShut {NoStop}%
\bibitem [{\citenamefont {Schulte}\ \emph {et~al.}(2005)\citenamefont
  {Schulte}, \citenamefont {Drenkelforth}, \citenamefont {Kruse}, \citenamefont
  {Ertmer}, \citenamefont {Arlt}, \citenamefont {Sacha}, \citenamefont
  {Zakrzewski},\ and\ \citenamefont {Lewenstein}}]{lewenstein-2005}%
  \BibitemOpen
  \bibfield  {author} {\bibinfo {author} {\bibfnamefont {T.}~\bibnamefont
  {Schulte}}, \bibinfo {author} {\bibfnamefont {S.}~\bibnamefont
  {Drenkelforth}}, \bibinfo {author} {\bibfnamefont {J.}~\bibnamefont {Kruse}},
  \bibinfo {author} {\bibfnamefont {W.}~\bibnamefont {Ertmer}}, \bibinfo
  {author} {\bibfnamefont {J.}~\bibnamefont {Arlt}}, \bibinfo {author}
  {\bibfnamefont {K.}~\bibnamefont {Sacha}}, \bibinfo {author} {\bibfnamefont
  {J.}~\bibnamefont {Zakrzewski}},\ and\ \bibinfo {author} {\bibfnamefont
  {M.}~\bibnamefont {Lewenstein}},\ }\bibfield  {title} {\bibinfo {title}
  {{Routes Towards Anderson-Like Localization of Bose-Einstein Condensates in
  Disordered Optical Lattices}},\ }\href
  {https://doi.org/10.1103/PhysRevLett.95.170411} {\bibfield  {journal}
  {\bibinfo  {journal} {Phys. Rev. Lett.}\ }\textbf {\bibinfo {volume} {95}},\
  \bibinfo {pages} {170411} (\bibinfo {year} {2005})}\BibitemShut {NoStop}%
\bibitem [{\citenamefont {Gavish}\ and\ \citenamefont
  {Castin}(2005)}]{castin-2005}%
  \BibitemOpen
  \bibfield  {author} {\bibinfo {author} {\bibfnamefont {U.}~\bibnamefont
  {Gavish}}\ and\ \bibinfo {author} {\bibfnamefont {Y.}~\bibnamefont
  {Castin}},\ }\bibfield  {title} {\bibinfo {title} {{Matter-Wave Localization
  in Disordered Cold Atom Lattices}},\ }\href
  {https://doi.org/10.1103/PhysRevLett.95.020401} {\bibfield  {journal}
  {\bibinfo  {journal} {Phys. Rev. Lett.}\ }\textbf {\bibinfo {volume} {95}},\
  \bibinfo {pages} {020401} (\bibinfo {year} {2005})}\BibitemShut {NoStop}%
\bibitem [{\citenamefont {Falco}\ \emph {et~al.}(2007)\citenamefont {Falco},
  \citenamefont {Pelster},\ and\ \citenamefont {Graham}}]{graham-2007}%
  \BibitemOpen
  \bibfield  {author} {\bibinfo {author} {\bibfnamefont {G.~M.}\ \bibnamefont
  {Falco}}, \bibinfo {author} {\bibfnamefont {A.}~\bibnamefont {Pelster}},\
  and\ \bibinfo {author} {\bibfnamefont {R.}~\bibnamefont {Graham}},\
  }\bibfield  {title} {\bibinfo {title} {{Collective oscillations in trapped
  Bose-Einstein-condensed gases in the presence of weak disorder}},\ }\href
  {https://doi.org/10.1103/PhysRevA.76.013624} {\bibfield  {journal} {\bibinfo
  {journal} {Phys. Rev. A}\ }\textbf {\bibinfo {volume} {76}},\ \bibinfo
  {pages} {013624} (\bibinfo {year} {2007})}\BibitemShut {NoStop}%
\bibitem [{\citenamefont {Zobay}(2006)}]{zobay-2006}%
  \BibitemOpen
  \bibfield  {author} {\bibinfo {author} {\bibfnamefont {O.}~\bibnamefont
  {Zobay}},\ }\bibfield  {title} {\bibinfo {title} {{Condensation temperature
  of interacting Bose gases with and without disorder}},\ }\href
  {https://doi.org/10.1103/PhysRevA.73.023616} {\bibfield  {journal} {\bibinfo
  {journal} {Phys. Rev. A}\ }\textbf {\bibinfo {volume} {73}},\ \bibinfo
  {pages} {023616} (\bibinfo {year} {2006})}\BibitemShut {NoStop}%
\bibitem [{\citenamefont {Li}\ and\ \citenamefont {{S\'a}~de
  Melo}(2011)}]{sa-de-melo-2011}%
  \BibitemOpen
  \bibfield  {author} {\bibinfo {author} {\bibfnamefont {H.}~\bibnamefont
  {Li}}\ and\ \bibinfo {author} {\bibfnamefont {C.~A.~R.}\ \bibnamefont
  {{S\'a}~de Melo}},\ }\bibfield  {title} {\bibinfo {title} {{Evolution from
  Bardeen–Cooper–Schrieffer to Bose–Einstein condensate superfluidity in
  the presence of disorder}},\ }\href
  {https://doi.org/doi:10.1088/1367-2630/13/5/055012} {\bibfield  {journal}
  {\bibinfo  {journal} {New Journal of Physics}\ }\textbf {\bibinfo {volume}
  {13}},\ \bibinfo {pages} {055012} (\bibinfo {year} {2011})}\BibitemShut
  {NoStop}%
\bibitem [{\citenamefont {Sanchez-Palencia}\ \emph {et~al.}(2007)\citenamefont
  {Sanchez-Palencia}, \citenamefont {Cl\'ement}, \citenamefont {Lugan},
  \citenamefont {Bouyer}, \citenamefont {Shlyapnikov},\ and\ \citenamefont
  {Aspect}}]{aspect-2007}%
  \BibitemOpen
  \bibfield  {author} {\bibinfo {author} {\bibfnamefont {L.}~\bibnamefont
  {Sanchez-Palencia}}, \bibinfo {author} {\bibfnamefont {D.}~\bibnamefont
  {Cl\'ement}}, \bibinfo {author} {\bibfnamefont {P.}~\bibnamefont {Lugan}},
  \bibinfo {author} {\bibfnamefont {P.}~\bibnamefont {Bouyer}}, \bibinfo
  {author} {\bibfnamefont {G.~V.}\ \bibnamefont {Shlyapnikov}},\ and\ \bibinfo
  {author} {\bibfnamefont {A.}~\bibnamefont {Aspect}},\ }\bibfield  {title}
  {\bibinfo {title} {{Anderson Localization of Expanding Bose-Einstein
  Condensates in Random Potentials}},\ }\href
  {https://doi.org/10.1103/PhysRevLett.98.210401} {\bibfield  {journal}
  {\bibinfo  {journal} {Phys. Rev. Lett.}\ }\textbf {\bibinfo {volume} {98}},\
  \bibinfo {pages} {210401} (\bibinfo {year} {2007})}\BibitemShut {NoStop}%
\bibitem [{\citenamefont {Roati}\ \emph {et~al.}(2008)\citenamefont {Roati},
  \citenamefont {D'Errico}, \citenamefont {Fallani}, \citenamefont {Fattori},
  \citenamefont {Fort}, \citenamefont {Zaccanti}, \citenamefont {Modugno},
  \citenamefont {Modugno},\ and\ \citenamefont {Inguscio}}]{inguscio-2008}%
  \BibitemOpen
  \bibfield  {author} {\bibinfo {author} {\bibfnamefont {G.}~\bibnamefont
  {Roati}}, \bibinfo {author} {\bibfnamefont {C.}~\bibnamefont {D'Errico}},
  \bibinfo {author} {\bibfnamefont {L.}~\bibnamefont {Fallani}}, \bibinfo
  {author} {\bibfnamefont {M.}~\bibnamefont {Fattori}}, \bibinfo {author}
  {\bibfnamefont {C.}~\bibnamefont {Fort}}, \bibinfo {author} {\bibfnamefont
  {M.}~\bibnamefont {Zaccanti}}, \bibinfo {author} {\bibfnamefont
  {G.}~\bibnamefont {Modugno}}, \bibinfo {author} {\bibfnamefont
  {M.}~\bibnamefont {Modugno}},\ and\ \bibinfo {author} {\bibfnamefont
  {M.}~\bibnamefont {Inguscio}},\ }\bibfield  {title} {\bibinfo {title}
  {{Anderson localization of a non-interacting Bose--Einstein condensate}},\
  }\href {https://doi.org/10.1038/nature07071} {\bibfield  {journal} {\bibinfo
  {journal} {Nature}\ }\textbf {\bibinfo {volume} {453}},\ \bibinfo {pages}
  {895} (\bibinfo {year} {2008})}\BibitemShut {NoStop}%
\bibitem [{\citenamefont {Nagler}\ \emph {et~al.}(2020)\citenamefont {Nagler},
  \citenamefont {Will}, \citenamefont {Hiebel}, \citenamefont {Barbosa},
  \citenamefont {Koch}, \citenamefont {Fleischhauer},\ and\ \citenamefont
  {Widera}}]{widera-2020}%
  \BibitemOpen
  \bibfield  {author} {\bibinfo {author} {\bibfnamefont {B.}~\bibnamefont
  {Nagler}}, \bibinfo {author} {\bibfnamefont {M.}~\bibnamefont {Will}},
  \bibinfo {author} {\bibfnamefont {S.}~\bibnamefont {Hiebel}}, \bibinfo
  {author} {\bibfnamefont {S.}~\bibnamefont {Barbosa}}, \bibinfo {author}
  {\bibfnamefont {J.}~\bibnamefont {Koch}}, \bibinfo {author} {\bibfnamefont
  {M.}~\bibnamefont {Fleischhauer}},\ and\ \bibinfo {author} {\bibfnamefont
  {A.}~\bibnamefont {Widera}},\ }\bibfield  {title} {\bibinfo {title}
  {{Ultracold Bose gases in disorder potentials with spatiotemporal
  dynamics}},\ }\href {https://arxiv.org/abs/2007.11523} {\bibfield  {journal}
  {\bibinfo  {journal} {arXiv:2007.11523}\ } (\bibinfo {year}
  {2020})}\BibitemShut {NoStop}%
\bibitem [{\citenamefont {Barbosa}\ \emph {et~al.}(2024)\citenamefont
  {Barbosa}, \citenamefont {Kiefer-Emmanouilidis}, \citenamefont {Lang},
  \citenamefont {Koch},\ and\ \citenamefont {Widera}}]{widera-2024a}%
  \BibitemOpen
  \bibfield  {author} {\bibinfo {author} {\bibfnamefont {S.}~\bibnamefont
  {Barbosa}}, \bibinfo {author} {\bibfnamefont {M.}~\bibnamefont
  {Kiefer-Emmanouilidis}}, \bibinfo {author} {\bibfnamefont {F.}~\bibnamefont
  {Lang}}, \bibinfo {author} {\bibfnamefont {J.}~\bibnamefont {Koch}},\ and\
  \bibinfo {author} {\bibfnamefont {A.}~\bibnamefont {Widera}},\ }\bibfield
  {title} {\bibinfo {title} {{Characterizing localization effects in an
  ultracold disordered Fermi gas by diffusion analysis}},\ }\href
  {https://doi.org/10.1103/PhysRevResearch.6.033039} {\bibfield  {journal}
  {\bibinfo  {journal} {Phys. Rev. Res.}\ }\textbf {\bibinfo {volume} {6}},\
  \bibinfo {pages} {033039} (\bibinfo {year} {2024})}\BibitemShut {NoStop}%
\bibitem [{\citenamefont {Koch}\ \emph {et~al.}(2024)\citenamefont {Koch},
  \citenamefont {Barbosa}, \citenamefont {Lang},\ and\ \citenamefont
  {Widera}}]{widera-2024b}%
  \BibitemOpen
  \bibfield  {author} {\bibinfo {author} {\bibfnamefont {J.}~\bibnamefont
  {Koch}}, \bibinfo {author} {\bibfnamefont {S.}~\bibnamefont {Barbosa}},
  \bibinfo {author} {\bibfnamefont {F.}~\bibnamefont {Lang}},\ and\ \bibinfo
  {author} {\bibfnamefont {A.}~\bibnamefont {Widera}},\ }\bibfield  {title}
  {\bibinfo {title} {Stability and sensitivity of interacting fermionic
  superfluids to quenched disorder},\ }\href
  {https://doi.org/10.1038/s41467-024-51903-8} {\bibfield  {journal} {\bibinfo
  {journal} {Nature Communications}\ }\textbf {\bibinfo {volume} {15}},\
  \bibinfo {pages} {9292} (\bibinfo {year} {2024})}\BibitemShut {NoStop}%
\bibitem [{\citenamefont {Hebbe~Madhusudhana}\ \emph
  {et~al.}(2021)\citenamefont {Hebbe~Madhusudhana}, \citenamefont {Scherg},
  \citenamefont {Kohlert}, \citenamefont {Bloch},\ and\ \citenamefont
  {Aidelsburger}}]{aidelsburger-2021a}%
  \BibitemOpen
  \bibfield  {author} {\bibinfo {author} {\bibfnamefont {B.}~\bibnamefont
  {Hebbe~Madhusudhana}}, \bibinfo {author} {\bibfnamefont {S.}~\bibnamefont
  {Scherg}}, \bibinfo {author} {\bibfnamefont {T.}~\bibnamefont {Kohlert}},
  \bibinfo {author} {\bibfnamefont {I.}~\bibnamefont {Bloch}},\ and\ \bibinfo
  {author} {\bibfnamefont {M.}~\bibnamefont {Aidelsburger}},\ }\bibfield
  {title} {\bibinfo {title} {{Benchmarking a Novel Efficient Numerical Method
  for Localized 1D Fermi-Hubbard Systems on a Quantum Simulator}},\ }\href
  {https://doi.org/10.1103/PRXQuantum.2.040325} {\bibfield  {journal} {\bibinfo
   {journal} {PRX Quantum}\ }\textbf {\bibinfo {volume} {2}},\ \bibinfo {pages}
  {040325} (\bibinfo {year} {2021})}\BibitemShut {NoStop}%
\bibitem [{\citenamefont {Scherg}\ \emph {et~al.}(2021)\citenamefont {Scherg},
  \citenamefont {Kohlert}, \citenamefont {Sala}, \citenamefont {Pollmann},
  \citenamefont {Hebbe~Madhusudhana}, \citenamefont {Bloch},\ and\
  \citenamefont {Aidelsburger}}]{aidelsburger-2021b}%
  \BibitemOpen
  \bibfield  {author} {\bibinfo {author} {\bibfnamefont {S.}~\bibnamefont
  {Scherg}}, \bibinfo {author} {\bibfnamefont {T.}~\bibnamefont {Kohlert}},
  \bibinfo {author} {\bibfnamefont {P.}~\bibnamefont {Sala}}, \bibinfo {author}
  {\bibfnamefont {F.}~\bibnamefont {Pollmann}}, \bibinfo {author}
  {\bibfnamefont {B.}~\bibnamefont {Hebbe~Madhusudhana}}, \bibinfo {author}
  {\bibfnamefont {I.}~\bibnamefont {Bloch}},\ and\ \bibinfo {author}
  {\bibfnamefont {M.}~\bibnamefont {Aidelsburger}},\ }\bibfield  {title}
  {\bibinfo {title} {{Observing non-ergodicity due to kinetic constraints in
  tilted Fermi-Hubbard chains}},\ }\href
  {https://doi.org/10.1038/s41467-021-24726-0} {\bibfield  {journal} {\bibinfo
  {journal} {Nature Communications}\ }\textbf {\bibinfo {volume} {12}},\
  \bibinfo {pages} {4490} (\bibinfo {year} {2021})}\BibitemShut {NoStop}%
\bibitem [{\citenamefont {Yue}\ \emph {et~al.}(2020)\citenamefont {Yue},
  \citenamefont {S\'a~de Melo},\ and\ \citenamefont
  {Spielman}}]{spielman-2020}%
  \BibitemOpen
  \bibfield  {author} {\bibinfo {author} {\bibfnamefont {Y.}~\bibnamefont
  {Yue}}, \bibinfo {author} {\bibfnamefont {C.~A.~R.}\ \bibnamefont {S\'a~de
  Melo}},\ and\ \bibinfo {author} {\bibfnamefont {I.~B.}\ \bibnamefont
  {Spielman}},\ }\bibfield  {title} {\bibinfo {title} {{Enhanced transport of
  spin-orbit-coupled Bose gases in disordered potentials}},\ }\href
  {https://doi.org/10.1103/PhysRevA.102.033325} {\bibfield  {journal} {\bibinfo
   {journal} {Phys. Rev. A}\ }\textbf {\bibinfo {volume} {102}},\ \bibinfo
  {pages} {033325} (\bibinfo {year} {2020})}\BibitemShut {NoStop}%
\bibitem [{\citenamefont {Nagler}\ \emph {et~al.}(2022)\citenamefont {Nagler},
  \citenamefont {Barbosa}, \citenamefont {Koch}, \citenamefont {Orso},\ and\
  \citenamefont {Widera}}]{widera-2022}%
  \BibitemOpen
  \bibfield  {author} {\bibinfo {author} {\bibfnamefont {B.}~\bibnamefont
  {Nagler}}, \bibinfo {author} {\bibfnamefont {S.}~\bibnamefont {Barbosa}},
  \bibinfo {author} {\bibfnamefont {J.}~\bibnamefont {Koch}}, \bibinfo {author}
  {\bibfnamefont {G.}~\bibnamefont {Orso}},\ and\ \bibinfo {author}
  {\bibfnamefont {A.}~\bibnamefont {Widera}},\ }\bibfield  {title} {\bibinfo
  {title} {{Observing the loss and revival of long-range phase coherence
  through disorder quenches}},\ }\href
  {https://doi.org/10.1073/pnas.2111078118} {\bibfield  {journal} {\bibinfo
  {journal} {PNAS}\ }\textbf {\bibinfo {volume} {119}},\ \bibinfo {pages}
  {e2111078118} (\bibinfo {year} {2022})}\BibitemShut {NoStop}%
\bibitem [{\citenamefont {Bai}\ \emph {et~al.}(2026)\citenamefont {Bai},
  \citenamefont {Dardia}, \citenamefont {Shimasaki},\ and\ \citenamefont
  {Weld}}]{weld-2026}%
  \BibitemOpen
  \bibfield  {author} {\bibinfo {author} {\bibfnamefont {Y.}~\bibnamefont
  {Bai}}, \bibinfo {author} {\bibfnamefont {A.~R.}\ \bibnamefont {Dardia}},
  \bibinfo {author} {\bibfnamefont {T.}~\bibnamefont {Shimasaki}},\ and\
  \bibinfo {author} {\bibfnamefont {D.~M.}\ \bibnamefont {Weld}},\ }\bibfield
  {title} {\bibinfo {title} {{Exploring Light-Induced Phases of 2D Materials in
  a Modulated 1D Quasicrystal}},\ }\href {https://doi.org/10.1103/37vk-qz71}
  {\bibfield  {journal} {\bibinfo  {journal} {Phys. Rev. X}\ }\textbf {\bibinfo
  {volume} {16}},\ \bibinfo {pages} {011036} (\bibinfo {year}
  {2026})}\BibitemShut {NoStop}%
\bibitem [{\citenamefont {Hatsugai}(1997)}]{hatsugai-1997}%
  \BibitemOpen
  \bibfield  {author} {\bibinfo {author} {\bibfnamefont {Y.}~\bibnamefont
  {Hatsugai}},\ }\bibfield  {title} {\bibinfo {title} {{Topological aspects of
  the quantum Hall effect}},\ }\href
  {https://doi.org/10.1088/0953-8984/9/12/003} {\bibfield  {journal} {\bibinfo
  {journal} {Journal of Physics: Condensed Matter}\ }\textbf {\bibinfo {volume}
  {9}},\ \bibinfo {pages} {2507} (\bibinfo {year} {1997})}\BibitemShut
  {NoStop}%
\bibitem [{\citenamefont {Thouless}\ \emph {et~al.}(1982)\citenamefont
  {Thouless}, \citenamefont {Kohmoto}, \citenamefont {Nightingale},\ and\
  \citenamefont {den Nijs}}]{den-nijs-1982}%
  \BibitemOpen
  \bibfield  {author} {\bibinfo {author} {\bibfnamefont {D.~J.}\ \bibnamefont
  {Thouless}}, \bibinfo {author} {\bibfnamefont {M.}~\bibnamefont {Kohmoto}},
  \bibinfo {author} {\bibfnamefont {M.~P.}\ \bibnamefont {Nightingale}},\ and\
  \bibinfo {author} {\bibfnamefont {M.}~\bibnamefont {den Nijs}},\ }\bibfield
  {title} {\bibinfo {title} {{Quantized Hall Conductance in a Two-Dimensional
  Periodic Potential}},\ }\href {https://doi.org/10.1103/PhysRevLett.49.405}
  {\bibfield  {journal} {\bibinfo  {journal} {Phys. Rev. Lett.}\ }\textbf
  {\bibinfo {volume} {49}},\ \bibinfo {pages} {405} (\bibinfo {year}
  {1982})}\BibitemShut {NoStop}%
\bibitem [{\citenamefont {Kohmoto}(1985)}]{kohmoto-1985}%
  \BibitemOpen
  \bibfield  {author} {\bibinfo {author} {\bibfnamefont {M.}~\bibnamefont
  {Kohmoto}},\ }\bibfield  {title} {\bibinfo {title} {{Topological invariant
  and the quantization of the Hall conductance}},\ }\href
  {https://doi.org/https://doi.org/10.1016/0003-4916(85)90148-4} {\bibfield
  {journal} {\bibinfo  {journal} {Annals of Physics}\ }\textbf {\bibinfo
  {volume} {160}},\ \bibinfo {pages} {343} (\bibinfo {year}
  {1985})}\BibitemShut {NoStop}%
\bibitem [{\citenamefont {Hatsugai}(1993)}]{yasuhiro-1993}%
  \BibitemOpen
  \bibfield  {author} {\bibinfo {author} {\bibfnamefont {Y.}~\bibnamefont
  {Hatsugai}},\ }\bibfield  {title} {\bibinfo {title} {{Chern number and edge
  states in the integer quantum Hall effect}},\ }\href
  {https://doi.org/10.1103/PhysRevLett.71.3697} {\bibfield  {journal} {\bibinfo
   {journal} {Phys. Rev. Lett.}\ }\textbf {\bibinfo {volume} {71}},\ \bibinfo
  {pages} {3697} (\bibinfo {year} {1993})}\BibitemShut {NoStop}%
\bibitem [{\citenamefont {Kane}\ and\ \citenamefont
  {Mele}(2005{\natexlab{a}})}]{mele-2005a}%
  \BibitemOpen
  \bibfield  {author} {\bibinfo {author} {\bibfnamefont {C.~L.}\ \bibnamefont
  {Kane}}\ and\ \bibinfo {author} {\bibfnamefont {E.~J.}\ \bibnamefont
  {Mele}},\ }\bibfield  {title} {\bibinfo {title} {{${Z}_{2}$ Topological Order
  and the Quantum Spin Hall Effect}},\ }\href
  {https://doi.org/10.1103/PhysRevLett.95.146802} {\bibfield  {journal}
  {\bibinfo  {journal} {Phys. Rev. Lett.}\ }\textbf {\bibinfo {volume} {95}},\
  \bibinfo {pages} {146802} (\bibinfo {year} {2005}{\natexlab{a}})}\BibitemShut
  {NoStop}%
\bibitem [{\citenamefont {Kane}\ and\ \citenamefont
  {Mele}(2005{\natexlab{b}})}]{mele-2005b}%
  \BibitemOpen
  \bibfield  {author} {\bibinfo {author} {\bibfnamefont {C.~L.}\ \bibnamefont
  {Kane}}\ and\ \bibinfo {author} {\bibfnamefont {E.~J.}\ \bibnamefont
  {Mele}},\ }\bibfield  {title} {\bibinfo {title} {Quantum spin hall effect in
  graphene},\ }\href {https://doi.org/10.1103/PhysRevLett.95.226801} {\bibfield
   {journal} {\bibinfo  {journal} {Phys. Rev. Lett.}\ }\textbf {\bibinfo
  {volume} {95}},\ \bibinfo {pages} {226801} (\bibinfo {year}
  {2005}{\natexlab{b}})}\BibitemShut {NoStop}%
\bibitem [{\citenamefont {Bernevig}\ and\ \citenamefont
  {Zhang}(2006)}]{shou-cheng-2006}%
  \BibitemOpen
  \bibfield  {author} {\bibinfo {author} {\bibfnamefont {B.~A.}\ \bibnamefont
  {Bernevig}}\ and\ \bibinfo {author} {\bibfnamefont {S.-C.}\ \bibnamefont
  {Zhang}},\ }\bibfield  {title} {\bibinfo {title} {Quantum spin hall effect},\
  }\href {https://doi.org/10.1103/PhysRevLett.96.106802} {\bibfield  {journal}
  {\bibinfo  {journal} {Phys. Rev. Lett.}\ }\textbf {\bibinfo {volume} {96}},\
  \bibinfo {pages} {106802} (\bibinfo {year} {2006})}\BibitemShut {NoStop}%
\bibitem [{\citenamefont {Lin}\ \emph {et~al.}(2009)\citenamefont {Lin},
  \citenamefont {Compton}, \citenamefont {Jim{\'e}nez-Garc{\'i}a},
  \citenamefont {Porto},\ and\ \citenamefont {Spielman}}]{spielman-2009}%
  \BibitemOpen
  \bibfield  {author} {\bibinfo {author} {\bibfnamefont {Y.-J.}\ \bibnamefont
  {Lin}}, \bibinfo {author} {\bibfnamefont {R.~L.}\ \bibnamefont {Compton}},
  \bibinfo {author} {\bibfnamefont {K.}~\bibnamefont {Jim{\'e}nez-Garc{\'i}a}},
  \bibinfo {author} {\bibfnamefont {J.~V.}\ \bibnamefont {Porto}},\ and\
  \bibinfo {author} {\bibfnamefont {I.~B.}\ \bibnamefont {Spielman}},\
  }\bibfield  {title} {\bibinfo {title} {{Synthetic magnetic fields for
  ultracold neutral atoms}},\ }\href {https://doi.org/10.1038/nature08609}
  {\bibfield  {journal} {\bibinfo  {journal} {Nature}\ }\textbf {\bibinfo
  {volume} {462}},\ \bibinfo {pages} {628} (\bibinfo {year}
  {2009})}\BibitemShut {NoStop}%
\bibitem [{\citenamefont {Aidelsburger}\ \emph {et~al.}(2013)\citenamefont
  {Aidelsburger}, \citenamefont {Atala}, \citenamefont {Lohse}, \citenamefont
  {Barreiro}, \citenamefont {Paredes},\ and\ \citenamefont
  {Bloch}}]{bloch-2013}%
  \BibitemOpen
  \bibfield  {author} {\bibinfo {author} {\bibfnamefont {M.}~\bibnamefont
  {Aidelsburger}}, \bibinfo {author} {\bibfnamefont {M.}~\bibnamefont {Atala}},
  \bibinfo {author} {\bibfnamefont {M.}~\bibnamefont {Lohse}}, \bibinfo
  {author} {\bibfnamefont {J.~T.}\ \bibnamefont {Barreiro}}, \bibinfo {author}
  {\bibfnamefont {B.}~\bibnamefont {Paredes}},\ and\ \bibinfo {author}
  {\bibfnamefont {I.}~\bibnamefont {Bloch}},\ }\bibfield  {title} {\bibinfo
  {title} {{Realization of the Hofstadter Hamiltonian with Ultracold Atoms in
  Optical Lattices}},\ }\href {https://doi.org/10.1103/PhysRevLett.111.185301}
  {\bibfield  {journal} {\bibinfo  {journal} {Phys. Rev. Lett.}\ }\textbf
  {\bibinfo {volume} {111}},\ \bibinfo {pages} {185301} (\bibinfo {year}
  {2013})}\BibitemShut {NoStop}%
\bibitem [{\citenamefont {Aidelsburger}\ \emph {et~al.}(2015)\citenamefont
  {Aidelsburger}, \citenamefont {Lohse}, \citenamefont {Schweizer},
  \citenamefont {Atala}, \citenamefont {Barreiro}, \citenamefont
  {Nascimb{\`e}ne}, \citenamefont {Cooper}, \citenamefont {Bloch},\ and\
  \citenamefont {Goldman}}]{goldman-2015}%
  \BibitemOpen
  \bibfield  {author} {\bibinfo {author} {\bibfnamefont {M.}~\bibnamefont
  {Aidelsburger}}, \bibinfo {author} {\bibfnamefont {M.}~\bibnamefont {Lohse}},
  \bibinfo {author} {\bibfnamefont {C.}~\bibnamefont {Schweizer}}, \bibinfo
  {author} {\bibfnamefont {M.}~\bibnamefont {Atala}}, \bibinfo {author}
  {\bibfnamefont {J.~T.}\ \bibnamefont {Barreiro}}, \bibinfo {author}
  {\bibfnamefont {S.}~\bibnamefont {Nascimb{\`e}ne}}, \bibinfo {author}
  {\bibfnamefont {N.~R.}\ \bibnamefont {Cooper}}, \bibinfo {author}
  {\bibfnamefont {I.}~\bibnamefont {Bloch}},\ and\ \bibinfo {author}
  {\bibfnamefont {N.}~\bibnamefont {Goldman}},\ }\bibfield  {title} {\bibinfo
  {title} {{Measuring the Chern number of Hofstadter bands with ultracold
  bosonic atoms}},\ }\href {https://doi.org/10.1038/nphys3171} {\bibfield
  {journal} {\bibinfo  {journal} {Nature Physics}\ }\textbf {\bibinfo {volume}
  {11}},\ \bibinfo {pages} {162} (\bibinfo {year} {2015})}\BibitemShut
  {NoStop}%
\bibitem [{\citenamefont {Lin}\ \emph {et~al.}(2011)\citenamefont {Lin},
  \citenamefont {Jim{\'e}nez-Garc{\'i}a},\ and\ \citenamefont
  {Spielman}}]{spielman-2011}%
  \BibitemOpen
  \bibfield  {author} {\bibinfo {author} {\bibfnamefont {Y.-J.}\ \bibnamefont
  {Lin}}, \bibinfo {author} {\bibfnamefont {K.}~\bibnamefont
  {Jim{\'e}nez-Garc{\'i}a}},\ and\ \bibinfo {author} {\bibfnamefont {I.~B.}\
  \bibnamefont {Spielman}},\ }\bibfield  {title} {\bibinfo {title}
  {{Spin--orbit-coupled Bose--Einstein condensates}},\ }\href
  {https://doi.org/10.1038/nature09887} {\bibfield  {journal} {\bibinfo
  {journal} {Nature}\ }\textbf {\bibinfo {volume} {471}},\ \bibinfo {pages}
  {83} (\bibinfo {year} {2011})}\BibitemShut {NoStop}%
\bibitem [{\citenamefont {Cheuk}\ \emph {et~al.}(2012)\citenamefont {Cheuk},
  \citenamefont {Sommer}, \citenamefont {Hadzibabic}, \citenamefont {Yefsah},
  \citenamefont {Bakr},\ and\ \citenamefont {Zwierlein}}]{zwierlein-2012}%
  \BibitemOpen
  \bibfield  {author} {\bibinfo {author} {\bibfnamefont {L.~W.}\ \bibnamefont
  {Cheuk}}, \bibinfo {author} {\bibfnamefont {A.~T.}\ \bibnamefont {Sommer}},
  \bibinfo {author} {\bibfnamefont {Z.}~\bibnamefont {Hadzibabic}}, \bibinfo
  {author} {\bibfnamefont {T.}~\bibnamefont {Yefsah}}, \bibinfo {author}
  {\bibfnamefont {W.~S.}\ \bibnamefont {Bakr}},\ and\ \bibinfo {author}
  {\bibfnamefont {M.~W.}\ \bibnamefont {Zwierlein}},\ }\bibfield  {title}
  {\bibinfo {title} {{Spin-Injection Spectroscopy of a Spin-Orbit Coupled Fermi
  Gas}},\ }\href {https://doi.org/10.1103/PhysRevLett.109.095302} {\bibfield
  {journal} {\bibinfo  {journal} {Phys. Rev. Lett.}\ }\textbf {\bibinfo
  {volume} {109}},\ \bibinfo {pages} {095302} (\bibinfo {year}
  {2012})}\BibitemShut {NoStop}%
\bibitem [{\citenamefont {Yau}\ and\ \citenamefont {S\'a~de
  Melo}(2019)}]{sa-de-melo-2019}%
  \BibitemOpen
  \bibfield  {author} {\bibinfo {author} {\bibfnamefont {M.~H.}\ \bibnamefont
  {Yau}}\ and\ \bibinfo {author} {\bibfnamefont {C.~A.~R.}\ \bibnamefont
  {S\'a~de Melo}},\ }\bibfield  {title} {\bibinfo {title} {{Chern-number
  spectrum of ultracold fermions in optical lattices tuned independently via
  artificial magnetic, Zeeman, and spin-orbit fields}},\ }\href
  {https://doi.org/10.1103/PhysRevA.99.043625} {\bibfield  {journal} {\bibinfo
  {journal} {Phys. Rev. A}\ }\textbf {\bibinfo {volume} {99}},\ \bibinfo
  {pages} {043625} (\bibinfo {year} {2019})}\BibitemShut {NoStop}%
\bibitem [{\citenamefont {Hasan}\ \emph {et~al.}(2022)\citenamefont {Hasan},
  \citenamefont {Madasu}, \citenamefont {Rathod}, \citenamefont {Kwong},\ and\
  \citenamefont {Wilkowski}}]{wilkowski-2022}%
  \BibitemOpen
  \bibfield  {author} {\bibinfo {author} {\bibfnamefont {M.}~\bibnamefont
  {Hasan}}, \bibinfo {author} {\bibfnamefont {C.~S.}\ \bibnamefont {Madasu}},
  \bibinfo {author} {\bibfnamefont {K.~D.}\ \bibnamefont {Rathod}}, \bibinfo
  {author} {\bibfnamefont {C.~C.}\ \bibnamefont {Kwong}},\ and\ \bibinfo
  {author} {\bibfnamefont {D.}~\bibnamefont {Wilkowski}},\ }\bibfield  {title}
  {\bibinfo {title} {{Evolution of an ultracold gas in a non-Abelian gauge
  field: finite temperature effect}},\ }\href
  {https://doi.org/10.1070/QEL18071} {\bibfield  {journal} {\bibinfo  {journal}
  {Quantum Electronics}\ }\textbf {\bibinfo {volume} {52}},\ \bibinfo {pages}
  {532} (\bibinfo {year} {2022})}\BibitemShut {NoStop}%
\bibitem [{\citenamefont {Alluf}\ and\ \citenamefont {S\'a~de
  Melo}(2025)}]{sa-de-melo-2025}%
  \BibitemOpen
  \bibfield  {author} {\bibinfo {author} {\bibfnamefont {B.}~\bibnamefont
  {Alluf}}\ and\ \bibinfo {author} {\bibfnamefont {C.~A.~R.}\ \bibnamefont
  {S\'a~de Melo}},\ }\bibfield  {title} {\bibinfo {title} {{Duality breaking,
  mobility edges, and the connection between topological Aubry-Andr\'e and
  quantum Hall insulators in atomic wires with fermions}},\ }\href
  {https://doi.org/10.1103/PhysRevA.111.023322} {\bibfield  {journal} {\bibinfo
   {journal} {Phys. Rev. A}\ }\textbf {\bibinfo {volume} {111}},\ \bibinfo
  {pages} {023322} (\bibinfo {year} {2025})}\BibitemShut {NoStop}%
\bibitem [{\citenamefont {Mardonov}\ \emph {et~al.}(2015)\citenamefont
  {Mardonov}, \citenamefont {Modugno},\ and\ \citenamefont
  {Sherman}}]{sherman-2015}%
  \BibitemOpen
  \bibfield  {author} {\bibinfo {author} {\bibfnamefont {S.}~\bibnamefont
  {Mardonov}}, \bibinfo {author} {\bibfnamefont {M.}~\bibnamefont {Modugno}},\
  and\ \bibinfo {author} {\bibfnamefont {E.~Y.}\ \bibnamefont {Sherman}},\
  }\bibfield  {title} {\bibinfo {title} {{Dynamics of Spin-Orbit Coupled
  Bose-Einstein Condensates in a Random Potential}},\ }\href
  {https://doi.org/10.1103/PhysRevLett.115.180402} {\bibfield  {journal}
  {\bibinfo  {journal} {Phys. Rev. Lett.}\ }\textbf {\bibinfo {volume} {115}},\
  \bibinfo {pages} {180402} (\bibinfo {year} {2015})}\BibitemShut {NoStop}%
\bibitem [{\citenamefont {Alluf}\ and\ \citenamefont {S\'a~de
  Melo}(2023)}]{sa-de-melo-2023}%
  \BibitemOpen
  \bibfield  {author} {\bibinfo {author} {\bibfnamefont {B.}~\bibnamefont
  {Alluf}}\ and\ \bibinfo {author} {\bibfnamefont {C.~A.~R.}\ \bibnamefont
  {S\'a~de Melo}},\ }\bibfield  {title} {\bibinfo {title} {{Controlling
  Anderson localization of a Bose-Einstein condensate via spin-orbit coupling
  and Rabi fields in bichromatic lattices}},\ }\href
  {https://doi.org/10.1103/PhysRevA.107.033312} {\bibfield  {journal} {\bibinfo
   {journal} {Phys. Rev. A}\ }\textbf {\bibinfo {volume} {107}},\ \bibinfo
  {pages} {033312} (\bibinfo {year} {2023})}\BibitemShut {NoStop}%
\bibitem [{\citenamefont {Fukuhara}\ \emph {et~al.}(2007)\citenamefont
  {Fukuhara}, \citenamefont {Takasu}, \citenamefont {Kumakura},\ and\
  \citenamefont {Takahashi}}]{takahashi-2007}%
  \BibitemOpen
  \bibfield  {author} {\bibinfo {author} {\bibfnamefont {T.}~\bibnamefont
  {Fukuhara}}, \bibinfo {author} {\bibfnamefont {Y.}~\bibnamefont {Takasu}},
  \bibinfo {author} {\bibfnamefont {M.}~\bibnamefont {Kumakura}},\ and\
  \bibinfo {author} {\bibfnamefont {Y.}~\bibnamefont {Takahashi}},\ }\bibfield
  {title} {\bibinfo {title} {{Degenerate Fermi Gases of Ytterbium}},\ }\href
  {https://doi.org/10.1103/PhysRevLett.98.030401} {\bibfield  {journal}
  {\bibinfo  {journal} {Phys. Rev. Lett.}\ }\textbf {\bibinfo {volume} {98}},\
  \bibinfo {pages} {030401} (\bibinfo {year} {2007})}\BibitemShut {NoStop}%
\bibitem [{\citenamefont {Cazalilla}\ \emph {et~al.}(2009)\citenamefont
  {Cazalilla}, \citenamefont {Ho},\ and\ \citenamefont {Ueda}}]{ueda-2009}%
  \BibitemOpen
  \bibfield  {author} {\bibinfo {author} {\bibfnamefont {M.~A.}\ \bibnamefont
  {Cazalilla}}, \bibinfo {author} {\bibfnamefont {A.~F.}\ \bibnamefont {Ho}},\
  and\ \bibinfo {author} {\bibfnamefont {M.}~\bibnamefont {Ueda}},\ }\bibfield
  {title} {\bibinfo {title} {{Ultracold gases of ytterbium: ferromagnetism and
  Mott states in an SU(6) Fermi system}},\ }\href
  {https://doi.org/10.1088/1367-2630/11/10/103033} {\bibfield  {journal}
  {\bibinfo  {journal} {New Journal of Physics}\ }\textbf {\bibinfo {volume}
  {11}},\ \bibinfo {pages} {103033} (\bibinfo {year} {2009})}\BibitemShut
  {NoStop}%
\bibitem [{\citenamefont {Taie}\ \emph {et~al.}(2010)\citenamefont {Taie},
  \citenamefont {Takasu}, \citenamefont {Sugawa}, \citenamefont {Yamazaki},
  \citenamefont {Tsujimoto}, \citenamefont {Murakami},\ and\ \citenamefont
  {Takahashi}}]{takahashi-2010}%
  \BibitemOpen
  \bibfield  {author} {\bibinfo {author} {\bibfnamefont {S.}~\bibnamefont
  {Taie}}, \bibinfo {author} {\bibfnamefont {Y.}~\bibnamefont {Takasu}},
  \bibinfo {author} {\bibfnamefont {S.}~\bibnamefont {Sugawa}}, \bibinfo
  {author} {\bibfnamefont {R.}~\bibnamefont {Yamazaki}}, \bibinfo {author}
  {\bibfnamefont {T.}~\bibnamefont {Tsujimoto}}, \bibinfo {author}
  {\bibfnamefont {R.}~\bibnamefont {Murakami}},\ and\ \bibinfo {author}
  {\bibfnamefont {Y.}~\bibnamefont {Takahashi}},\ }\bibfield  {title} {\bibinfo
  {title} {{Realization of a
  $\mathrm{SU}(2)\ifmmode\times\else\texttimes\fi{}\mathrm{SU}(6)$ System of
  Fermions in a Cold Atomic Gas}},\ }\href
  {https://doi.org/10.1103/PhysRevLett.105.190401} {\bibfield  {journal}
  {\bibinfo  {journal} {Phys. Rev. Lett.}\ }\textbf {\bibinfo {volume} {105}},\
  \bibinfo {pages} {190401} (\bibinfo {year} {2010})}\BibitemShut {NoStop}%
\bibitem [{\citenamefont {Taie}\ \emph {et~al.}(2012)\citenamefont {Taie},
  \citenamefont {Yamazaki}, \citenamefont {Sugawa},\ and\ \citenamefont
  {Takahashi}}]{takahashi-2012}%
  \BibitemOpen
  \bibfield  {author} {\bibinfo {author} {\bibfnamefont {S.}~\bibnamefont
  {Taie}}, \bibinfo {author} {\bibfnamefont {R.}~\bibnamefont {Yamazaki}},
  \bibinfo {author} {\bibfnamefont {S.}~\bibnamefont {Sugawa}},\ and\ \bibinfo
  {author} {\bibfnamefont {Y.}~\bibnamefont {Takahashi}},\ }\bibfield  {title}
  {\bibinfo {title} {{An SU(6) Mott insulator of an atomic Fermi gas realized
  by large-spin Pomeranchuk cooling}},\ }\href
  {https://doi.org/10.1038/nphys2430} {\bibfield  {journal} {\bibinfo
  {journal} {Nature Physics}\ }\textbf {\bibinfo {volume} {8}},\ \bibinfo
  {pages} {825} (\bibinfo {year} {2012})}\BibitemShut {NoStop}%
\bibitem [{\citenamefont {Pagano}\ \emph {et~al.}(2014)\citenamefont {Pagano},
  \citenamefont {Mancini}, \citenamefont {Cappellini}, \citenamefont
  {Lombardi}, \citenamefont {Sch{\"a}fer}, \citenamefont {Hu}, \citenamefont
  {Liu}, \citenamefont {Catani}, \citenamefont {Sias}, \citenamefont
  {Inguscio},\ and\ \citenamefont {Fallani}}]{fallani-2014}%
  \BibitemOpen
  \bibfield  {author} {\bibinfo {author} {\bibfnamefont {G.}~\bibnamefont
  {Pagano}}, \bibinfo {author} {\bibfnamefont {M.}~\bibnamefont {Mancini}},
  \bibinfo {author} {\bibfnamefont {G.}~\bibnamefont {Cappellini}}, \bibinfo
  {author} {\bibfnamefont {P.}~\bibnamefont {Lombardi}}, \bibinfo {author}
  {\bibfnamefont {F.}~\bibnamefont {Sch{\"a}fer}}, \bibinfo {author}
  {\bibfnamefont {H.}~\bibnamefont {Hu}}, \bibinfo {author} {\bibfnamefont
  {X.-J.}\ \bibnamefont {Liu}}, \bibinfo {author} {\bibfnamefont
  {J.}~\bibnamefont {Catani}}, \bibinfo {author} {\bibfnamefont
  {C.}~\bibnamefont {Sias}}, \bibinfo {author} {\bibfnamefont {M.}~\bibnamefont
  {Inguscio}},\ and\ \bibinfo {author} {\bibfnamefont {L.}~\bibnamefont
  {Fallani}},\ }\bibfield  {title} {\bibinfo {title} {{A one-dimensional liquid
  of fermions with tunable spin}},\ }\href {https://doi.org/10.1038/nphys2878}
  {\bibfield  {journal} {\bibinfo  {journal} {Nature Physics}\ }\textbf
  {\bibinfo {volume} {10}},\ \bibinfo {pages} {198} (\bibinfo {year}
  {2014})}\BibitemShut {NoStop}%
\bibitem [{\citenamefont {Hofrichter}\ \emph {et~al.}(2016)\citenamefont
  {Hofrichter}, \citenamefont {Riegger}, \citenamefont {Scazza}, \citenamefont
  {H\"ofer}, \citenamefont {Fernandes}, \citenamefont {Bloch},\ and\
  \citenamefont {F\"olling}}]{folling-2016}%
  \BibitemOpen
  \bibfield  {author} {\bibinfo {author} {\bibfnamefont {C.}~\bibnamefont
  {Hofrichter}}, \bibinfo {author} {\bibfnamefont {L.}~\bibnamefont {Riegger}},
  \bibinfo {author} {\bibfnamefont {F.}~\bibnamefont {Scazza}}, \bibinfo
  {author} {\bibfnamefont {M.}~\bibnamefont {H\"ofer}}, \bibinfo {author}
  {\bibfnamefont {D.~R.}\ \bibnamefont {Fernandes}}, \bibinfo {author}
  {\bibfnamefont {I.}~\bibnamefont {Bloch}},\ and\ \bibinfo {author}
  {\bibfnamefont {S.}~\bibnamefont {F\"olling}},\ }\bibfield  {title} {\bibinfo
  {title} {{Direct Probing of the Mott Crossover in the $\mathrm{SU}(N)$
  Fermi-Hubbard Model}},\ }\href {https://doi.org/10.1103/PhysRevX.6.021030}
  {\bibfield  {journal} {\bibinfo  {journal} {Phys. Rev. X}\ }\textbf {\bibinfo
  {volume} {6}},\ \bibinfo {pages} {021030} (\bibinfo {year}
  {2016})}\BibitemShut {NoStop}%
\bibitem [{\citenamefont {DeSalvo}\ \emph {et~al.}(2010)\citenamefont
  {DeSalvo}, \citenamefont {Yan}, \citenamefont {Mickelson}, \citenamefont
  {Martinez~de Escobar},\ and\ \citenamefont {Killian}}]{killian-2010}%
  \BibitemOpen
  \bibfield  {author} {\bibinfo {author} {\bibfnamefont {B.~J.}\ \bibnamefont
  {DeSalvo}}, \bibinfo {author} {\bibfnamefont {M.}~\bibnamefont {Yan}},
  \bibinfo {author} {\bibfnamefont {P.~G.}\ \bibnamefont {Mickelson}}, \bibinfo
  {author} {\bibfnamefont {Y.~N.}\ \bibnamefont {Martinez~de Escobar}},\ and\
  \bibinfo {author} {\bibfnamefont {T.~C.}\ \bibnamefont {Killian}},\
  }\bibfield  {title} {\bibinfo {title} {{Degenerate Fermi Gas of
  $^{87}\mathrm{Sr}$}},\ }\href
  {https://doi.org/10.1103/PhysRevLett.105.030402} {\bibfield  {journal}
  {\bibinfo  {journal} {Phys. Rev. Lett.}\ }\textbf {\bibinfo {volume} {105}},\
  \bibinfo {pages} {030402} (\bibinfo {year} {2010})}\BibitemShut {NoStop}%
\bibitem [{\citenamefont {Tey}\ \emph {et~al.}(2010)\citenamefont {Tey},
  \citenamefont {Stellmer}, \citenamefont {Grimm},\ and\ \citenamefont
  {Schreck}}]{schreck-2010}%
  \BibitemOpen
  \bibfield  {author} {\bibinfo {author} {\bibfnamefont {M.~K.}\ \bibnamefont
  {Tey}}, \bibinfo {author} {\bibfnamefont {S.}~\bibnamefont {Stellmer}},
  \bibinfo {author} {\bibfnamefont {R.}~\bibnamefont {Grimm}},\ and\ \bibinfo
  {author} {\bibfnamefont {F.}~\bibnamefont {Schreck}},\ }\bibfield  {title}
  {\bibinfo {title} {{Double-degenerate Bose-Fermi mixture of strontium}},\
  }\href {https://doi.org/10.1103/PhysRevA.82.011608} {\bibfield  {journal}
  {\bibinfo  {journal} {Phys. Rev. A}\ }\textbf {\bibinfo {volume} {82}},\
  \bibinfo {pages} {011608(R)} (\bibinfo {year} {2010})}\BibitemShut {NoStop}%
\bibitem [{\citenamefont {Stellmer}\ \emph {et~al.}(2011)\citenamefont
  {Stellmer}, \citenamefont {Grimm},\ and\ \citenamefont
  {Schreck}}]{schreck-2011}%
  \BibitemOpen
  \bibfield  {author} {\bibinfo {author} {\bibfnamefont {S.}~\bibnamefont
  {Stellmer}}, \bibinfo {author} {\bibfnamefont {R.}~\bibnamefont {Grimm}},\
  and\ \bibinfo {author} {\bibfnamefont {F.}~\bibnamefont {Schreck}},\
  }\bibfield  {title} {\bibinfo {title} {{Detection and manipulation of nuclear
  spin states in fermionic strontium}},\ }\href
  {https://doi.org/10.1103/PhysRevA.84.043611} {\bibfield  {journal} {\bibinfo
  {journal} {Phys. Rev. A}\ }\textbf {\bibinfo {volume} {84}},\ \bibinfo
  {pages} {043611} (\bibinfo {year} {2011})}\BibitemShut {NoStop}%
\bibitem [{\citenamefont {Stellmer}\ \emph {et~al.}(2014)\citenamefont
  {Stellmer}, \citenamefont {Schreck},\ and\ \citenamefont
  {Killian}}]{killian-2014}%
  \BibitemOpen
  \bibfield  {author} {\bibinfo {author} {\bibfnamefont {S.}~\bibnamefont
  {Stellmer}}, \bibinfo {author} {\bibfnamefont {F.}~\bibnamefont {Schreck}},\
  and\ \bibinfo {author} {\bibfnamefont {T.~C.}\ \bibnamefont {Killian}},\
  }\bibfield  {title} {\bibinfo {title} {Degenerate quantum gases of
  strontium},\ }in\ \href {https://doi.org/10.1142/9789814590174_0001} {\emph
  {\bibinfo {booktitle} {Annual Review of Cold Atoms and Molecules}}}\
  (\bibinfo  {publisher} {World Scientific},\ \bibinfo {address} {Singapore},\
  \bibinfo {year} {2014})\ Chap.~\bibinfo {chapter} {1}, pp.\ \bibinfo {pages}
  {1--80}\BibitemShut {NoStop}%
\bibitem [{\citenamefont {Taie}\ \emph {et~al.}(2022)\citenamefont {Taie},
  \citenamefont {Ibarra-Garc{\'i}a-Padilla}, \citenamefont {Nishizawa},
  \citenamefont {Takasu}, \citenamefont {Kuno}, \citenamefont {Wei},
  \citenamefont {Scalettar}, \citenamefont {Hazzard},\ and\ \citenamefont
  {Takahashi}}]{takahashi-2022}%
  \BibitemOpen
  \bibfield  {author} {\bibinfo {author} {\bibfnamefont {S.}~\bibnamefont
  {Taie}}, \bibinfo {author} {\bibfnamefont {E.}~\bibnamefont
  {Ibarra-Garc{\'i}a-Padilla}}, \bibinfo {author} {\bibfnamefont
  {N.}~\bibnamefont {Nishizawa}}, \bibinfo {author} {\bibfnamefont
  {Y.}~\bibnamefont {Takasu}}, \bibinfo {author} {\bibfnamefont
  {Y.}~\bibnamefont {Kuno}}, \bibinfo {author} {\bibfnamefont {H.-T.}\
  \bibnamefont {Wei}}, \bibinfo {author} {\bibfnamefont {R.~T.}\ \bibnamefont
  {Scalettar}}, \bibinfo {author} {\bibfnamefont {K.~R.~A.}\ \bibnamefont
  {Hazzard}},\ and\ \bibinfo {author} {\bibfnamefont {Y.}~\bibnamefont
  {Takahashi}},\ }\bibfield  {title} {\bibinfo {title} {{Observation of
  antiferromagnetic correlations in an ultracold SU(N) Hubbard model}},\ }\href
  {https://doi.org/10.1038/s41567-022-01725-6} {\bibfield  {journal} {\bibinfo
  {journal} {Nature Physics}\ }\textbf {\bibinfo {volume} {18}},\ \bibinfo
  {pages} {1356} (\bibinfo {year} {2022})}\BibitemShut {NoStop}%
\bibitem [{\citenamefont {Tusi}\ \emph {et~al.}(2022)\citenamefont {Tusi},
  \citenamefont {Franchi}, \citenamefont {Livi}, \citenamefont {Baumann},
  \citenamefont {Orenes}, \citenamefont {Del~Re}, \citenamefont {Barfknecht},
  \citenamefont {Zhou}, \citenamefont {Inguscio}, \citenamefont {Cappellini},
  \citenamefont {Capone}, \citenamefont {Catani},\ and\ \citenamefont
  {Fallani}}]{fallani-2022}%
  \BibitemOpen
  \bibfield  {author} {\bibinfo {author} {\bibfnamefont {D.}~\bibnamefont
  {Tusi}}, \bibinfo {author} {\bibfnamefont {L.}~\bibnamefont {Franchi}},
  \bibinfo {author} {\bibfnamefont {L.~F.}\ \bibnamefont {Livi}}, \bibinfo
  {author} {\bibfnamefont {K.}~\bibnamefont {Baumann}}, \bibinfo {author}
  {\bibfnamefont {D.~B.}\ \bibnamefont {Orenes}}, \bibinfo {author}
  {\bibfnamefont {L.}~\bibnamefont {Del~Re}}, \bibinfo {author} {\bibfnamefont
  {R.~E.}\ \bibnamefont {Barfknecht}}, \bibinfo {author} {\bibfnamefont
  {T.}~\bibnamefont {Zhou}}, \bibinfo {author} {\bibfnamefont {M.}~\bibnamefont
  {Inguscio}}, \bibinfo {author} {\bibfnamefont {G.}~\bibnamefont
  {Cappellini}}, \bibinfo {author} {\bibfnamefont {M.}~\bibnamefont {Capone}},
  \bibinfo {author} {\bibfnamefont {J.}~\bibnamefont {Catani}},\ and\ \bibinfo
  {author} {\bibfnamefont {L.}~\bibnamefont {Fallani}},\ }\bibfield  {title}
  {\bibinfo {title} {{Flavour-selective localization in interacting lattice
  fermions}},\ }\href {https://doi.org/10.1038/s41567-022-01726-5} {\bibfield
  {journal} {\bibinfo  {journal} {Nature Physics}\ }\textbf {\bibinfo {volume}
  {18}},\ \bibinfo {pages} {1201} (\bibinfo {year} {2022})}\BibitemShut
  {NoStop}%
\bibitem [{\citenamefont {Kurkcuoglu}\ and\ \citenamefont {S\'a~de
  Melo}(2018)}]{sa-de-melo-2019b}%
  \BibitemOpen
  \bibfield  {author} {\bibinfo {author} {\bibfnamefont {D.~M.}\ \bibnamefont
  {Kurkcuoglu}}\ and\ \bibinfo {author} {\bibfnamefont {C.~A.~R.}\ \bibnamefont
  {S\'a~de Melo}},\ }\bibfield  {title} {\bibinfo {title} {{Color superfluidity
  of neutral ultracold fermions in the presence of color-flip and color-orbit
  fields}},\ }\href {https://doi.org/10.1103/PhysRevA.97.023632} {\bibfield
  {journal} {\bibinfo  {journal} {Phys. Rev. A}\ }\textbf {\bibinfo {volume}
  {97}},\ \bibinfo {pages} {023632} (\bibinfo {year} {2018})}\BibitemShut
  {NoStop}%
\bibitem [{\citenamefont {Yau}\ and\ \citenamefont {S\'a~de
  Melo}(2022)}]{sa-de-melo-2022}%
  \BibitemOpen
  \bibfield  {author} {\bibinfo {author} {\bibfnamefont {M.~H.}\ \bibnamefont
  {Yau}}\ and\ \bibinfo {author} {\bibfnamefont {C.~A.~R.}\ \bibnamefont
  {S\'a~de Melo}},\ }\bibfield  {title} {\bibinfo {title} {{Quantum Hall
  response of SU(3) fermions}},\ }\href
  {https://doi.org/10.1103/PhysRevA.106.053313} {\bibfield  {journal} {\bibinfo
   {journal} {Phys. Rev. A}\ }\textbf {\bibinfo {volume} {106}},\ \bibinfo
  {pages} {053313} (\bibinfo {year} {2022})}\BibitemShut {NoStop}%
\bibitem [{\citenamefont {Schumacher}\ \emph {et~al.}(2026)\citenamefont
  {Schumacher}, \citenamefont {M{\"a}kinen}, \citenamefont {Ji}, \citenamefont
  {T.~Assump{\c{c}}{\~a}o}, \citenamefont {Chen}, \citenamefont {Huang},
  \citenamefont {Vivanco},\ and\ \citenamefont {Navon}}]{navon-2026}%
  \BibitemOpen
  \bibfield  {author} {\bibinfo {author} {\bibfnamefont {G.~L.}\ \bibnamefont
  {Schumacher}}, \bibinfo {author} {\bibfnamefont {J.~T.}\ \bibnamefont
  {M{\"a}kinen}}, \bibinfo {author} {\bibfnamefont {Y.}~\bibnamefont {Ji}},
  \bibinfo {author} {\bibfnamefont {G.~G.}\ \bibnamefont
  {T.~Assump{\c{c}}{\~a}o}}, \bibinfo {author} {\bibfnamefont {J.}~\bibnamefont
  {Chen}}, \bibinfo {author} {\bibfnamefont {S.}~\bibnamefont {Huang}},
  \bibinfo {author} {\bibfnamefont {F.~J.}\ \bibnamefont {Vivanco}},\ and\
  \bibinfo {author} {\bibfnamefont {N.}~\bibnamefont {Navon}},\ }\bibfield
  {title} {\bibinfo {title} {{Observation of anomalous decay of a polarized
  three-component Fermi gas}},\ }\href
  {https://doi.org/10.1038/s41467-025-65183-3} {\bibfield  {journal} {\bibinfo
  {journal} {Nature Communications}\ }\textbf {\bibinfo {volume} {17}},\
  \bibinfo {pages} {174} (\bibinfo {year} {2026})}\BibitemShut {NoStop}%
\bibitem [{\citenamefont {Mongkolkiattichai}\ \emph {et~al.}(2025)\citenamefont
  {Mongkolkiattichai}, \citenamefont {Liu}, \citenamefont {Dasgupta},
  \citenamefont {Hazzard},\ and\ \citenamefont {Schauss}}]{schauss-2025}%
  \BibitemOpen
  \bibfield  {author} {\bibinfo {author} {\bibfnamefont {J.}~\bibnamefont
  {Mongkolkiattichai}}, \bibinfo {author} {\bibfnamefont {L.}~\bibnamefont
  {Liu}}, \bibinfo {author} {\bibfnamefont {S.}~\bibnamefont {Dasgupta}},
  \bibinfo {author} {\bibfnamefont {K.~R.~A.}\ \bibnamefont {Hazzard}},\ and\
  \bibinfo {author} {\bibfnamefont {P.}~\bibnamefont {Schauss}},\ }\bibfield
  {title} {\bibinfo {title} {{Quantum gas microscopy of three-flavor Hubbard
  systems}},\ }\href {https://arxiv.org/abs/2503.05687} {\bibfield  {journal}
  {\bibinfo  {journal} {arXiv:2503.05687}\ } (\bibinfo {year}
  {2025})}\BibitemShut {NoStop}%
\bibitem [{\citenamefont {Madasu}\ \emph {et~al.}(2025)\citenamefont {Madasu},
  \citenamefont {Mitra}, \citenamefont {Gabardos}, \citenamefont {Rathod},
  \citenamefont {Zanon-Willette}, \citenamefont {Miniatura}, \citenamefont
  {Chevy}, \citenamefont {Kwong},\ and\ \citenamefont
  {Wilkowski}}]{wilkowski-2025}%
  \BibitemOpen
  \bibfield  {author} {\bibinfo {author} {\bibfnamefont {C.~S.}\ \bibnamefont
  {Madasu}}, \bibinfo {author} {\bibfnamefont {C.}~\bibnamefont {Mitra}},
  \bibinfo {author} {\bibfnamefont {L.}~\bibnamefont {Gabardos}}, \bibinfo
  {author} {\bibfnamefont {K.~D.}\ \bibnamefont {Rathod}}, \bibinfo {author}
  {\bibfnamefont {T.}~\bibnamefont {Zanon-Willette}}, \bibinfo {author}
  {\bibfnamefont {C.}~\bibnamefont {Miniatura}}, \bibinfo {author}
  {\bibfnamefont {F.}~\bibnamefont {Chevy}}, \bibinfo {author} {\bibfnamefont
  {C.~C.}\ \bibnamefont {Kwong}},\ and\ \bibinfo {author} {\bibfnamefont
  {D.}~\bibnamefont {Wilkowski}},\ }\bibfield  {title} {\bibinfo {title}
  {{Experimental realization of a SU(3) color-orbit coupling in an ultracold
  gas}},\ }\href {https://doi.org/10.1038/s41467-025-63142-6} {\bibfield
  {journal} {\bibinfo  {journal} {Nature Communications}\ }\textbf {\bibinfo
  {volume} {16}},\ \bibinfo {pages} {8448} (\bibinfo {year}
  {2025})}\BibitemShut {NoStop}%
\bibitem [{\citenamefont {Zhang}\ and\ \citenamefont {S\'a~de
  Melo}(2025)}]{sa-de-melo-2025b}%
  \BibitemOpen
  \bibfield  {author} {\bibinfo {author} {\bibfnamefont {X.}~\bibnamefont
  {Zhang}}\ and\ \bibinfo {author} {\bibfnamefont {C.~A.~R.}\ \bibnamefont
  {S\'a~de Melo}},\ }\bibfield  {title} {\bibinfo {title} {{Effects of
  spin–orbit coupling and Rabi fields in Tomonaga–Luttinger liquids:
  current status and open questions}},\ }\href
  {https://doi.org/10.5802/crphys.254} {\bibfield  {journal} {\bibinfo
  {journal} {Comptes Rendus Physique}\ }\textbf {\bibinfo {volume} {26}},\
  \bibinfo {pages} {483} (\bibinfo {year} {2025})}\BibitemShut {NoStop}%
\bibitem [{\citenamefont {Aubry}\ and\ \citenamefont
  {Andr{\'e}}(1980)}]{andre-1980}%
  \BibitemOpen
  \bibfield  {author} {\bibinfo {author} {\bibfnamefont {S.}~\bibnamefont
  {Aubry}}\ and\ \bibinfo {author} {\bibfnamefont {G.}~\bibnamefont
  {Andr{\'e}}},\ }\bibfield  {title} {\bibinfo {title} {{Analyticity breaking
  and Anderson localization in incommensurate lattices}},\ }\href@noop {}
  {\bibfield  {journal} {\bibinfo  {journal} {Ann. Israel Phys. Soc}\ }\textbf
  {\bibinfo {volume} {3}},\ \bibinfo {pages} {133} (\bibinfo {year}
  {1980})}\BibitemShut {NoStop}%
\bibitem [{\citenamefont {Yau}\ and\ \citenamefont {Sá~de
  Melo}(2021)}]{sa-de-melo-2021}%
  \BibitemOpen
  \bibfield  {author} {\bibinfo {author} {\bibfnamefont {M.~H.}\ \bibnamefont
  {Yau}}\ and\ \bibinfo {author} {\bibfnamefont {C.~A.~R.}\ \bibnamefont
  {Sá~de Melo}},\ }\bibfield  {title} {\bibinfo {title} {{SU(3) vs. SU(2)
  fermions in optical lattices: Color-Hall vs. spin-Hall topological
  insulators}},\ }\href {https://doi.org/10.1209/0295-5075/135/16001}
  {\bibfield  {journal} {\bibinfo  {journal} {Europhysics Letters}\ }\textbf
  {\bibinfo {volume} {135}},\ \bibinfo {pages} {16001} (\bibinfo {year}
  {2021})}\BibitemShut {NoStop}%
\bibitem [{\citenamefont {Yao}\ \emph {et~al.}(2019)\citenamefont {Yao},
  \citenamefont {Khoudli}, \citenamefont {Bresque},\ and\ \citenamefont
  {Sanchez-Palencia}}]{sanchez-palencia-2019}%
  \BibitemOpen
  \bibfield  {author} {\bibinfo {author} {\bibfnamefont {H.}~\bibnamefont
  {Yao}}, \bibinfo {author} {\bibfnamefont {A.}~\bibnamefont {Khoudli}},
  \bibinfo {author} {\bibfnamefont {L.}~\bibnamefont {Bresque}},\ and\ \bibinfo
  {author} {\bibfnamefont {L.}~\bibnamefont {Sanchez-Palencia}},\ }\bibfield
  {title} {\bibinfo {title} {{Critical Behavior and Fractality in Shallow
  One-Dimensional Quasiperiodic Potentials}},\ }\href
  {https://doi.org/10.1103/PhysRevLett.123.070405} {\bibfield  {journal}
  {\bibinfo  {journal} {Phys. Rev. Lett.}\ }\textbf {\bibinfo {volume} {123}},\
  \bibinfo {pages} {070405} (\bibinfo {year} {2019})}\BibitemShut {NoStop}%
\bibitem [{\citenamefont {Yau}\ and\ \citenamefont {S\'a~de
  Melo}(2021)}]{sa-de-melo-2021b}%
  \BibitemOpen
  \bibfield  {author} {\bibinfo {author} {\bibfnamefont {M.~H.}\ \bibnamefont
  {Yau}}\ and\ \bibinfo {author} {\bibfnamefont {C.~A.~R.}\ \bibnamefont
  {S\'a~de Melo}},\ }\bibfield  {title} {\bibinfo {title} {{Eigenspectrum,
  Chern numbers and phase diagrams of ultracold color-orbit-coupled SU(3)
  fermions in optical lattices}},\ }\href
  {https://doi.org/10.1103/PhysRevA.103.043302} {\bibfield  {journal} {\bibinfo
   {journal} {Phys. Rev. A}\ }\textbf {\bibinfo {volume} {103}},\ \bibinfo
  {pages} {043302} (\bibinfo {year} {2021})}\BibitemShut {NoStop}%
\bibitem [{\citenamefont {Hofstadter}(1976)}]{hofstadter-1976}%
  \BibitemOpen
  \bibfield  {author} {\bibinfo {author} {\bibfnamefont {D.~R.}\ \bibnamefont
  {Hofstadter}},\ }\bibfield  {title} {\bibinfo {title} {{Energy levels and
  wave functions of Bloch electrons in rational and irrational magnetic
  fields}},\ }\href {https://doi.org/10.1103/PhysRevB.14.2239} {\bibfield
  {journal} {\bibinfo  {journal} {Phys. Rev. B}\ }\textbf {\bibinfo {volume}
  {14}},\ \bibinfo {pages} {2239} (\bibinfo {year} {1976})}\BibitemShut
  {NoStop}%
\bibitem [{\citenamefont {Claro}\ and\ \citenamefont
  {Wannier}(1979)}]{wannier-1979}%
  \BibitemOpen
  \bibfield  {author} {\bibinfo {author} {\bibfnamefont {F.~H.}\ \bibnamefont
  {Claro}}\ and\ \bibinfo {author} {\bibfnamefont {G.~H.}\ \bibnamefont
  {Wannier}},\ }\bibfield  {title} {\bibinfo {title} {{Magnetic subband
  structure of electrons in hexagonal lattices}},\ }\href
  {https://doi.org/10.1103/PhysRevB.19.6068} {\bibfield  {journal} {\bibinfo
  {journal} {Phys. Rev. B}\ }\textbf {\bibinfo {volume} {19}},\ \bibinfo
  {pages} {6068} (\bibinfo {year} {1979})}\BibitemShut {NoStop}%
\end{thebibliography}%

\end{document}